\documentclass[preprint,3p,10pt]{elsarticle}
\usepackage[utf8]{inputenc}

\usepackage[english,francais]{babel}

\journal{CR Physique}

\usepackage{xparse}
\usepackage{xspace}
\usepackage{xargs}
\usepackage{ifthen}

\usepackage[x11names,svgnames]{xcolor}

\usepackage[T1]{fontenc}
\usepackage{csquotes}
\usepackage[kerning=true,babel=true]{microtype} % babel=true

\usepackage[bitstream-charter]{mathdesign}

\usepackage{amsmath}
\usepackage{stmaryrd}
\usepackage{mathtools}
\usepackage{dsfont}
\makeatletter
\DeclareFontFamily{OMX}{MnSymbolE}{}
\DeclareSymbolFont{MnLargeSymbols}{OMX}{MnSymbolE}{m}{n}
\SetSymbolFont{MnLargeSymbols}{bold}{OMX}{MnSymbolE}{b}{n}
\DeclareFontShape{OMX}{MnSymbolE}{m}{n}{
    <-6>  MnSymbolE5
   <6-7>  MnSymbolE6
   <7-8>  MnSymbolE7
   <8-9>  MnSymbolE8
   <9-10> MnSymbolE9
  <10-12> MnSymbolE10
  <12->   MnSymbolE12
}{}
\DeclareFontShape{OMX}{MnSymbolE}{b}{n}{
    <-6>  MnSymbolE-Bold5
   <6-7>  MnSymbolE-Bold6
   <7-8>  MnSymbolE-Bold7
   <8-9>  MnSymbolE-Bold8
   <9-10> MnSymbolE-Bold9
  <10-12> MnSymbolE-Bold10
  <12->   MnSymbolE-Bold12
}{}

\let\llangle\@undefined
\let\rrangle\@undefined
\DeclareMathDelimiter{\llangle}{\mathopen}%
                     {MnLargeSymbols}{'164}{MnLargeSymbols}{'164}
\DeclareMathDelimiter{\rrangle}{\mathclose}%
                     {MnLargeSymbols}{'171}{MnLargeSymbols}{'171}
\makeatother

\usepackage{braket}
\usepackage{mathtools}
\usepackage{eucal}
\usepackage{upgreek} 

\usepackage[version=3]{mhchem}
\usepackage{graphicx}

\usepackage[unicode]{hyperref}
\hypersetup{
    colorlinks=true,
}

\newcommandx{\Iverson}[1]{\ensuremath{\left[ #1 \right] }}

\let\Cross\times %\wedge
\let\Tensor\otimes

\newcommand\RealField{\mathbb{R}}
\newcommand\ComplexField{\mathbb{C}}
\newcommand\IntegerRing{\mathbb{Z}}
\newcommandx\PermutationGroup[1]{\ensuremath{\mathfrak{S}_{#1}}}
\newcommand\CyclicGroup[1]{\ensuremath{\mathbb{Z}_{#1}}}
\newcommandx\GeneralLinearGroup[2][2={}]{
\ifthenelse{\equal{#2}{}}{
\ensuremath{\text{GL}(#1)}
}{
\ensuremath{\text{GL}(#1,#2)}
}\xspace
}
\newcommandx\SpecialLinearGroup[2][2={}]{
\ifthenelse{\equal{#2}{}}{
\ensuremath{\text{SL}(#1)}
}{
\ensuremath{\text{SL}(#1,#2)}
}\xspace
}
\newcommand\UnitaryGroup[1]{\ensuremath{\text{U}(#1)}\xspace}
\newcommand\SpecialUnitaryGroup[1]{\ensuremath{\text{SU}(#1)}\xspace}

\newcommandx\ContinuityClass[3][2={},3={}]{
\ifthenelse{\equal{#2}{}}{
    \ensuremath{\mathcal{C}^{#1}}
  }{
    \ifthenelse{\equal{#3}{}}{
      \ensuremath{\mathcal{C}^{#1}(#2)}
    }{
      \ensuremath{\mathcal{C}^{#1}(#2,#3)}
  }
}\xspace
}
\newcommandx\IntegerPart[1]{\ensuremath{\text{E}\left[ #1 \right]}}
\newcommand\Abs[1]{\ensuremath{\left| {#1} \right|}}
\newcommand\dd{\xspace\ensuremath{\text{d}}\xspace}
\newcommand\ee{\ensuremath{\text{e}}}
\newcommand\IdentityMatrix{\ensuremath{\mathds{1}}}
\newcommandx\PauliMatrix[2][1={\sigma}]{\xspace\ensuremath{
\ifthenelse{\equal{#2}{}}{
#1
}{
#1_{#2}
}
}\xspace}
\newcommandx\ExteriorAlgebra[2][1={}]{\xspace\ensuremath{
\ifthenelse{\equal{#1}{}}{
\Lambda #2
}{
\Lambda^{#1} #2
}
}\xspace}
\newcommand\Transpose[1]{\ensuremath{{#1}^{\text{T}}}}
\DeclareMathOperator{\Tr}{Tr}
\newcommandx\Norm[1]{\ensuremath{\left\lVert #1 \right\rVert}}
\newcommandx\norm[1]{\ensuremath{\lVert #1 \rVert}}

\newcommand\ii{\ensuremath{\text{i}}}
\newcommand\Conjugate[1]{\ensuremath{{#1}^{\star}}}
\newcommand\Adjoint[1]{\ensuremath{{#1}^{\dag}}}
\newcommand\Commutator[2]{\ensuremath{\left[ #1 , #2 \right]}}
\newcommand\commutator[2]{\ensuremath{[ #1 , #2 ]}}
\newcommand\anticommutator[2]{\ensuremath{\{ #1 , #2 \}}}
\newcommand\Op[1]{\ensuremath{\hat{#1}}}
\newcommand\PlanckConstantReduced{\ensuremath{\hbar}}

\newcommand\ElementaryCharge{\ensuremath{\text{e}}}

\DeclareMathOperator{\Degree}{deg}

\newcommand{\except}[1]{\ensuremath{\widehat{#1}}}

\newcommand{\Torus}[1]{\ensuremath{\mathbb{T}^{#1}}}
\newcommand{\Sphere}[1]{\ensuremath{\mathbb{S}^{#1}}} % TODO mettre ça partout

\newcommand{\TR}{\ensuremath{\Theta}}

\newcommand{\trb}{\ensuremath{\vartheta}}

\newcommand{\Parity}{\ensuremath{\mathcal{P}}}

\newcommand{\HighSymmetryPoints}{\ensuremath{\Lambda}}

\newcommand{\HighSymmetryPoint}{\ensuremath{\lambda}}

\newcommand{\ExteriorPower}[1]{\ensuremath{\Uplambda^{#1}}}

\newcommand{\QuaternionRing}{\ensuremath{\xspace\mathbb{H}\xspace}}
\newcommand{\jj}{\ensuremath{\text{j}}}
\newcommand{\kk}{\ensuremath{\text{k}}}

\newcommand{\ComplexConjugate}{\ensuremath{\mathcal{K}}}

\newcommand\SpinPauliMatrix[1]{\PauliMatrix[s]{#1}}
\newcommand\BandPauliMatrix[1]{\PauliMatrix[\sigma]{#1}}

\newcommand{\FirstQuantizedHamiltonian}{\ensuremath{\mathcal{H}}}

\newcommand{\Ztwo}{\texorpdfstring{\ensuremath{\CyclicGroup{2}}}{Z\texttwoinferior}\xspace}

\newcommand{\vect}[1]{\ensuremath{\vec{#1}}}

\newcommand{\modTwo}{\ensuremath{\text{(mod. $2$)}}}

\newcommand{\KMTRMatrix}{\ensuremath{\xspace m \xspace}} % KM
\newcommand{\SewingMatrix}{\ensuremath{\xspace w \xspace}} % FK

\newcommand{\ZtwoInvariantModTwo}{\ensuremath{\xspace \nu \xspace}}

\newcommand{\EBZ}{\ensuremath{\text{EBZ}}}
\newcommand{\PTRC}{\ensuremath{I}}

\newcommand{\FilledBandsBundle}{\ensuremath{\mathcal{V}}}
\newcommand{\FilledBandsFiber}[1]{\ensuremath{\mathcal{V}_{#1}}}
\newcommand{\KroneckerDelta}{\ensuremath{\delta}}

\newcommand{\swm}{\SewingMatrix}
\newcommand{\swf}{\ensuremath{f}}

\NewDocumentCommand{\GammaMatrix}{mg}{
\IfNoValueTF{#2}{
\Gamma_{#1}
}{
\Gamma_{#1 #2}
}
}

\NewDocumentCommand{\TRIClosedCurve}{O{} m}{\ensuremath{\mathcal{C}_{#2}^{#1}}}

\DeclareMathOperator{\sign}{sign}
\DeclareMathOperator{\Pf}{Pf}

\newcommand{\ZtwoInvariantIPKM}{\ensuremath{\Delta}}

\newcommand{\AllBandsFiber}[1]{\mathcal{H}_{#1}}

\begin{document}
\hypersetup{
    citecolor=PaleGreen4!80!black,
    linkcolor=DarkRed, 
    urlcolor=DarkSeaGreen4!90!black,
}

%  You can place here the title of the dossier, if you know it,
%     firstly in English, then in French
\centerline{Topological insulators/Isolants topologiques}
\begin{frontmatter}

% Title, authors and addresses

% use the thanksref command within \title, \author or \address for footnotes;
% use the ead command for the email address,
% and the form \ead[url] for the home page:
% \title{Title\thanksref{label1}}
% \thanks[label1]{}
% \author{Name\thanksref{label2}}
% \ead{email address}
% \ead[url]{home page}
% \thanks[label2]{}
% \address{Address\thanksref{label3}}
% \thanks[label3]{}
\selectlanguage{english}
\title{An introduction to topological insulators\\ {\it Introduction aux isolants topologiques}}

% use optional labels to link authors explicitly to addresses:
% \author[label1,label2]{}
% \address[label1]{}
% \address[label2]{}
% If all authors are at the same address, the [label1] can be suppressed

\selectlanguage{english}
\author[ENS Lyon]{Michel Fruchart}
\ead{michel.fruchart@ens-lyon.fr}
\author[ENS Lyon]{David Carpentier}
\ead{david.carpentier@ens-lyon.fr}
%\author[ENS Lyon]{Krzysztof Gawędzki}
%\ead{krysztof.gawedzki@ens-lyon.fr}

\address[ENS Lyon]{Laboratoire de physique, École normale supérieure de Lyon (UMR CNRS 5672), 46, allée d'Italie,  69007 Lyon, France}

\begin{abstract}
Electronic bands in crystals are described by an ensemble of Bloch wave functions indexed by momenta defined in the first Brillouin Zone, and their associated energies. In an insulator, an energy gap around the chemical potential separates valence bands from conduction bands. The ensemble of valence bands is then a well defined object, which can possess non-trivial or twisted  topological properties. In the case of a twisted topology, the insulator is called a topological insulator. We introduce this notion of topological order in insulators as an obstruction to define the Bloch wave functions over the whole 
Brillouin Zone using a single phase convention. Several simple historical models displaying a topological order in dimension two are considered. 
Various expressions of the corresponding topological index are finally discussed. 

\vskip 1\baselineskip

\selectlanguage{francais}
\noindent
\textbf{R\'esum\'e}
\vskip 0.5\baselineskip

\noindent
Les bandes électroniques dans un cristal sont définies par un ensemble de fonctions d'onde de Bloch dépendant du moment défini dans la première zone de Brillouin, ainsi que des énergies associées. Dans un isolant, les bandes de valence sont séparées des bandes de conduction par un gap en énergie. L'ensemble des bandes de valence est alors un objet bien défini, qui peut en particulier posséder une topologie non triviale. Lorsque cela se produit, l'isolant correspondant est appelé isolant topologique. 
Nous introduisons cette notion d'ordre topologique d'une bande comme une obstruction à la définition des fonctions d'ondes de Bloch à l'aide d'une convention de phase unique. Plusieurs modèles simples d'isolants topologiques en dimension deux sont considérés. 
Différentes expressions des indices topologiques correspondants sont finalement discutées. 
\end{abstract}

% Ordre Topologique; Isolant; Th\'eorie des Bandes; Fibr\'e Vectoriel; Renversement du Temps
\begin{keyword}
%% keywords here, in the form: keyword \sep keyword
\selectlanguage{english}
topological insulator \sep 
%band theory \sep 
topological band theory \sep 
%quantum Hall effect \sep 
quantum anomalous Hall effect \sep 
quantum spin Hall effect \sep 
Chern insulator \sep 
Kane--Mele insulator

\vskip 0.5\baselineskip

\selectlanguage{francais}
{\it Mots-clés:} \hspace{1ex}
isolant topologique \sep 
%théorie des bandes \sep 
théorie des bandes topologique \sep 
%effet Hall quantique \sep 
effet Hall quantique anomal \sep 
effet Hall quantique de spin \sep 
isolant de Chern \sep 
isolant de Kane--Mele

%% MSC codes here, in the form: \MSC code \sep code
%% or \MSC[2008] code \sep code (2000 is the default)

\end{keyword}

\end{frontmatter}

\selectlanguage{english}
% main text

%%%%%
\section{Introduction}

Topological insulators are phases of matter characterized by an order of a new kind, which is not fit into the standard symmetry breaking paradigm. 
Instead these new phases are described by a \emph{global} quantity which does not depend on the details of the system - a so-called topological order. More precisely, their ensemble of  valence bands possess a non-standard topological property. 
  A band insulator is a material which has a well-defined set of valence bands separated by an energy gap from a well-defined set of conduction bands. The  object of interest in the study of topological order in insulators is the ensemble of valence bands, which is   
 unambiguously well defined for an insulator.
The question underlying the topological classification of insulators is whether all insulating phases are equivalent to each other, 
{\it i.e.} whether their ensemble of valence bands can be continuously transformed into each other  without closing the gap. 
Topological insulators correspond to insulating materials whose valence bands possess non-standard topological properties. 
  Related to their classification is the determination of topological indices which will differentiate standard insulators from the different types of topological insulators. 
  A canonical example of such a topological index is the Euler--Poincaré characteristic of a two-dimensional manifold \cite{Nakahara}.  This index 
 counts the number of \enquote{holes} in the manifold. Two manifolds with the same Euler characteristic can be continuously  deformed into each other, which is not possible for manifolds with different Euler characteristics. 
 
 The existence of topological order in an insulator induces unique characteristic experimental signatures. 
 The most universal and remarkable consequence of a nontrivial bulk topology is the existence of gapless edge or surface states; in other words, the surface of the
  topological insulator is necessarily metallic. 
 An informal argument explaining those surface states is as follows. The vacuum as well as most conventional insulating crystals are topologically trivial. At the interface between such a standard insulator and a topological insulator, it is not possible for the \enquote{band structure} to interpolate continuously between a topological insulator and the vacuum without closing the gap. This
 forces the gap to close at this interface leading to metallic states of topological origin. 
 
This kind of topological phase ordering first arose in condensed matter in the context of the integer quantum Hall effect. This phase, discovered in 1980
by Klaus von Klitzing {\it et al.} \cite{KlitzingDordaPepper1980}, is reached when electrons trapped in a two-dimensional interface between semi-conductors are 
submitted to a strong transverse magnetic field. Quantized plateaux appear for the Hall conductivity while the longitudinal 
resistance simultaneously vanishes \cite{Doucot:2004}.  In the bulk of the sample, the electronic states are distributed in Landau levels with a large gap between them. The quantization of the Hall conductivity can be attributed  
 within standard linear response theory to a topological property of these bulk Landau levels, the so-called first Chern 
number of the bands located below the chemical potential \cite{ThoulessKohmotoNightingaleNijs1982}. From this point of view the robustness of the phase manifested 
in the high precision of the Hall conductivity plateau is an expression of the topological nature of the related order, which 
by definition is insensitive to perturbations. The existence of robust edge states is another manifestation of this topological ordering. The quantized Hall conductivity can be alternatively accounted for by the ballistic transport properties of the edge states. 

Note that in the initial work of Thouless {\it et al.} \cite{ThoulessKohmotoNightingaleNijs1982}, this topological ordering was described as a property of electronic Bloch bands of electrons on a lattice, and was only generalized later to free electrons on a planar interface. The topological property of the ensemble of Bloch states of a valence band can be 
inferred by the explicit determination of these Bloch states. In a non-trivial or twisted insulator, one faces an impossibility or
{\it obstruction} to define electronic Bloch states over the whole band using a single phase convention: at least two different phase conventions are required, as opposed to the usual case. This obstruction is a direct manifestation of the non-trivial topology or twist of the corresponding  band. 
 It was realized in 1988 by D. Haldane \cite{Haldane88} that while this type of order was specific to two dimensional insulators, it did not 
 require a strong magnetic field, but only time reversal symmetry breaking. 
This author considered a model of electrons on a bipartite lattice (graphene), with time-reversal symmetry broken explicitly but 
without any net magnetic flux through the lattice. The phase diagram consists then of 
three insulating phases, {\it i.e.} with a finite gap separating the conduction from the valence bands. These insulators only differs
by their topological property, quantized by a Chern number. The analogous phases of matter are now denoted Chern topological 
insulators, or anomalous quantum Hall effect. 
 Such a phase was recently discovered experimentally \cite{Chang:2013}. 

 Within the field of topological characterization of insulators a breakthrough occurred with the seminal work of C. 
Kane and G. Mele \cite{KaneMele2005,KaneMele2005B}. 
These authors considered the effect of a strong spin--orbit interaction on electronic bands of graphene. 
They discovered that in such a two dimensional system where the spin of electrons cannot be neglected in determining the band structure, the constraints imposed by time-reversal symmetry could lead to a new topological order and  
associated metallic edge states of a new kind. 
 While the quantum Hall effect arises in electronic system without any symmetry and is characterized by a Chern number, 
 this new topological phase is possible only in presence of time-reversal symmetry, and is characterized  
 by a new $\mathbb{Z}_2$ index. It was called a quantum spin Hall phase. 
This discovery triggered a huge number of theoretical and experimental works on the topological 
properties of time reversal symmetric spin-dependent valence bands and the associated surface states and physical signatures. 
Soon after the initial Kane and Mele papers, A. Bernevig, T. Hughes and S.C. Zhang proposed a realistic realization of this phase in
 HgTe quantum wells \cite{BHZ2006}. They identified a possible mechanism for the appearance of this $\mathbb{Z}_2$  topological order through  the inversion of order
  of bulk bands around one point in the Brillouin zone. 
    This phase was discovered experimentally in the group of L. Molenkamp who conducted two-terminal and 
  multi-probe transport experiments to demonstrate the 
 existence of the edge states associated with the $\mathbb{Z}_2$ order \cite{Konig2007,Roth:2009}. 
 
 In 2007, three theoretical groups extended the 
 expression of the $\mathbb{Z}_2$  topological index to three dimensions: it was then realized that three dimensional insulating materials and not
  only quasi-two dimensional systems could display a topological order \cite{FuKaneMele2007,MooreBalents2007,Roy2009b}. 
Several classes of materials, including the Bismuth compounds BiSb, Bi$_2$Se$_3$ and Bi$_2$Te$_3$, and strained 
HgTe were discovered to be three-dimensional topological insulators \cite{HasanKane2010,Qi:2011,Bernevig}. 
The hallmark of the $\mathbb{Z}_2$ topological order in $d=3$ is the existence of 
surface states with a linear dispersion and obeying the Dirac equation. The unique existence of these Dirac states  as well as their associated spin polarization spinning around the Dirac point have 
 been probed by experimental surface techniques including Angle-Resolved PhotoEmission (ARPES) and Scanning Tunneling Microscopy (STM).  Their presence in several materials has been confirmed by numerous studies, while a clear signature of their existence on transport experiments has proven to be 
 more difficult to obtain. 
Note that such Dirac dispersion relations for topological surface states arise around 
 a single (or an odd number of) Dirac points in the Brillouin zone, as opposed to real two dimensional materials like graphene where
  these Dirac points can only occur in pairs.

 The purpose of the present paper is to introduce pedagogically the notion of topological order in insulators as a bulk property, {\it i.e.} as a property of the
  ensemble of Bloch wave functions of the valence bands. For the sake of clarity we will discuss simple examples in dimension $d=2$ only, 
 instead of 
  focusing on generals definitions.  As a consequence of this pedagogical choice, we  will omit a discussion of  the physical consequences of this topological order, 
 most  notably the physical properties of Dirac surface states of interest experimentally, as well as other kind topological order in, {\it e.g.}, 
  superconductors. The reader interested by these aspects can turn towards existing reviews \cite{HasanKane2010,Qi:2011,Bernevig,Volovik:2003}. Note that a different notion of topological order was introduced in e.g. \cite{Wen2004}, which differs from the property of topological insulators discussed in this review.
  
 In the part which follows, 
 we will define more precisely the object of study. In a following part (section \ref{sec:ChernInsulators}), we will describe the simplest model of a Chern insulator, i.e. in a two-bands system. This will give us an excuse to define the Berry curvature and the Chern number, and to comprehend the nontrivial topology as an ``obstruction'' to properly define electronic wavefunctions. As the understanding of the more recent and subtle \Ztwo topological order was carved by its discoverers in a strong analogy with the Chern topological order, these concepts will equip us for the third part (section \ref{sec:Z2Insulators}), where we will develop simple models to understand the \Ztwo insulators as well as the different expressions of the \Ztwo invariant characterizing them.

\section{Bloch bundles and topology}
\label{sec:GeneralConsiderationsBlochBundles}

The aim of this first part is to define more precisely the object of this paper, namely the notion of topological order of an ensemble of valence bands in an insulator. We will first review a very simple example of nontrivial bundle: the Möbius strip, before defining the notion of valence Bloch bundle in an insulator. 

\subsection{The simplest twisted bundle: a Möbius strip}
\label{sec:MobiusStrip}

A vector bundle $\pi:E\rightarrow B$ is specified
by a projection $\pi$ from the bundle space $E$ to the base space $B$.
The fiber $F_x=\pi^{-1}(x)$ above each point of the base $x\in B$
is assumed to be isomorphic to a fixed typical fiber $F$. 
 The fibers $F_x$ and $F$ possesses a 
vector space structure assumed to be preserved by the isomorphism $F_x \cong F$.
Hence, the vector bundle $E$ indeed looks locally like the cartesian product $B \times F$.
The bundle is called trivial if this also holds globally, i.e. $E$ and $B \times F$ are isomorphic. When it is not the case, the vector bundle is said to be nontrivial, or twisted (see \cite{Baez,Nakahara} for details).
As a consequence, a $n$-dimensional vector bundle is trivial iff it has a basis of never-vanishing global sections
{\it i.e.} iff it has a set of $n$ global sections which at each point form a basis of the fiber \cite{HatcherVBKT}.
On the contrary, the \emph{obstruction} to define a basis of never-vanishing global sections (or basis of the fibers) will signal a twisted topology of a vector bundle. 
In the following, we will rely on this property to identify a non-trivial topology of a vector bundle when studying simple models.

\newcommand{\NorthOpen}{\ensuremath{\text{N}}}
\newcommand{\SouthOpen}{\ensuremath{\text{S}}}

\begin{figure}[htb]
\centering
\includegraphics[width=5cm]{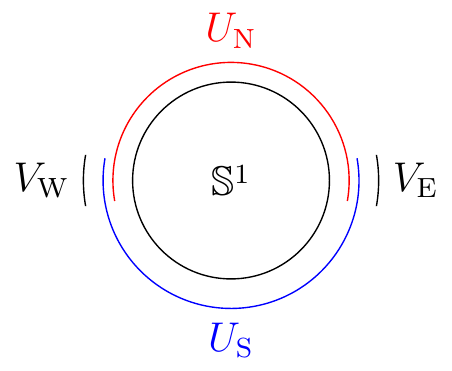}
\caption{Schematic view of the open covering $(U_{\NorthOpen}, U_{\SouthOpen})$ of \Sphere{1}, with the intersections $V_{\text{E}}$ and $V_{\text{W}}$ of the open sets. }
\label{fig:OpenCoveringS1}
\end{figure}

\begin{figure}[htb]
\centering
\includegraphics[width=10cm]{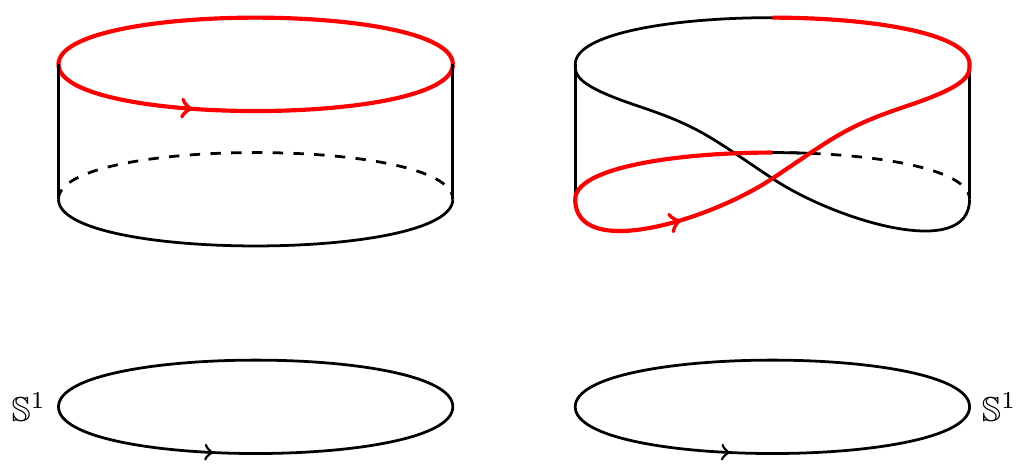}
\caption{A cylinder (left) is a trivial bundle (with no twist), whereas a Möbius strip (right) is a nontrivial bundle (with twist). Here, we have used the typical fiber $F=[-1,1]$ instead of $\RealField$ to get a compact manifold that is easier to draw.}
\label{fig:MobiusBandCylinder}
\end{figure}

To provide an intuitive picture of nontrivial bundles, we will consider a simple example: the Möbius bundle \cite{Nakahara}.
Let us consider as the base manifold the circle $\Sphere{1}$, and let $U_{\NorthOpen} = (0 - \epsilon, \pi + \epsilon)$ and $U_{\SouthOpen} = (- \pi- \epsilon, 0 + \epsilon)$ with $\epsilon > 0$ be an open covering of $\Sphere{1} \simeq [0, 2 \pi]$, parameterized by the angle $\theta \in \Sphere{1}$ (see Fig.~\ref{fig:OpenCoveringS1}). Take the typical fiber to be the line $F=\RealField$, parameterized by $t \in F$, and take as a structure group the two-elements group $\CyclicGroup{2} = \left\{ -1, 1 \right\}$. 
To construct a fibre bundle $\pi: E \to \Sphere{1}$ over \Sphere{1}, we have to glue together the products $U_{\NorthOpen} \Cross F$ and $U_{\SouthOpen} \Cross F$. 
The intersection of the two open sets of the covering is $U_{\NorthOpen} \cap U_{\SouthOpen} = V_{\text{E}} \cup V_{\text{W}}$ with $V_{\text{E}} = (- \epsilon, \epsilon)$ and $V_{\text{W}} = (\pi - \epsilon, \pi + \epsilon)$. The transition functions $t_{\NorthOpen \, \SouthOpen}(\theta)$ can be either $t \mapsto t$ or $t \mapsto -t$. If we choose both transition functions equal: 
\begin{equation}
t_{\NorthOpen \, \SouthOpen}(\theta \in V_{\text{E}}): t \mapsto t
\qquad
\text{and}
\qquad
t_{\NorthOpen \, \SouthOpen}(\theta \in V_{\text{W}}): t \mapsto t
\end{equation}
\noindent the bundle $\pi: E \to \Sphere{1}$ is a trivial (nontwisted) bundle, which is a cylinder (Fig.~\ref{fig:MobiusBandCylinder}, left). However, is we choose different transition functions on each side: 
\begin{equation}
t_{\NorthOpen \, \SouthOpen}(\theta \in V_{\text{E}}): t \mapsto t
\qquad
\text{and}
\qquad
t_{\NorthOpen \, \SouthOpen}(\theta \in V_{\text{W}}): t \mapsto - t
\end{equation}
\noindent the bundle is not trivial (it is twisted), and is the Möbius bundle (Fig.~\ref{fig:MobiusBandCylinder}, right). This illustrates the relation between the triviality of the bundle and the choice of the transition function $t_{\NorthOpen \, \SouthOpen}$. When  the 
bundle can be continuously deformed such that the transition functions be always the identity function the bundle will be trivial.

Let us illustrate on this example 
another property of a twisted bundle : the obstruction to define a basis of never-vanishing global sections in a twisted bundle. 
 First, notice that as $\RealField$ is a one-dimensional vector space, the Möbius bundle is  a one-dimensional (line) real vector bundle. 
 Let $s$ be a global section of the Möbius bundle. After one full turn from a generic position $\theta$, we have crossed one transition function $t \mapsto t$ and one transition function $t \mapsto -t$ so we have $s(\theta + 2 \pi) = - s(\theta)$. 
 Hence $s=0$ everywhere : the only global section on the Möbius bundle is 
 the zero section. 
 There is no global section of the Möbius bundle (except the zero section), so this bundle is indeed nontrivial.

%Finally, notice that a subbundle of a trivial bundle can be nontrivial: an example of this case is the Möbius bundle on \Sphere{1} as a subbundle of the trivial bundle $\Sphere{1} \Cross \RealField^{2}$.

\subsection{Bloch bundles}
\label{sec:TrivialityTotalBundle}

\begin{figure}[htb]
\centering
\includegraphics[width=8cm]{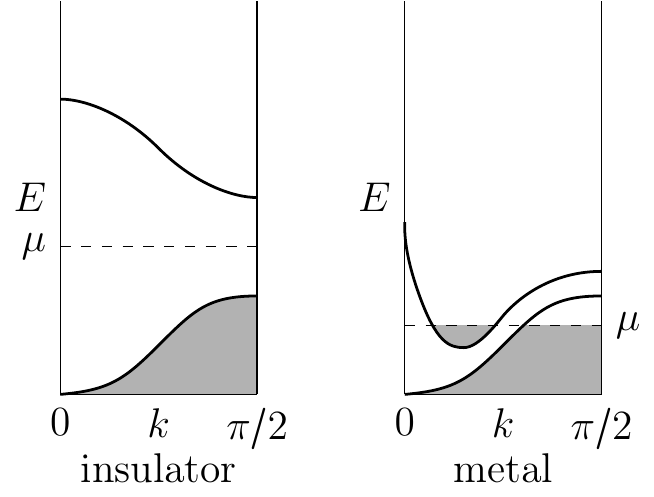}
\caption{Schematic band structures of an insulator (left) and a metal (right). The variable $k$ corresponds to the coordinate on some generic curve on the Brillouin torus. }
\label{fig:MetalInsulator}
\end{figure}

We consider a $d$-dimensional crystal in a tight-binding approach. We will describe its electronic properties using a single electron Hamiltonian, {\it i.e.} neglecting interaction effects. Hence, from now on, we only focus on first-quantized one-particle Hamiltonians. 
 The discrete real space lattice periodicity of this Hamiltonian reflects itself into the nature of its eigenstates, which are Bloch wavefunctions indexed by a quasi-momentum $k$. This quasi-momentum $k$ is restricted to the first Brillouin zone of the initial lattice: it is defined up to a reciprocal lattice vector $G$. 
 Hence this Brillouin zone has the topology of a $d$-dimensional torus \Torus{d}, which we call the Brillouin torus. From the initial 
 Hamiltonian, we deduce for each value of this quasi-momentum $k$  
 a \enquote{Bloch Hamiltonian} $H(k)$ acting on a $2n$-dimensional Hilbert space, 
 which accounts for the $2n$ electronic degrees of freedom in the unit cell (e.g. sites, orbitals, or spin).
 Associated with this Bloch Hamiltonian are its Bloch eigenstates and eigen-energies 
 $E_\alpha(k)$, $\alpha=1,\dots,2n$. The evolution of each $E_\alpha(k)$ as $k$ evolves in the Brillouin torus defines a band. 
 An insulator corresponds to the situation where a gap in energy separates the empty bands above the gap, from the 
 filled bands or valence bands below the gap (see Fig.~\ref{fig:MetalInsulator}). 
 In this situation, when the chemical potential lies inside the gap, electronic states of the crystal cannot be excited by a small perturbation such as the application of the difference of potential: no current can be created. 
 The ground state of such an insulator is determined from the ensemble of single particle 
 eigenstates corresponding to the filled bands. These eigenstates are defined for each valence band, and for each point $k$ 
 of the Brillouin torus, up to a phase. 
 The corresponding fiber bundle over the Brillouin zone defined from the eigenstates of the valence bands is the object of study in the present paper. 
 
Bloch Hamiltonians $H(k)$ define for each $k$ 
Hermitian operators on the effective Hilbert space $\mathcal{H}_k\cong \ComplexField^{2n}$
at $k$.
The collection of  spaces $\mathcal{H}_k$ forms a vector bundle on the
base space $\Torus{d}$. This vector bundle  happens to be always trivial, hence isomorphic to $\Torus{d} \Cross \ComplexField^{2n}$, 
at least for low dimensions of space $d\leq 3$ (this is due to the vanishing of the 
total Berry curvature, see \cite{Simon1983,Panati2007}). 
This means that we may assume that the Bloch Hamiltonians $H(k)$ are $k$-dependent Hermitian $2n\times 2n$
matrices defined so that $H(k)=H(k+G)$ for $G$ in the reciprocal lattice 
(note that this does not always correspond to common conventions in particular on multi-partite lattices, see {\it e.g.} \cite{BenaMontambaux2009})
  
%  To be more precise, a Bloch Hamiltonian defines a map from the Brillouin torus to Hermitian matrices 
% %
%\begin{equation}
%\begin{split}
%H \;:\; & \Torus{d} \to \Hermitian{2n} \\
%        & k \mapsto H(k)
%\end{split}
%\end{equation}
%
%\noindent that maps a point $k$ of Brillouin torus to an Hermitian matrix $H(k) \in \Hermitian{2n}$, which acts linearly on the effective Hilbert  space $\AllBandsFiber{k} \simeq \ComplexField^{2n}$ at $k$. 
%In other words, it is a section of the bundle of fiber $\mathcal{H}_{k} \simeq \ComplexField^{2n}$ on the base space \Torus{d}.
%This $2n$-dimensional vector bundle is the bundle of all bands, both filled and empty, which happens to be always trivial: 
%it is $\Torus{d} \times \ComplexField^{2n}$, where $d$ is the dimension of the space, and $2n$ the number of bands. As a consequence,  as this complete Bloch bundle $\Torus{d} \Cross \ComplexField^{2n}$ is trivial, it is always possible to find a global basis of this bundle, i.e. $2n$ never-vanishing global sections of the bundle which are linearly independent at each point. Moreover, the Hamiltonian is a section of the trivial bundle $\Torus{d} \Cross \Hermitian{2n}$. This means it is always possible to globally define the Hamiltonian, i.e. to have a periodic Bloch Hamiltonian which satisfies $H(k+G)=H(k)$ for any reciprocal lattice vector $G$, or equivalently to choose Bloch eigenstates which are periodic on the Torus . 

In an insulator, there are at least two well-defined subbundles of this complete trivial bundle: the valence bands bundle, which corresponds to all the filled bands, under the energy gap, and the conduction bands bundle, which corresponds to all the empty bands, over the energy gap. In the context of topological insulators, we want to characterize the topology of the valence bands bundle, which underlies the ground state properties of the insulators. In a topological insulator  this valence bands subbundle possesses a twisted topology while the complete bundle is trivial.

In the following, we will discuss two different kinds of topological orders. In the first one, we will discuss Chern insulators (section \ref{sec:ChernInsulators}): no symmetry constraints are imposed on the Bloch bundle, and in particular 
time-reversal invariance is broken. In the second part, we will discuss \Ztwo insulators (section \ref{sec:Z2Insulators}): 
here, time-reversal invariance is preserved. In a time-reversal invariant system, the bundle of filled bands and the bundle of empty bands happen to be separately trivial. However, the time-reversal invariance  adds additional constraints on the bundle: even if the filled bands bundle is always trivial as a vector bundle when time-reversal invariance is present, it is not always trivial in a way which preserves a structure compatible with the time-reversal operator.

\section{Chern topological insulators}
\label{sec:ChernInsulators}
\subsection{Introduction}

The first example of a topological insulator is the quantum Hall effect (QHE) discovered in 1980 by von Klitzing et al. \cite{KlitzingDordaPepper1980}. Two years later,  Thouless, Kohmoto, Nightingale, and de Nijs (TKNN) \cite{ThoulessKohmotoNightingaleNijs1982} showed 
that QHE in a two-dimensional electron gas in a strong magnetic field is related to a topological property of the filled band (see also \cite{Simon1983,AvronSeilerSimon1983}). Namely, the Hall conductance is quantized, and proportional to a topological invariant of the filled band named Chern number (hence the name Chern insulator). Haldane \cite{Haldane88} has generalized this argument to a system with time-reversal breaking without a net magnetic flux, hence without Landau levels. This kind of Chern insulator, which has recently been observed experimentally \cite{Chang:2013}, is called quantum anomalous Hall effect. Chern insulators, with or without a net magnetic flux, only exist in two dimensions.

% Although not realistic, Haldane model offers the simplest description of a Chern insulator.

\subsection{The simplest model: a two-bands insulator}
\label{sec:TwoLevelSystem}

The simplest insulator possesses two bands, one above and one below the band gap.
Such an insulator can generically be described as a two-level system, which corresponds to a two-dimensional Hilbert space 
$\mathcal{H}_{k} \simeq \ComplexField^{2}$
 at each point of the Brillouin torus, on which acts a Bloch Hamiltonian continuously defined on the Brillouin torus. 
Hence $H(k)$ can be written as a $2 \times 2$ Hermitian matrix, parameterized by the real functions $h_{\mu}(z)$: 
\begin{equation}
H(k) = 
\begin{pmatrix}
h_0 + h_z & h_x - \ii h_y \\ 
h_x + \ii h_y & h_0 - h_z
\end{pmatrix}, 
\end{equation}
which can re rewritten on the basis of Pauli matrices\footnote{We use the usual convention that a greek index starts at $0$ whereas a latin index starts at $1$.} plus the identity matrix~$\PauliMatrix{0} = \IdentityMatrix$ as:
\begin{equation}
H(k) = h^{\mu}(k) \PauliMatrix{\mu} = h_0(k) \IdentityMatrix + \vec{h}(k) \cdot \vec{\PauliMatrix{}}
\label{eq:GenericTwoLevelHamiltonian}
\end{equation}

\noindent  In the following, we always assumed that the coefficients $h_{\mu}$ are well defined on Brillouin torus, i.e. are periodic. The spectral theorem ensures that~$H(k)$ has two orthogonal normalized eigenvectors~$u_{\pm}(k)$ with eigenvalues~$\epsilon_{\pm}(k)$, which satisfy:

\begin{equation}
H(k) \, u_{\pm}(k) = \epsilon_{\pm}(k) \, u_{\pm}(k). 
\end{equation}

\noindent Using $\Tr(H)=2 h_0$ and $\det(H)=h_{0}^{2} -h^2$ with $h(k)= \norm{\vec{h}(k)}=\sqrt{h_{x}^2(k)+h_{y}^2(k)+h_{z}^2(k)}$ 
we obtain the energy eigenvalues: 

\begin{equation}
\epsilon_{\pm}(k) = h_{0}(k) \pm h(k)
\end{equation}

\noindent The corresponding normalized eigenvectors are, up to a phase:

\begin{equation}
u_{\pm}(k) = 
\left(
1 + 
\frac{h_{z} + \epsilon_{\pm}^{2}}{h_x^2+h_y^2}
\right)^{-\frac12} \,
\begin{pmatrix}
\displaystyle\frac{\epsilon_{\pm}}{h_x + \ii h_y}  \\ 
1
\end{pmatrix} 
\label{eq:TwoLevel:Eigenvectors}
\end{equation}

The energy shift of both energies has no effect on topological properties, provided the system remains insulating. To simplify the discussion, let us take $h_0 = 0$. Therefore, the system is insulating provided $h(k)$ never vanishes on the whole Brillouin torus, which we enforce in the following. As 
we focus only on the topological behavior of the filled band, which is now well-defined, we only consider the filled eigenvector 
$u_{-}(k)$ in the following.

% Yet, we keep in mind that the topological properties of the empty band are indeed reflected in the topological properties of the filled one (they are opposite), because the sum of the Berry curvatures of all bands is always zero (see section~\ref{sec:TrivialityTotalBundle}, p.~\pageref{sec:TrivialityTotalBundle}).

\subsection{An obstruction to continuously define the eigenstates}
\label{sec:ChernObstruction}

The filled band of the two-bands insulator is described by a map that assigns a filled eigenvector $u_{-}(k)$ to each point of the Brillouin torus: it defines a one-dimensional complex vector bundle on the torus. When this vector bundle is trivial, this map can be chosen to be continuous on the whole Brillouin torus: this corresponds to the standard situation where a choice of phase for the Bloch eigenstate at a given point $k_0$ of the Brillouin torus can be continuously extrapolated to the whole torus.  
When it is not trivial, there is an obstruction to do so. 

To clarify this notion of obstruction, let us first notice that the Hamiltonian (\ref{eq:GenericTwoLevelHamiltonian}) 
 is parameterized by a three-dimensional real vector $\vect{h}$. The energy shift $h_{0}$ does not affect the topological properties of the system and has been discarded. In spherical coordinates, this vector reads:

\begin{equation}
    \vect{h} = h \, \begin{pmatrix}
        \sin \theta \, \cos \varphi \\
        \sin \theta \, \sin \varphi \\
        \cos \theta
    \end{pmatrix}. 
\end{equation}

\noindent With these coordinates, we rewrite the filled eigenvector \eqref{eq:TwoLevel:Eigenvectors} as:

\begin{equation}
u_{-}(\vect{h}) = 
\begin{pmatrix}
- \sin \frac{\theta}{2} \\ 
\ee^{\ii \varphi} \, \cos \frac{\theta}{2}
\end{pmatrix} 
\label{eq:FilledEigenvectorObstruction}
\end{equation}

\begin{figure}
\centering
\includegraphics[width=3cm]{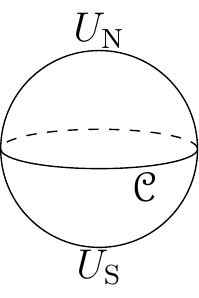}

\caption{Open covering $(U_{\NorthOpen}, U_{\SouthOpen})$ of the sphere \Sphere{2}. The intersection $\mathcal{C}$ of the open sets is topologically a circle \Sphere{1} and can be viewed as the boundary of either of the open sets.}
\label{fig:OpenCoveringSphere}
\end{figure}

We notice that the norm $h=\norm{\vec{h}}$ of the parameter vector $\vect{h}$ does not affect the eigenvector. Therefore, 
the parameter space is a $2$-sphere \Sphere{2}. We will first see in the following 
that there is always an obstruction to define a continuous eigenvector $u_{-}(\vect{h}/h)$ on the sphere, or in other words, that the corresponding vector bundle on the sphere \Sphere{2} is not trivial. 
%More precisely, two local trivializations are needed to describe it. 
Hence, we will realize that the original vector bundle on Brillouin torus (the pullback bundle by $\vec{h}$ of the bundle on the sphere) is only nontrivial when the map $k \mapsto \vect{h}(k)$ covers the whole sphere.

In the limit $\theta \to 0$, the eigenvector \eqref{eq:FilledEigenvectorObstruction} is not well defined because it has an ill-defined phase. We could change our phase convention and, {\it e.g.}, 
multiply the eigenvector (\ref{eq:FilledEigenvectorObstruction}) 
by $\ee^{\ii \varphi}$, but this would only move the ill-defined limit to $\theta \to \pi$. 
It turns out that it is not possible to get rid of this singularity and define a continuous eigenvector on the whole sphere. This behaviour unveils the nontrivial topology of a vector bundle on the sphere, discovered by Dirac and Hopf in 1931 \cite{Dirac1931}: at least two local trivializations are needed to describe a vector bundle on the sphere 
\cite{ChrusscinnskiJamiolkowski,Nakahara}. 

Let us choose an open covering $(U_{\NorthOpen}, U_{\SouthOpen})$ of the sphere, the two open sets being the north hemisphere $U_{\NorthOpen}$ and the south hemisphere $U_{\SouthOpen}$, chosen so that they have a nonzero intersection homotopy equivalent to the equator circle (Fig.~\ref{fig:OpenCoveringSphere}). We define local trivializations of the filled band bundle by

\begin{equation}
u_{-}^{\SouthOpen}(\vect{h}) = 
\begin{pmatrix}
- \sin \frac{\theta}{2} \\ 
\ee^{\ii \varphi} \, \cos \frac{\theta}{2}
\end{pmatrix}
\qquad
\text{and}
\qquad 
u_{-}^{\NorthOpen}(\vect{h}) = 
\begin{pmatrix}
- \ee^{-\ii \varphi} \, \sin \frac{\theta}{2} \\ 
\cos \frac{\theta}{2}
\end{pmatrix}
\end{equation}

\noindent Indeed, $u_{-}^{\NorthOpen}$ is correctly defined on $U_{\NorthOpen}$ (resp. $u_{-}^{\SouthOpen}$ on $U_{\SouthOpen}$), but neither are well defined on the whole sphere. 
The intersection $\mathcal{C} = U_{\NorthOpen} \cap U_{\SouthOpen}$ can be reduced to a circle, and can be viewed as the
 boundary e.g. of $U_{\NorthOpen}$, i.e. $\mathcal{C} = \partial U_{\NorthOpen} \simeq \Sphere{1}$. 
The transition function from the trivialization on $U_{\NorthOpen}$ to the trivialization on $U_{\SouthOpen}$ is phase change on the equator, {\it i.e.} a map $t_{\NorthOpen \, \SouthOpen}: \mathcal{C} \to \UnitaryGroup{1}$

\begin{equation}
    t_{\NorthOpen \, \SouthOpen} = \ee^{\ii \varphi}
    \label{eq:ChernTransitionFunction}
\end{equation}

%we could choose another open covering, e.g. we could take $U_{\NorthOpen}$ to be $\Sphere{2}$ minus the south pole, 
% and $U_{\SouthOpen}$ to be some neighborhood of the south pole. 

Now let us recall that the Bloch electronic states are described by a bundle on the Brillouin torus. 
If the map~$\vect{h}/h$ from the Brillouin torus to the parameter manifold does not completely cover the sphere (taking into account the orientations, see below), there will be no obstruction to globally define eigenstates of the Bloch Hamiltonian, by smoothly deforming the pulled-back transition function $h^{\star} t_{\NorthOpen \, \SouthOpen} = t_{\NorthOpen \, \SouthOpen} \circ h $ to the identity. %This corresponds to the case where $W=0$.
On the contrary, if $\vec{h}/h$ does completely cover the sphere, the topology of the Bloch bundle is not trivial: it is never possible to deform the transition function to the identity. Those statements are indeed made quantitative through the introduction of 
 the notion of Chern number.

\subsection{Berry curvature and Chern number}
\label{sec:BerryCurvatureChernNumber}

As the complete Bloch bundle (with filled and empty bands) is always trivial (see section~\ref{sec:TrivialityTotalBundle}), 
we can indifferently study the topological properties of the filled band or of the empty one: 
the topology of the filled band will reflect the topological properties of the empty one. 
%To simplify, we will speak about the topology of a \emph{band} for the topological properties of the corresponding bundle.
In the context of Bloch bundles, topological properties of the filled band are characterized by its Chern class (see \cite[ch.~2]{ChrusscinnskiJamiolkowski} as well as \cite[ch.~10]{Nakahara} for a general introduction to the Berry phase, connection and curvature). %See also \cite{Frankel}.
%
%The Chern classes are defined as the coefficients in the development in homogeneous invariant polynomials of an invariant polynomial called total Chern class \cite{Nakahara}. As such, 
The Chern classes are an intrinsic characterization of the considered bundle, and do not depend on a specific connection on it. 
However, in the context of condensed matter and of Bloch systems, the Berry connection appears as a natural and 
particularly useful choice \cite{Berry1984}.

The Berry curvature $A$ associated with the filled band $k \to u_{-}(k)$ is the 1-form defined by\footnote{Note that Berry curvature is sometimes defined with an additional $\ii$ factor, in which case it is purely imaginary.}

\begin{equation}
A = \frac{1}{\ii} \; \braket{ u_{-} | \dd \, u_{-} } = - \frac{1}{\ii} \; \braket{ \dd \, u_{-} | u_{-} }
\label{eq:BerryConnectionDefinition}
\end{equation}

\noindent where $\dd$ is the exterior derivative. The Berry curvature is then

\begin{equation}
F = \dd A
\label{eq:BerryCurvatureDefinition}
\end{equation}

In the following, we consider only two-dimensional systems. In this case, the vector bundles are characterized by the first Chern number $c_{1}$, that can be computed as an integral of Berry curvature $F$ over the Brillouin torus: 
\begin{equation}
c_{1} = \frac{1}{2 \pi} \, \int_{\mathrlap{\text{BZ}}} \ F .
\label{eq:ChernNumberDefinition}
\end{equation}

This integral of a $2$-form is only defined on a $2$-dimensional surface, and the Chern number characterises insulators in dimension $d=2$ only. More generally, a quantum hall insulator only exists in even dimensions.

In order to relate the discussion of Chern insulators to the pictorial example of the Möbius strip (sec.~\ref{sec:MobiusStrip}, p.~\pageref{sec:MobiusStrip}), we now express the first Chern number as the winding number of the 
transition function $t_{\NorthOpen \, \SouthOpen}$ introduced in sec.~\ref{sec:ChernObstruction}. Indeed, 

\begin{align}
c_{1}
=
\frac{1}{2 \pi} \, \int_{\mathrlap{\text{BZ}}} F
&= 
\frac{1}{2 \pi} \, \left[
\int_{\mathrlap{h^{-1}(U_{\NorthOpen})}} F
\qquad
+
\int_{\mathrlap{h^{-1}(U_{\SouthOpen})}} F
\qquad
\right]
\nonumber \\
&= 
\frac{1}{2 \pi} \, \left[
\int_{\mathrlap{\partial h^{-1}(U_{\NorthOpen})}} h^{\star} A_{\NorthOpen}
\qquad
+
\int_{\mathrlap{\partial h^{-1}(U_{\SouthOpen})}} h^{\star} A_{\SouthOpen}
\qquad
\right]
\nonumber \\
&= 
\frac{1}{2 \pi} \, 
\int_{\mathrlap{\partial h^{-1}(U_{\NorthOpen})}} \left( h^{\star} A_{\NorthOpen} - h^{\star} A_{\SouthOpen} \right)
\label{eq:ChernSphere}
\end{align}

\noindent where $h^{\star} A = A \circ h$ represents the connection form defined on the sphere pulled back by the map $h$ to be defined on the 
torus\footnote{The attentive reader will notice that we actually considered a map $h$ from the sphere to the sphere when assuming that 
$h^{-1}(U_{\NorthOpen})$ and $h^{-1}(U_{\SouthOpen})$ defines a open covering of the manifold we consider. To be more precise, we should consider two 
maps : from the torus (BZ) to the sphere, and the map $h$ from the sphere to the sphere. For the sake of simplicity, we have implicitly assumed in 
writing eq.~\eqref{eq:ChernSphere} that the first map from the torus to the sphere was topologically trivial. 
}, 
and we used  that $\partial h^{-1}(U_{\NorthOpen})$ and $\partial h^{-1}(U_{\SouthOpen})$ have opposite orientations. We can now relate the Berry connections on the two hemispheres with the transition function, by using \eqref{eq:BerryConnectionDefinition} with the transition function \eqref{eq:ChernTransitionFunction}, we get

\begin{equation}
A_{\NorthOpen} = t_{\NorthOpen \, \SouthOpen}^{-1} \, A_{\SouthOpen} \, t_{\NorthOpen \, \SouthOpen} + t_{\NorthOpen \, \SouthOpen}^{-1} \, \frac{\dd}{\ii} \, t_{\NorthOpen \, \SouthOpen}
= A_{\SouthOpen} + \dd \varphi
\end{equation}

\noindent so that

\begin{equation}
A_{\NorthOpen} - A_{\SouthOpen} = \dd \varphi = \frac{1}{\ii} \; \dd \log (t_{\NorthOpen \, \SouthOpen}) .
\end{equation}

\noindent Finally, we obtain the expression of the Chern number as the winding of the transition function

\begin{equation}
c_{1} = \frac{1}{2 \pi} \, \int_{\text{BZ}} F
= 
\frac{1}{2 \pi \ii} \, \int_{\partial h^{-1}(U_{\NorthOpen})} \, \dd \log (t_{\NorthOpen \, \SouthOpen} \circ h) .
\end{equation}

Hence, the filled eigenvector as well as the associated Berry connection are  well defined on the whole Brillouin torus 
only if the transition function $t_{\NorthOpen \, \SouthOpen} \circ h: \partial h^{-1}(U_{\NorthOpen}) \to \UnitaryGroup{1}$ 
can be continuously deformed to the identity, {\it i.e.} when it does not wind around the circle, which corresponds to a trivial Chern class $c_{1}=0$. The first Chern number is therefore the winding number of the transition function ; when the Chern class is not trivial, it is not possible to deform the transition function to the identity. Notice that a nonzero first Chern number can be seen as an obstruction to Stokes theorem, as its expression \eqref{eq:ChernNumberDefinition} in terms of the Berry curvature would vanish if we could write \enquote{$F = \dd A$} on the whole torus.

%A simple image is that when $c_{1}=0$, the bundle is trivial, like the cylinder in section~\ref{sec:MobiusStrip} (p.~\pageref{sec:MobiusStrip}). However, when~$c_{1} \neq 0$, the \UnitaryGroup{1} phase winds around the circle $\mathcal{C}$ and the bundle is twisted, like the Möbius bundle.

 % In the general case, it is only well-defined on local trivializations of the bundle, which is why the first Chern number is not always zero\footnote{If Berry curvature is well-defined on the whole Brillouin torus, as we have $F = \dd A$, we can use Stokes theorem with $\dd F = \dd^2 A = 0$ or $\partial \mathrlap{\text{BZ}} = \emptyset$. }. 

Let us now come back to our two-band model with the parameterization \eqref{eq:GenericTwoLevelHamiltonian} 
 where we have omitted the part proportional to the identity ($h_0 \IdentityMatrix$) :
\begin{equation}
H(k) = \vec{h}(k) \cdot \vec{\PauliMatrix{}}
\end{equation}

\noindent In this case, the curvature 2-form takes the form \cite{Berry1984}: 
\begin{equation}
F = \frac{1}{4} \, \epsilon^{i j k} \, h^{-3} \, h_{i} \, \dd h_{j} \wedge \dd h_{k} , 
\label{eq:BerryCurvatureTwoLevelSystem}
\end{equation}

\noindent and the first Chern number reads:
\begin{equation}
c_{1}
= \frac{1}{2 \pi} \int_{\text{BZ}} F
\;
= 
\;
\frac{1}{2 \pi} \int_{\text{BZ}} \frac{1}{4} \epsilon^{i j k} \, \Norm{h}^{-3} \, h_{i} \, \dd h_{j} \wedge \dd h_{k} .
\label{eq:TwoLevel:ChernNumberIntrinsic}
\end{equation}

As $\vec{h}(\vec{k})$ depends on the two components $k_{x}$ et $k_{y}$ of the wavevector, we have:
%%we can also rewrite it as

\begin{equation}
\dd h_{j} = \frac{\partial h_{j}}{\partial k_{a}} \dd k^{a}
\qquad
\text{et}
\qquad
\dd h_{j} \wedge \dd h_{k} = \frac{\partial h_{j}}{\partial k_{a}} \, \frac{\partial h_{k}}{\partial k_{b}} \; \dd k_{a} \wedge \dd k_{b}
\end{equation}

\noindent The curvature $F$ can be written in the more practical form:

\begin{equation}
F =  \frac{1}{4} \, \epsilon^{i j k} \, \Norm{h}^{-3} \, h_{i} \, \frac{\partial h_{j}}{\partial k_{a}} \, \frac{\partial h_{k}}{\partial k_{b}} \; \dd k_{a} \wedge \dd k_{b}
=
\frac{1}{2}
\,
\frac{\vec{h}}{\Norm{h}^3}
\cdot
\left(
\frac{\partial \vec{h}}{\partial k_x}
\Cross
\frac{\partial \vec{h}}{\partial k_y}
\right)
\; \dd k_{x} \wedge \dd k_{y} , 
\label{eq:BerryCurvatureUsefullExpression}
\end{equation}

\noindent corresponding to the first Chern number 

\begin{equation}
c_{1}
= \frac{1}{4 \pi}  \int_{\text{BZ}}
\,
\frac{\vec{h}}{\Norm{h}^3}
\cdot
\left(
\frac{\partial \vec{h}}{\partial k_x}
\Cross
\frac{\partial \vec{h}}{\partial k_y}
\right)
\; \dd k_{x} \wedge \dd k_{y}
\label{eq:ChernNumberAsIntegral}
\end{equation}

One recognises in the expression \eqref{eq:TwoLevel:ChernNumberIntrinsic} 
the index of the map $\vec{h}$ (see \ref{app:IndexTheoryWindingNumber}). Hence, the first Chern number 
$c_{1} = \Degree(\vec{h}, 0)$. This identification provides a geometrical interpretation of the Chern number in the case of two-band insulators.  When $k$ spreads over Brillouin torus, $\vec{h}$ describes a closed surface $\Sigma$. The Chern number can then be viewed as
\begin{itemize}
\item the (normalized) flux of a magnetic monopole located at the origin through the surface~$\Sigma$
\item the number of times the surface $\Sigma$ wraps around the origin (in particular, it is zero if the origin is \enquote{outside} $\Sigma$ ; more precisely it is the homotopy class of $\Sigma$ in the punctured space $\mathbb{R}^ 3 - 0$) 
\item the number of (algebraically counted) intersections of a ray coming from the origin with $\Sigma$, which is the method used in  \cite{SticletPiechonFuchsKukuginSimon2012}. 
\end{itemize}

%%%%%
\subsection{Haldane's model}
\label{sec:HaldaneModel}

\subsubsection{General considerations}

In this section, we consider an explicit example of such a two band model displaying a topological insulating phase, namely the 
model proposed by Haldane \cite{Haldane88}. 
Besides its description using the semi-metallic graphene, Haldane's model describes a whole class of simple two bands insulating phases with possibly a nontrivial topological structure, and proposes a description of one of the simplest examples of a topological insulator, namely a Chern insulator.
In this model, both inversion symmetry and time-reversal symmetry are simultaneously broken in a sheet of graphene.
 Inversion symmetry is broken by assigning different on-site energies to the two inequivalent sublattices of the honeycomb lattice,
  while time-reversal invariance is lifted by local magnetic fluxes organized so that the net flux per unit cell vanishes. 
 Therefore, the first neighbors hopping amplitudes are not affected by the magnetic fluxes, whereas the second neighbors hopping 
 amplitudes acquire an Aharonov--Bohm phase.

\subsubsection{Notations}

\begin{figure}[htb]
\begin{minipage}{.45\textwidth}
  \centering
\includegraphics[width=4cm]{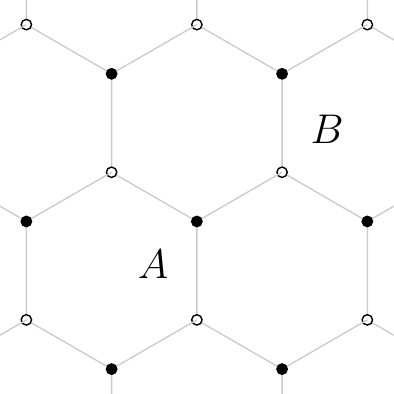}
\end{minipage}
\begin{minipage}{.45\textwidth}
  \centering
\includegraphics[width=4cm]{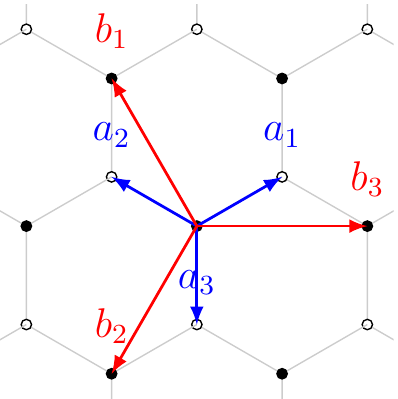}
\end{minipage}
\caption{Honeycomb lattice used in Haldane's model}
\label{fig:HoneycombLattice}
\end{figure}

We consider a tight-binding model of spinless electrons on a two-dimensional hexagonal (honeycomb) lattice. Crucially, this lattice is not a Bravais lattice, 
%\footnote{The honeycomb lattice is the canonical example of a lattice which is not a Bravais lattice. Here is a way to convince oneself of that. If two nearest neighbors are related by a vector $\vect{v}$, the lattice is not invariant by a translation of $\vect{v}$. To choose a bipartite lattice is of essential importance in the Haldane model, as it is what leads to a two-level system.}
 and the cristal is described as a triangular Bravais lattice with two non-equivalent atoms in a unit cell, hence its description requires a two-level Hamiltonian. 
  Let us denote by $A$ and $B$ the two inequivalent sublattices corresponding to those atoms (See Fig.~\ref{fig:HoneycombLattice}). 
 The lattice parameter $a$, defined as the shortest distance between nearest neighbors, sets the unit of length: $a=1$. 

The vectors between nearest neighbors, {\it i.e.} between sites of different sublattices $A$ and~$B$, are:
\begin{equation}
a_{1} = \begin{pmatrix}
\sqrt{3}/2 \\
1/2
\end{pmatrix}
\qquad
a_{2} = \begin{pmatrix}
-\sqrt{3}/2 \\
1/2
\end{pmatrix}
\qquad
a_{3} = \begin{pmatrix}
0 \\
-1
\end{pmatrix} = - (a_{1} + a_{2}), 
\end{equation}

\noindent whereas the vectors between second-nearest neighbors belonging to the same sublattice are: 

\begin{equation}
b_{1} = a_{2} - a_{3} = \begin{pmatrix}
-\sqrt{3}/2 \\
3/2
\end{pmatrix}
\qquad
b_{2} = a_{3} - a_{1} = \begin{pmatrix}
-\sqrt{3}/2 \\
-3/2
\end{pmatrix}
\qquad
b_{3} = a_{1} - a_{2} = \begin{pmatrix}
\sqrt{3} \\
0
\end{pmatrix}. 
\end{equation}

\noindent Two of those vectors will serve as base vectors of the Bravais lattice ; we will choose $b_{1}$ and $b_{2}$.
The reciprocal lattice is then spanned by the two vectors $b^{\star}_{1}$ and $b^{\star}_{2}$ which satisfy:

\begin{equation}
b_{i} b^{\star}_{j} = 2 \pi \, \delta_{i j}, 
\end{equation}

\noindent so

\begin{equation}
b^{\star}_{1} = 2 \pi \, \begin{pmatrix}
- 1 / \sqrt{3} \\
1/3
\end{pmatrix}
\qquad
b^{\star}_{2} = 2 \pi \, \begin{pmatrix}
- 1 / \sqrt{3} \\
- 1/3
\end{pmatrix}. 
\end{equation}

\noindent Two points of the Brillouin zone are of particular interest in graphene, corresponding to the origin of the low-energy Dirac dispersion relations. 
They are defined by: 

\begin{equation}
K = \frac{1}{2} \, \left( b^{\star}_{1} + b^{\star}_{2} \right)
\qquad
\text{and}
\qquad
K' = -K . 
\label{eq:DiracPoints}
\end{equation}

In the following, $G_{m n} = m b^{\star}_{1} + n b^{\star}_{2}$ denotes an arbitrary reciprocal lattice vector ($n,m \in \IntegerRing$). 
%, labelled by
%
%\begin{equation}
%G_{m n} = m b^{\star}_{1} + n b^{\star}_{2}
%\qquad
%\text{($n,m \in \IntegerRing$)} . 
%\end{equation}

\subsubsection{Haldane's Hamiltonian}

The first quantized Hamiltonian of Haldane's model can be written as:

\begin{equation}
    \Op{H} = 
    t \, \sum_{\langle \, i, j \,  \rangle} \ket{i} \bra{ \,j} 
    +
    t_{2} \, \sum_{\llangle \, i, j \, \rrangle} \ket{i} \bra{\,j}
    +
    M \, \left[
    \sum_{i \in A} \ket{i}\bra{i}
    -
    \sum_{j \in B} \ket{j}\bra{j}
    \right]
\label{eq:HaldaneHamiltonianFirstQuantized}
\end{equation}

\noindent where $\ket{i}$ represents an electronic state localized at site $i$ (atomic orbital), 
$\langle \, i, j \, \rangle$ represents nearest neighbors lattice sites $i$ and $j$, 
$\llangle \, i, j \, \rrangle$ represents second nearest neighbors sites $i$ and $j$, 
$i \in A$ represents sites in the sublattice $A$ (resp. $i \in B$ in the sublattice $B$).
This Hamiltonian is composed of a first nearest neighbors hopping term with a hopping amplitude~$t$, 
a second neighbors hopping term with a hopping parameter $t_{2}$, and a last sublattice symmetry breaking  term 
with on-site energies $+M$ for sites of sublattice $A$, and $-M$ for sublattice $B$,  
which thus breaks inversion symmetry. 
Moreover, the Aharanov--Bohm phases due to the time-reversal breaking local magnetic fluxes are taken into account through 
 the Peierls substitution:

\begin{equation}
t_{i j} \to t_{i j} \exp \left( - \ii \frac{\ElementaryCharge}{\PlanckConstantReduced} \int_{\Gamma_{i j}} \vec{A} \cdot \dd \vec{\ell} \right)
\end{equation}

\noindent where $t_{i j}$ is the hopping parameter between sites $i$ and $j$, and  where $\Gamma_{i j}$ is the hop trajectory from 
site $i$ to site $j$ and $\vec{A}$ is a potential vector accounting for the presence of the magnetic flux. 
\begin{figure}
\centering
\includegraphics[width=15cm]{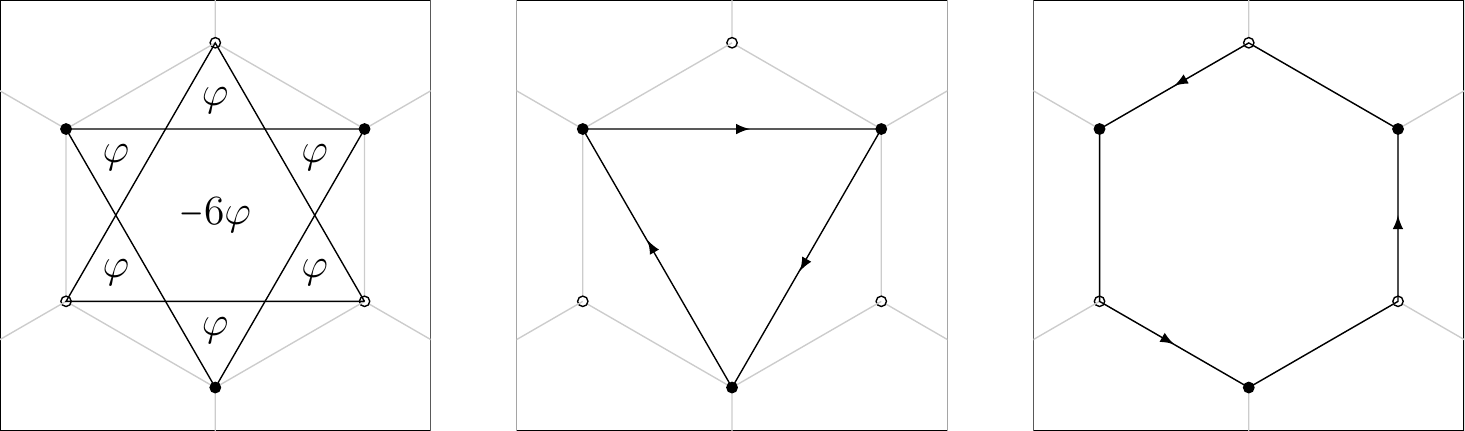}   
\caption{Example of a choice for magnetic flux in an Haldane cell (left). We have used $\varphi = \phi/2$ to simplify. Second-neighbors hopping corresponds to a nonzero flux (middle), whereas first-neighbors hopping gives a zero flux (right), so the total flux through a unit cell is zero.}
\label{fig:HaldaneMagneticFlux}
\end{figure}
In Haldane's model, magnetic fluxes are imposed such that the phase accumulated through a nearest neighbor $A \to B$ (or $B \to A$) hopping vanishes, whereas the phase accumulated through a second-neighbors hopping $A \to A$ or $B \to B$ is nonzero 
(see. Fig.~\ref{fig:HaldaneMagneticFlux} for a possible flux distribution).
 The Aharonov--Bohm phase gained through $A \to A$ hopping is opposite of the one gained through $B \to B$ hopping. 
 Notice that to use Peierls substitution, we choose a gauge for the vector potential;  \UnitaryGroup{1} gauge invariance of the model  is reflected 
 in the independence of results of this particular choice of the phases. The Peierls substitution amounts to the substitution: 

\begin{equation}
    t \to t
    \qquad
    \text{and}
    \qquad
    t_{2} \to t_{2} \ee^{\ii \phi}
\end{equation}

\noindent where the Aharonov--Bohm phase $\phi$ due to the local magnetic flux is taken as a parameter of the model.

The Fourier transform of the Hamiltonian \eqref{eq:HaldaneHamiltonianFirstQuantized} with Aharonov--Bohm phases leads to a $2 \times 2$ Bloch Hamiltonian in the $(A,B)$ sublattices basis:

\begin{equation}
\mathcal{H}(k) = h^{\mu}(k) \PauliMatrix{\mu}
\end{equation}

\noindent with

\begin{subequations}
\label{eq:Haldane_h}
\begin{align}
h_{0} &= 2 t_{2} \cos \phi \sum_{i=1}^{3} \cos( k \cdot b_{i} ) 
\quad ; \quad 
h_{z} = M - 2 t_{2} \sin \phi \sum_{i=1}^{3} \sin( k \cdot b_{i} ) 
\quad ;
\label{eq:Haldane_hz}
\\
h_{x} &= t \left[1 + \cos(k \cdot b_{1}) + \cos(k \cdot b_{2}) \right] 
\quad ; \quad 
h_{y} = t \left[\sin(k \cdot b_{1}) - \sin(k \cdot b_{2}) \right] 
\quad ; 
\label{eq:Haldane_hy}
\end{align}
\end{subequations}

\noindent with a convention where $\vec{h}$ is periodic: $\vec{h}(k + G_{m n})=\vec{h}(k)$.

\subsubsection{Phase diagram of Haldane's model}

\begin{figure}[th!]
\centering
\includegraphics[width=10cm]{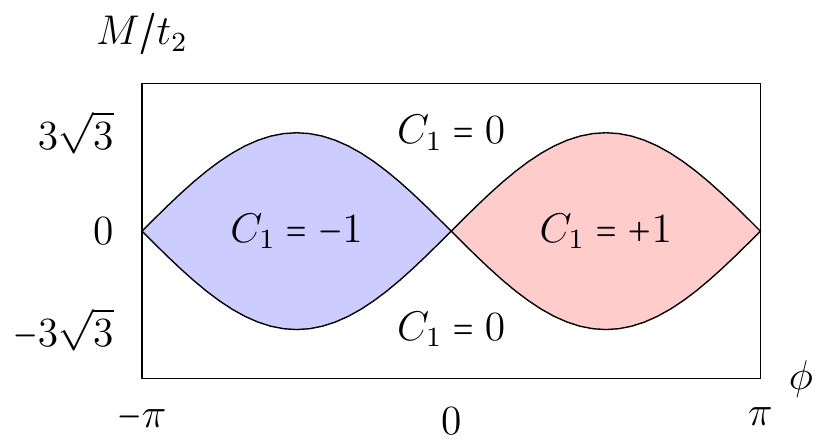}   
\caption{Phase diagram of the Haldane model, giving the first Chern number $c_{1}$ on the plane $(\phi, M/t_{2})$ (the manifold of parameters is 
$S^{1} \Cross \RealField$, variable $\phi$ being a phase).}
\label{fig:HaldanePhaseDiagram}
\end{figure}

To determine the phase diagram, let us find the points in the parameter space where the local gap closes 
(i.e.~$h=\norm{h} = 0$) at some points of the Brillouin torus. In graphene, which corresponds to $(M,\phi)=(0,0)$ in the diagram, the two energy bands are degenerate ($h=0$) 
at the Dirac points $K$ et $K'$ (see eq.~\eqref{eq:DiracPoints}). At a generic point of the diagram, this degeneracy is lifted, and the system is an insulator ($h\neq 0$), except when $\Abs{M} = 3 \sqrt{3} t_{2} \sin \phi$. 
 The corresponding line separates  four {\it a priori} different 
 insulating states, see Fig.~\ref{fig:HaldanePhaseDiagram}. 
Haldane  has shown that for $\Abs{M} > 3 \sqrt{3} t_{2} \sin \phi$, the Chern number of the filled band vanishes, which means that the corresponding insulator is topologically trivial. 
On the contrary, when $\Abs{M} < 3 \sqrt{3} t_{2} \sin \phi$, the Chern number is $\pm 1$ \cite{Haldane88}. This defines Haldane's phase diagram
 (Fig. \ref{fig:HaldanePhaseDiagram}); we will recover these results in section~\ref{sec:DeterminationChernNumbersPhaseDiagram}. 

Let us note that on the critical lines which separate insulating phases with different topologies, 
 there is a phase transition and the system is not insulating anymore: it is a semi-metal with low energy Dirac states. 
At the transition between $c_{1}=0$ and $c_{1}=1$, the gap closes at one Dirac point $K_{0 \, 1}(M, \phi),$ whereas at the transition between $c_{1}=0$ and $c_{1}=-1$, the gap closes at a different Dirac point $K_{0 \, -1}(M, \phi)$. These points evolve  continuously with the parameters $(M, \phi)$. From this perspective, 
the pure graphene $(M,\phi)=(0,0)$ corresponds to a bicritical transition where the gap closes simultaneously at 
both points $K_{0 \, 1}(0,0)$ and $K_{0 \, -1}(0,0)$ which are the two nonequivalent Dirac points \eqref{eq:DiracPoints}.

\subsubsection{Geometric interpretation}
\label{sec:HaldaneGeometricInterpretation}

\begin{figure}[t!]
\centering
\includegraphics[width=15cm]{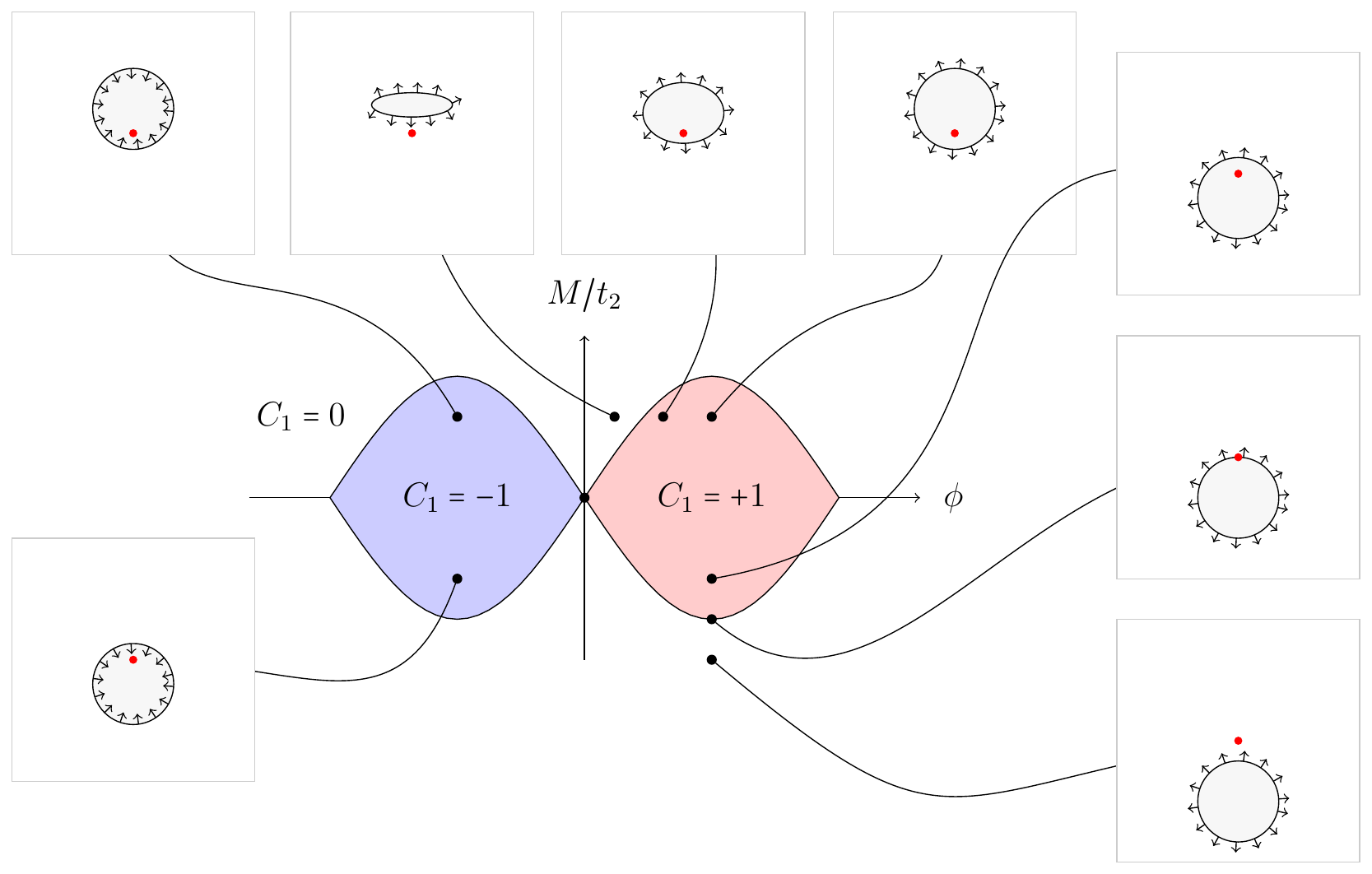}      
\caption{Pictorial view of the Haldane phase diagram, showing the transition between $0$ and $1$ Chern numbers. The system is topological when the Dirac monopole (red point) is inside the closed surface $\Sigma$ (gray spheroid). The orientation of the surface $\Sigma$ changes when $\phi$ changes sign. When $M$ varies, the center of the surface moves with respect to the Dirac monopole. }
\label{fig:diagrammePhaseHaldaneBerryMonopoles}
\end{figure}
\begin{figure}[ht!]
\centering
\includegraphics[width=8cm]{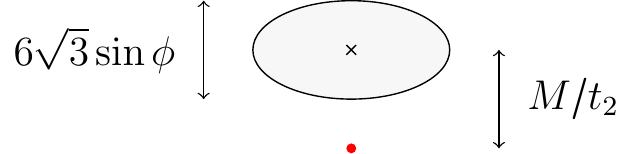}   
\caption{The simplified surface $\Sigma'$, depicted in two dimensions, and the origin (red circle). The $z$-axis is vertical, and goes upward. The surface $\Sigma'$ is an oblate or prolate spheroid, with height proportional to $\sin \phi$, with a coefficient that does not depend on $M/t_{2}$ nor on $\phi$. The center of the spheroid is at an height $M/t_{2}$. }
\label{fig:HaldaneSurface}
\end{figure}

As we have seen in section~\ref{sec:BerryCurvatureChernNumber}  
(see also \ref{app:IndexTheoryWindingNumber}), this Chern number admits a simple interpretation for a two-level Hamiltonian. 
Let us consider 
the closed surface $\Sigma$ defined by $\vec{h}(k)$  as $k$ runs through the Brillouin torus (see eq.~\eqref{eq:Haldane_h}). 
 The Chern number correspond to the total flux of the field created by a \enquote{Dirac monopole} located at the origin through this surface. 
  When the monopole is inside $\Sigma$, the flux is non zero and the phase is topologically nontrivial, 
  whereas when the monopole is outside $\Sigma$, the net flux vanishes and 
  the phase is trivial. 
Moreover,
% $h_{z}$ varies between $M - \alpha \sin\phi$ and $M + \alpha \sin\phi$ (see eq. \eqref{eq:Haldane_hz}) where $\alpha$ does not depend on $M/t_{2}$ neither on $\phi$, while 
as the components $h_{x}$ and $h_{y}$ do not depend on $M/t_{2}$ nor on $\phi$, it is interesting for illustration purpose to replace 
the surface $\Sigma$ by  a spheroid $\Sigma'$, with height $6\sqrt{3} \sin\phi$ 
 and center ($0,0,M/t_{2}$), as in figure \ref{fig:HaldaneSurface}.
%Hence to identify the phase transitions of Haldane's model we can reduce this surface $\Sigma$ to 
%\pounds a spheroid $\Sigma'$, with height $\alpha \sin\phi$ (where $\alpha$ is a constant) and center ($0,0,M/t_{2}$), as in figure \ref{fig:HaldaneSurface}.
%
From the evolution of this surface $\Sigma'$ as a function of $M/t_{2}$ and $\phi$, we identify easily 
 Haldane's phase diagram, see Fig.~\ref{fig:HaldanePhaseDiagram} and Fig.~\ref{fig:diagrammePhaseHaldaneBerryMonopoles}.
%  Indeed, the domain of $M/t_{2}$ such that the origin lies inside $\Sigma'$ becomes smaller as $\Sigma'$  shrinks with $\phi$. 
% Moreover, the orientation of  $\Sigma'$ changes when $\phi=0$, which explains the sign change of the Chern number.
%
%  for smaller 
%  
%   %it becomes more oblate ; 
% the surface $\Sigma'$ becomes more oblate when $\phi$ becomes smaller, and its orientation changes (the faces are switched) when $\phi=0$, which explains the sign change of the Chern number.
 %
At a phase transition between two insulators with different Chern numbers, the origin necessarily crosses the surface $\Sigma$, 
corresponding to a gap closing (i.e. $h=\norm{h}=0$) at least at one point on the Brillouin torus. Therefore, the topological phase transition is a semi-metallic phase.

\subsubsection{Determination of the Chern numbers in the phase diagram}
\label{sec:DeterminationChernNumbersPhaseDiagram}

To determine in a more rigorous manner the topological phase diagram of Haldane model requires the evaluation of 
 the Chern number $c_{1}1$ as a function of the parameters $(M/t_{2}, \phi)$. 
 A simple method \cite{SticletPiechonFuchsKukuginSimon2012}  consists in using the geometric interpretation of the Chern number as the number  of intersections between some ray coming from the origin and the oriented closed surface $\Sigma$ spanned by $h$  (see \ref{app:IndexTheoryWindingNumber}). 
 Alternatively, we can consider half of the number of intersections with a line instead of a ray. 
 A natural choice for this line is the $Oz$ axis. If $D$ is the set of pre-images by $h$ of those intersections, i.e. $D = h^{-1}(Oz \cap \Sigma)$, the Chern number is: 

\begin{equation}
c_{1} = \frac{1}{2} \, \sum_{k \in D} \sign \left[ 
h(k) \cdot n(k)
\right], 
\end{equation}

\noindent where $n(k)$ is the normal vector to $\Sigma$ at $k$ (where it is $\pm \hat{e}_{z}$ in the formula). We obtain (with a slight abuse of notation): 
%\footnote{This abuse is a legitimate one when we work on an orientable surface. 
%, on which an orientation, i.e. a non-vanishing top-form (i.e. a form of maximal - \enquote{top} - dimension). 
%Once the surface is oriented, one can define the sign of $F$ as the sign of its coefficient on the orienting form
% (all top-forms being proportional). For example, if the surface is oriented by $\dd k_{x} \wedge \dd k_{y}$ and that $F = F_{x y} \dd k_{x} \wedge \dd k_{y}$, we have $\sign(F) = \sign(F_{x y})$.}, we have 
%}
\begin{equation}
c_{1} = \frac{1}{2} \, \sum_{k \in D} \sign \left[ 
F(k)
\right], 
\end{equation}

\noindent where $F$ is the Berry curvature from eq.~\eqref{eq:BerryCurvatureUsefullExpression}. More 
explicitely, we obtain: 
\begin{equation}
c_{1} = \frac{1}{2} \, \sum_{k \in D} \sign \left[ 
h_{z}(k)
\right]
\,
\sign \left[ 
\left(
\frac{\partial \vec{h}}{\partial k_x}
\Cross
\frac{\partial \vec{h}}{\partial k_y}
\right)_{z}
\right], 
\end{equation}

\noindent  the second term accounting for the direction of the normal.
We now need to determine the set $D$, {\it i.e.}  the set of wavevectors $k$ such that $h_{x}(k)=h_{y}(k)=0$ 
(so that $\vec{h}$ lies the $z$ axis). 
As the components $h_x$ and $h_y$ of \eqref{eq:Haldane_h} are $M/t_{2}$ and $\phi$ independent, they are identical to those 
 in pure graphene (which is the point $(M/t_2,\phi)=(0,0)$), 
 these points are the Dirac points of graphene\footnote{As the Dirac points are on the boundary of the standard Brillouin zone, there appears six points on this Brillouin zone depicted in the place, but only two inequivalent ones on the torus.}: 

\begin{equation}
D = \left\lbrace
K = \begin{pmatrix}
- \frac{4 \pi}{3 \sqrt{3}} \\
0
\end{pmatrix}
\qquad
\text{et}
\qquad
K' = - K
\right\rbrace . 
\end{equation}

\noindent Hence the quantities we need to compute are the masses of the Dirac points:

\begin{equation}
h_{z}(K) = M - 3 \sqrt{3} t_{2} \sin \phi
\qquad
\text{and}
\qquad
h_{z}(K') = M + 3 \sqrt{3} t_{2} \sin \phi, 
\end{equation}

\noindent and the Chern number is thus given by:

\begin{equation}
c_{1} = \frac{1}{2} \left[
\sign\left( \frac{M}{t_{2}} + 3 \, \sqrt{3} \sin(\phi) \right)
 - 
\sign\left( \frac{M}{t_{2}} - 3 \, \sqrt{3} \sin(\phi) \right)
 \right], 
\end{equation}

\noindent which corresponds to the original result of \cite{Haldane88} (see also Fig.~\ref{fig:HaldanePhaseDiagram}).
Obviously this method is  specific neither to graphene  nor to Haldane's model; we can apply it very efficiently to two bands general models indexed by a vector $\vec{h}$, as it necessitate only to compute the sign of $h_{z}$ at  points
where $h_{x}$ and $h_{y}$ vanish. 

\subsubsection{Surface states}
\label{sec:ChernSurfaceStates}

One of the crucial consequences of a nontrivial bulk topology is the appearance of \emph{metallic edge states} at the surface of a topological insulator. An sketchy way to understand the appearance of these 
surface states is the following: as the Chern number is a topological quantity, it cannot change simply through a continuous transformation, but only at a phase transition associated with a gap closing. 
%Indeed, at the phase transition, the Chern number is not defined, which allows a topological change. 
Following \cite{HasanKane2010} and \cite{Fradkin}, we discuss the appearance of edge states due to the change in bulk topology. 

Let us start with a time-reversal invariant, parity-invariant system like graphene. The time-reversal symmetry implies 
$h_{z}(k)=h_{z}(-k)$, whereas the inversion symmetry implies $h_{z}(k)=-h_{z}(-k)$. 
Hence when both symmetries are present,  $h_{z}$ identically vanishes. 
A Dirac point is an isolated point $K$ of the Brillouin torus where the gap closes i.e. $h(K)=0$, so that the dispersion relation around this point is linear.
 Nielsen--Ninomiya's theorem \cite{Nielsen:1981} implies that Dirac points come in pairs in a time-reversal invariant system. Hence, the simplest case is one with two Dirac points $K$ and $K'$. This is the case of the Haldane model discussed in section~\ref{sec:HaldaneModel}.
In Haldane's model, the time-reversal invariance is lifted. The gap at the Dirac points opens because $h_{z}(K) \neq 0$, but we have still $h_{x}(K)=h_{y}(K)=0$. Hence, the Dirac points have gained a mass $m=h_{z}(K)$. Indeed, let us linearize the  Hamiltonian \eqref{eq:GenericTwoLevelHamiltonian} around a Dirac point $K$ by writing $k=K+q$ :
\begin{equation}
H_{\text{l}}(q) = \PlanckConstantReduced v_{\text{F}} \, q \cdot \PauliMatrix{\text{2d}} + m \, \PauliMatrix{z}
\label{eq:HI}
\end{equation}
with $q=(q_{x}, q_{y})$ and $\PauliMatrix{\text{2d}} = (\PauliMatrix{x}, \PauliMatrix{y})$, and $m=h_{z}(K)$. The linearization gives rise to a massive Dirac Hamiltonian with mass $m$. 
 In the following, we set $\PlanckConstantReduced v_{\text{F}} = 1$.
In Haldane's model, we have (see eq.\eqref{eq:Haldane_h}) 
\begin{equation}
m = h_{z}(K) = M - 3 \sqrt{3} t_{2} \sin \phi
\qquad
\text{and}
\qquad
m' = h_{z}(K') = M + 3 \sqrt{3} t_{2} \sin \phi
\end{equation}
\noindent As $c_{1}=(\sign m - \sign m')/2$, the masses $m$ and $m'$ of the Dirac points $K$ and $K'$ have the same sign in the trivial case, whereas they have opposite signs in the topological case.

\begin{figure}[ht]
\centering  
\includegraphics[width=8cm]{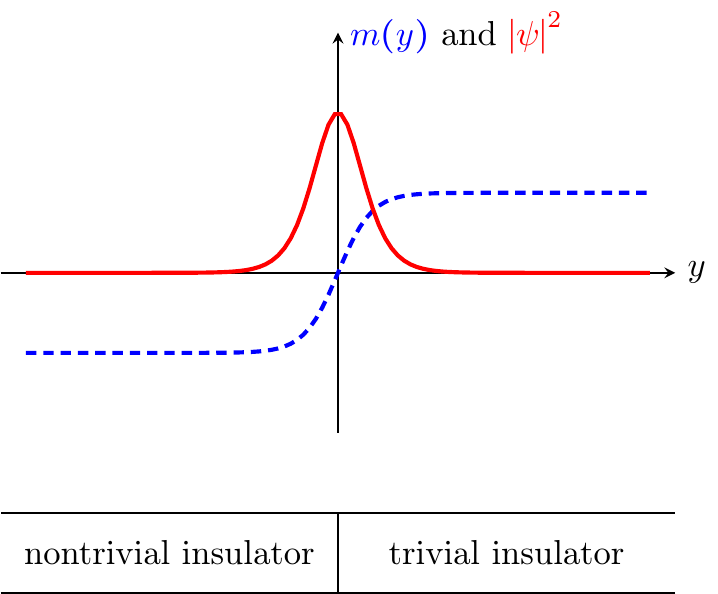}      
\caption{
Schematic view of edge states at a Chern--trivial insulator interface. The mass $m(y)$ (blue dashed line) and the wavefunction amplitude $\Abs{\psi}^2$ (red continuous line) are drawn along the coordinate $y$ orthogonal to the interface $y=0$. 
}
\label{fig:ChernEdge}
\end{figure}

Let us now  consider an interface at $y=0$ between a (nontrivial) Haldane insulator with a Chern number $c_{1}=1$ for $y<0$ and a (trivial) insulator with $c_{1}=0$ for $y>0$. 
Necessarily, one of the masses changes sign at the interface: $m(y<0)<0$ and $m(y>0)>0$, whereas the other one has a constant sign $m'>0$ (see Fig.~\ref{fig:ChernEdge}). 
It is then natural to set $m(0)=0$, which implies that the gap closes at the interface. A more precise analysis shows that there are indeed surface states \cite{HasanKane2010}.
As the mass $m$ depends on the position, it is more convenient to express the single-particle Hamiltonian in space representation. By inverting the Fourier transform in \eqref{eq:HI} (which amounts to the replacement $q \to - \ii \nabla$), we obtain the Hermitian Hamiltonian:
\begin{equation}
H_{\text{l}} = - \ii \nabla \cdot \PauliMatrix{\text{2d}} + m(y) \sigma_{z}
= \begin{pmatrix}
  m(y) & - \ii \partial_{x} - \partial_{y} \\
  - \ii \partial_{x} + \partial_{y} & -m(y)
\end{pmatrix}.
\end{equation}
\noindent In order to get separable PDE, let us rotate the basis with the unitary matrix:
\begin{equation}
  U = \frac{1}{\sqrt{2}} \; \begin{pmatrix}
    1 & 1 \\
    1 & -1
  \end{pmatrix}
\end{equation}
to obtain the Schrödinger equation: 
\begin{equation}
  U \cdot H_{\text{l}} \cdot U^{-1} \begin{pmatrix}
    \alpha \\
    \beta
  \end{pmatrix}
   = \begin{pmatrix}
    - \ii \partial_{x} & \partial_{y} + m(y) \\
    - \partial_{y} + m(y) & \ii \partial_{x}
  \end{pmatrix} 
  \begin{pmatrix}
    \alpha \\
    \beta
  \end{pmatrix}
  = E \, \begin{pmatrix}
    \alpha \\
    \beta
  \end{pmatrix}.
\end{equation}
\noindent This matrix equation corresponds to two separable PDE:
\begin{subequations}
\begin{align}
  (- \ii \partial_{x} - E) \, \alpha &= S_{1} = - (\partial_{y} + m(y)) \, \beta
  \\
  (\ii \partial_{x} - E) \, \beta &= S_{2} = - (- \partial_{y} + m(y)) \, \alpha
\end{align}
\end{subequations}
\noindent In order to obtain integrable solutions, the corresponding separations constants $S_{1}$ and $S_{2}$ must be zero. We can then solve separately for $\alpha$ and $\beta$.
For our choice of $m(y)$, there is only one normalizable solution, which reads in the original basis:

\begin{equation}
\psi_{q_{x}}(x,y) \propto \ee^{\ii q_{x} \, x} \, \exp \left[ - \int_{0}^{y} m(y') \, \dd{}y' \right] \; 
\begin{pmatrix}
    1 \\
    1
\end{pmatrix}
\end{equation}

\noindent and has an energy $E(q_{x}) = E_{\text{F}} + \PlanckConstantReduced v_{\text{F}} q_{x}$. 
This solution  is  localized transverse to the interface 
 where $m$ changes sign (see Fig.~\ref{fig:ChernEdge}). 
 The edge state crosses the Fermi energy at $q_{x}=0$, with a positive group velocity $v_{\text{F}}$ and thus corresponds to a ``chiral right moving'' edge state. When considering a transition from an insulator with the opposite Chern number to the vacuum, one would get a ``chiral left moving'' edge state.

\subsection{Models with higher Chern numbers}

\begin{figure}[ht]
\centering  
\includegraphics[width=5cm]{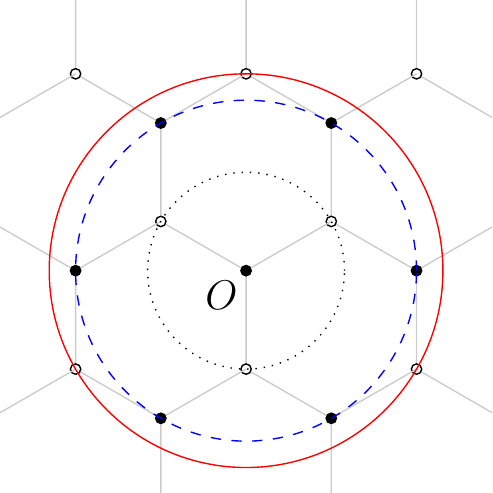}      
\caption{Let us consider a site at $O$. The nearest neighbors (from the opposite sublattice) are located on the dotted black circle. The second neigbors (from the same sublattice) are on the blue dashed circle. The third neighbors are on the continuous red circle.}
\label{fig:ThirdNeighbors}
\end{figure}

Topological phases  with higher Chern numbers can be incorporated into 
Haldane's model by considering interactions beyond second neighbors \cite{SticletPiechonFuchsKukuginSimon2012}. 
We briefly review an example with third nearest neighbors where the Chern number can take values $0, \pm 1, \pm 2$. 

We consider a Haldane's model with 
an additional interaction term between third nearest neighbors with a hopping amplitude 
$t_{3}$ (see Fig.~\ref{fig:ThirdNeighbors}). The vector $\vec{h}(k)$ parametrizing the effective two-band model is then
\begin{subequations}
\label{eq:H3N_h}
\begin{align}
h_{0} &= 2 t_{2} ~\cos \phi \sum_{i=1}^{3} \cos( k \cdot b_{i} ) ; 
 \label{eq:H3N_h0}\\
h_{x} &= t \left[1 + \cos(k \cdot b_{1}) + \cos(k \cdot b_{2}) \right]
+ t_{3} \left[ 2 \cos\left(k \cdot \left(b_{1} + b_{2}\right) \right) + \cos\left( k \cdot \left(b_{1} - b_{2}\right) \right) \right] ; 
\label{eq:H3N_hx}\\
h_{y} &= t \left[ \sin(k \cdot b_{1}) - \sin(k \cdot b_{2}) \right]
+ t_{3} ~\sin \left( k \cdot \left( b_{1} - b_{2} \right) \right) ; 
\label{eq:H3N_hy}\\
h_{z} &= M - 2 t_{2} ~\sin \phi \sum_{i=1}^{3} \sin( k \cdot b_{i} ) ; 
\label{eq:H3N_hz}
\end{align}
\end{subequations}
%
%\begin{figure}[ht]
%\centering  
%\includegraphics[width=8cm]{Figures/CRAS-figure-14}      
%\caption{Schematic phase diagram of the extended Haldane's model in the plane $(M/t_{2},\phi)$ for e.g. $t_{2}/t_{1}=0.5$ and $t_{3}/t_{1}=0.35$. \cite{SticletPiechonFuchsKukuginSimon2012}}
%\label{fig:ChernT3PhaseDiagram}
%\end{figure}

\begin{figure}[ht]
\centering
\includegraphics[width=8cm]{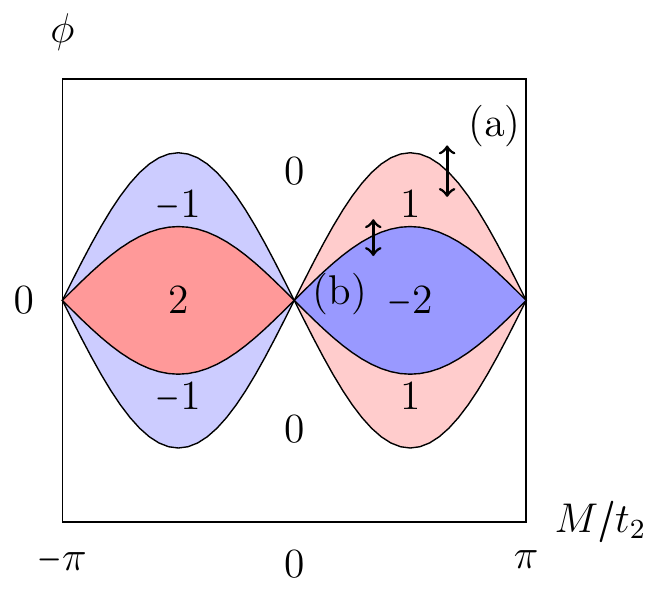}      
\caption{
Schematic phase diagram of the extended Haldane model in the plane $(M/t_{2},\phi)$ for, e.g., $t_{2}/t_{1}=0.5$ and $t_{3}/t_{1}=0.35$ \cite{SticletPiechonFuchsKukuginSimon2012}. Values of the Chern number are indicated as labels of the different phases. 
Arrows locate the topological transitions pictured in Fig.~\ref{fig:N2ModelTransitionZeroToOne}. }
\label{fig:ChernT3PhaseDiagram}
%\label{fig:ArrowedN2PhaseDiagram}
\end{figure}

\begin{figure}[ht]
\centering
\includegraphics[width=6cm]{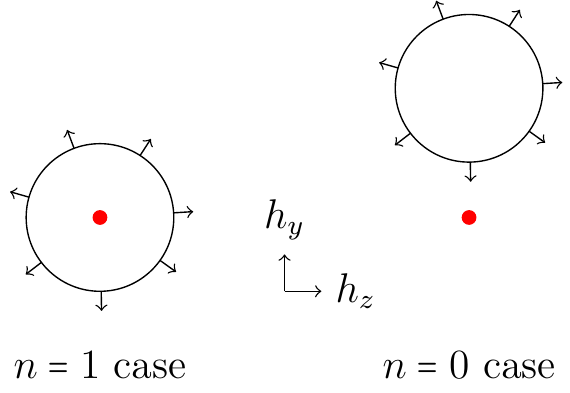}    
\includegraphics[width=6cm]{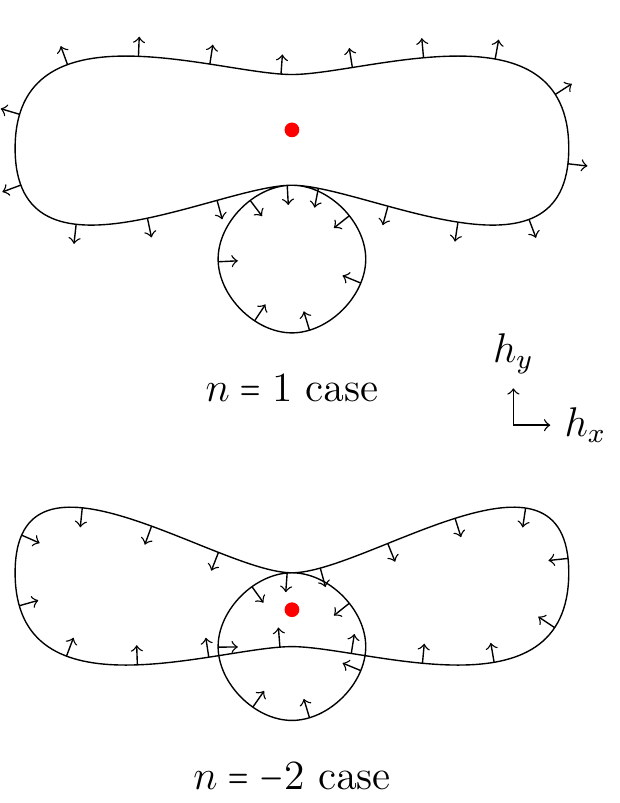}        
\caption{(Left) Transition $0 \leftrightarrow 1$ (arrow (a) in fig.~\ref{fig:ChernT3PhaseDiagram}), identical to the transition
 in Haldane's model.
(Right)
Transition $-2 \leftrightarrow 1$ (arrow (b) in fig.~\ref{fig:ChernT3PhaseDiagram}). When $M$ or $\phi$ vary, the surface $\Sigma$ is pushed in the $z$ direction. However, sections of the surface in the $xy$ plane differ at different heights $z$, implying a non-trivial  change of the Chern number.
}
\label{fig:N2ModelTransitionZeroToOne}
\end{figure}

For a correctly choosen domain of $t_{3}$, the Chern number can take the value $\Abs{c_{1}} = 2$. 
The phase diagram in the plane $(M/t_{2},\phi)$ is drawn in Fig.~\ref{fig:ChernT3PhaseDiagram}.
The geometric interpretation is essentially the same that in Haldane's model, but the surface~$\Sigma$ can now wrap multiple times around the origin, corresponding to higher Chern numbers. The subtleties of the surface play an important role in the transitions: to hint at its evolution, we consider sections in the planes~$xz$ and~$xy$. The results are presented in Figs.~\ref{fig:ChernT3PhaseDiagram} and \ref{fig:N2ModelTransitionZeroToOne}. 
%Depending on the values of the parameters, the evolution of the section of~$\Sigma$ with~$z$ is not always the same, but the transition mechanism does not change.

%
%\begin{figure}[ht]
%\centering
%\includegraphics[width=6cm]{Figures/CRAS-figure-17}      
%\caption{Transition $-2 \leftrightarrow 1$ (arrow (b) in fig.~\ref{fig:ChernT3PhaseDiagram}). When $M$ or $\phi$ vary, the surface $\Sigma$ is essentially moved in the $z$ direction. The subtelty is that the sections in the $xy$ plane differ at different height, and one goes from one of the situations drawn here to the other.}
%\label{fig:N2ModelTransitionMinus2ToOne}
%\end{figure}

%%%%%

\newpage

\section{\Ztwo topological insulators}
\label{sec:Z2Insulators}

\subsection{Introduction}

A different kind of topological insulator was discovered by Kane and Mele \cite{KaneMele2005} in 2005. 
Indeed, while the previous Chern topological order occurs in bands unconstrained by symmetries, and in particular by 
time reversal symmetry, the Kane--Mele topological order characterizes bands constrained by time-reversal symmetry. More precisely, it is a property of half-integer spin bands in the presence of  time-reversal symmetry. 
The topological property of the valence band arises only when the constraints imposed by time-reversal symmetry are enforced. 
Initially, Kane and Mele considered a model analogous to Haldane's model, but with spin $\frac12$ electrons and 
a spin-orbit interaction replacing the magnetic fluxes. This spin-orbit interaction is indeed time reversal symmetric. 
 Unlike the Chern insulator characterized by a Chern number that can take any integer value, 
 Kane and Mele's topological insulator is characterized by a \Ztwo index that can take only two values (e.g., $0$ and $1$ or $-1$ and $+1$). The corresponding two-dimensional phase, denoted quantum spin Hall effect (QSHE) 
 also exhibits edge states of different natures than for Chern insulators as they do not break time-reversal symmetry. 
 From these edge states point of view, 
  the \Ztwo index was interpreted as the parity of the number of time-reversal pairs of edge states. 
  In 2007, following a proposal by Bernevig, Hugues, and Zhang (BHZ) \cite{BHZ2006}, 
  the first experimental realization of the QSHE state was achieved in \ce{HgTe} quantum wells by the group of L. Molenkamp \cite{Konig2007}. 
  The same year, this topological order was generalized from two- to three-dimensional systems by three independent theoretical 
 groups \cite{FuKaneMele2007,MooreBalents2007,Roy2009b}. Hence, unlike Chern insulators, \Ztwo topological insulators also exist in three dimensions. This discovery triggered a  huge number of theoretical and experimental studies. 
 
 The purpose of this part of the article is to define the  \Ztwo topological index characterizing these new phases as a bulk property, in a manner analogous to the previous description of Chern insulators.  
To illustrate the \Ztwo topology, we wish, by analogy with Chern insulators, to exemplify explicitly the obstruction to globally define 
eigenvectors of the Bloch Hamiltonian that satisfy the time-reversal symmetry constraints. 
 For this purpose, we will consider simple tight-binding models that play an analogous role to the Haldane's model for the Chern topological order.
  These simplest models involve naturally the space inversion symmetry, both because of a drastic reduction of the number of free parameters and because of a very simple expression of the \Ztwo invariant (see section~\ref{sec:Z2:SewingMatrixInvariant}).
 For the sake of pedagogy, we will only consider two-dimensional \Ztwo insulators, which are 
  characterized by a topological index $(-1)^{\ZtwoInvariantModTwo}$. The system is trivial when $(-1)^{\ZtwoInvariantModTwo}=+1$, whereas it is nontrivial when $(-1)^{\ZtwoInvariantModTwo}=-1$.

%\subsection{General considerations}
%\label{sec:Z2:GeneralConsiderations}

\subsection{Time-reversal symmetry}
\label{sec:Z2:TimeReversal}

\subsubsection{The time-reversal operation}

Time-reversal operation amounts to the transformation in time $t \to -t$. 
As such, quantities like spatial position, energy, or electric field are even under time-reversal, whereas quantities like time, linear momentum, angular momentum, or magnetic field are odd under time-reversal operation. 
Within quantum mechanics, the time-reversal operation is described by an anti-unitary operator 
$\TR$ (which is allowed by Wigner's theorem) \cite{LeBellac,SakuraiModernQM}, that is to say 
(i) it is anti-linear, i.e.~$\TR(\alpha x) = \Conjugate{\alpha} \TR(x)$ for $\alpha \in \ComplexField$ and 
(ii) it satisfies $\Adjoint{\TR} \, \TR = \IdentityMatrix$, i.e.  $\Adjoint{\TR} = \TR^{-1}$.

 When spin degrees of freedom are included, time-reversal operation has to reverse the different spin expectation values: 
the corresponding standard representation of the time-reversal operator is \cite{SakuraiModernQM} $\TR = \ee^{- \ii \pi J_{y} / \PlanckConstantReduced } \; \ComplexConjugate$, where $J_{y}$ is the $y$ component of the spin operator, and 
$\ComplexConjugate$ is the complex conjugation (acting on the left). 
From this expression, the time-reversal operator appears to be a $\pi$ rotation in the spin space. 
%Note that the time-reversal operator is defined up to an arbitrary phase.
Therefore, and because the spin operator $\ee^{- \ii \pi J_{y} / \PlanckConstantReduced }$
is real and unaffected by $ \ComplexConjugate$, 
in an integer spin system, the time-reversal operator is involutive, i.e. $\TR^2 = \IdentityMatrix$. 
However,  for the  $\frac12$-integer spin system, this operation is anti-involutive: $\TR^2 = - \IdentityMatrix$. This property will have crucial consequences in the following. 
%For a system with an odd number of fermions (e.g. one), we have $\TR^{2} = - \IdentityMatrix$.
%
As usual, a first quantized single-particle Hamiltonian $\FirstQuantizedHamiltonian$ is time-reversal invariant if  it commutes with the time-reversal operator, {\it i.e.} $\Commutator{\FirstQuantizedHamiltonian}{\TR}=0$.

\subsubsection{Time-reversal symmetry in Bloch bands}

 In the following, we consider the band theory of electrons in crystals \cite{Madelung}, and hence we focus on the case of spin $\frac12$ particles, with $\TR^{2} = -  \IdentityMatrix$. 
%In the context of Bloch systems without interaction \cite{Madelung}, this first quantized Hamiltonian describes the dynamics of a single electron. Therefore, $\TR^{2} = - 1$.
%\footnote{Notice that the \enquote{total number} of electrons in the system has nothing to do with that. However, in real systems, the relevant spin representations are not always $J=1/2$ ; this can completely modify the problem if $J$ is an integer, and we do not consider this case.}. 
Focusing on non-interacting electrons, we can describe the electronic bands through a first-quantized Hamiltonian, or equivalently through the Fourier-transformed 
effective Bloch Hamiltonian $k \to H(k)$ defined on the Brillouin torus. 
In this context, the Bloch time-reversal operator $\TR$ will relate to the electronic Bloch states at $k$ and $-k$, 
{\it i.e.} it is an anti-unitary map from the fiber at $k$ to the fiber at $-k$ of the vector bundle on the Brillouin torus that represents the bands of the system. In a 
time-reversal invariant system, the Bloch Hamiltonians at $k$ and $-k$ satisfy:
\begin{equation}
H(-k) = \TR H(k) \TR^{-1}
\label{eq:TimeReversalBlochHamiltonian}. 
\end{equation}

As time-reversal operation maps a fiber at $k$ to a fiber at $-k$, it is useful to consider the application on the Brillouin torus that relates the corresponding momenta: $\trb \;:\; \Torus{2} \to \Torus{2}$, defined as $\trb \, k = -k$ on the torus, 
{\it i.e.} up to a lattice vector. 
The time-reversal operator is then viewed as a lift to this map $\trb$ on the total Bloch bundle 
$\Torus{2} \Cross \ComplexField^{2n}$ describing the electronic states of all bands. 
It can be represented by an unitary matrix $U_{\TR}$ which does not depend on the momentum $k$ on the Brillouin torus. 
Hence, it is a map:

\begin{equation}
\begin{split}
\Torus{2} \Cross \ComplexField^{2n} &\to \Torus{2} \Cross \ComplexField^{2n} \\
(k,v) &\mapsto (\trb k, \TR v) = (-k, U_{\TR} \, \ComplexConjugate \, v)
\end{split}
\end{equation} 

\noindent which sends the fiber of all bands $\AllBandsFiber{k} \simeq \ComplexField^{2n}$ at $k$ to the fiber $\AllBandsFiber{\trb k}$ at $\trb k=-k$. We sum that up by $\TR \;:\; \AllBandsFiber{k} \to \AllBandsFiber{\trb k}$. Notice that this implies that 
%$\Adjoint{\TR} \;:\; \AllBandsFiber{\trb k} \to \AllBandsFiber{k}$, that 
$\TR^{2} = - \IdentityMatrix$, indeed maps a fiber to itself. 
%, and that $\text{aut}_{\TR}$ maps endomorphisms of $\AllBandsFiber{k}$ to those of $\AllBandsFiber{\trb k}$.

In a time-reversal invariant system of spin $\frac12$ particles, the Berry curvature within valence bands is odd: $F_\alpha(k)=-F_\alpha(-k)$. Hence 
 the Chern number of the corresponding bands $\alpha$ vanishes: the valence vector bundle is always trivial from the point of 
 view of Chern indices. 
 It is only when the constraints imposed by time-reversal symmetry on the eigenstates are considered that a different kind of non-trivial  topology  can emerge.

\subsubsection{Kramers pairs}
\label{sec:Z2:Kramers}

\begin{figure}
%\begin{minipage}[t]{0.45\linewidth}
%\centering
%\includegraphics[width=6cm]{Figures/CRAS-figure-19}      
%
%\caption{Action of the space inversion on the graphene honeycomb lattice. The sublattices (black and white circles) are exchanged by the inversion operation (dotted lines), leaving the lattice globally invariant. The center of inversion is crossed out. The reader can try and see by himself that it is not possible to choose another center of inversion that preserves each sublattice separately.}
%\label{fig:GrapheneInversionSymmetry}
%\end{minipage}
%\hspace{0.5cm}
%\begin{minipage}[t]{0.45\linewidth}
\centering
\includegraphics[width=8cm]{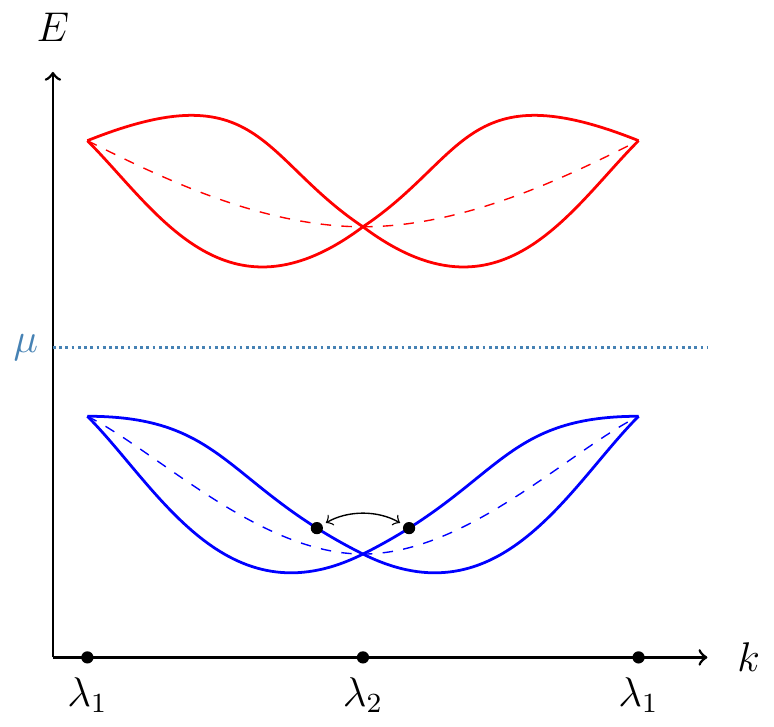}      
\caption{Typical energy spectrum of a time-reversal invariant system (continuous lines) on a closed loop of the Brillouin torus. At the TRIM $\lambda_{i}$, there is always a degeneracy of the filled bands (resp. empty bands). A Kramers pair is drawn as two black circles. When inversion symmetry is present, the filled bands (resp. empty bands) are everywhere degenerate (dashed lines).}
\label{fig:TypicalTimeReversalSpectrum}
%\end{minipage}
\end{figure}

Time reversal implies the existence of Kramers pairs of eigenstates:  
equation \eqref{eq:TimeReversalBlochHamiltonian} implies that the image by time-reversal of any eigenstate of the Bloch Hamiltonian $H(k)$ at $k$ is an eigenstate of the Bloch Hamiltonian $H(-k)$ at $-k$, with the same energy. 
This is the Kramers theorem \cite{SakuraiModernQM}. These two eigenstates, that a priori live in different fibers, are called Kramers partners. 
$\TR^2 = - \IdentityMatrix$ implies that these two  Kramers partners are orthogonal. Note that the orthogonality of these
 Kramers partners in different fibers has only a meaning if we embed these fibers in the complete trivial bundle $\Torus{2} \Cross \ComplexField^{2n}$ 
 corresponding to the whole state space of the Bloch Hamiltonian.

 \begin{figure}
\centering
\includegraphics[width=6cm]{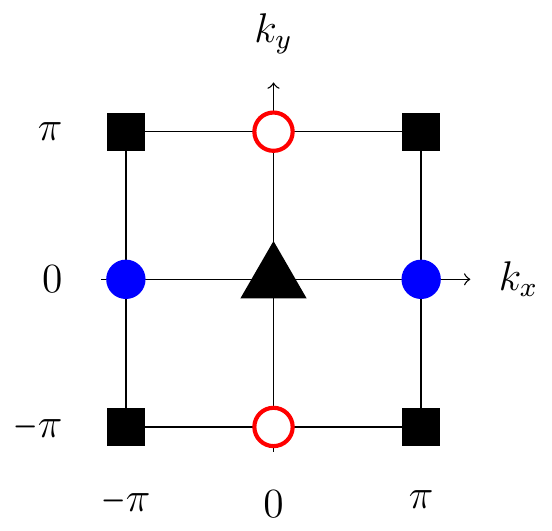}      
\caption{The four time-reversal invariant momenta in dimension $d=2$: $(0,0)$, $(\pi,0)$, $(0,\pi)$ et $(\pi, \pi)$. The Brillouin torus $\Torus{2}$ is represented as a primitive cell, whose sides must be glued together; points equivalent up to a lattice vector have been drawn with the same symbol.}
\label{fig:TRIM2D}
\end{figure}
 In view of the action of this time-reversal operation of the Bloch vector bundle, some points of the Brillouin torus appear 
 of high interest: the  points $\HighSymmetryPoint$ that are invariant under time-reversal \cite{FuKane2007}, {\it i.e.} 
 which verify $\HighSymmetryPoint = -\HighSymmetryPoint + G$ where $G$ is a reciprocal lattice vector.
 Therefore, they are the points $\HighSymmetryPoint = G / 2$, with $G$ a reciprocal lattice vector (see Fig.~\ref{fig:TRIM2D} for 
 the example of a two dimensional square lattice). 
  These points, usually named time-reversal invariant momenta (TRIM), or high-symmetry points, are fixed points of $\trb$
 and play an important role in time-reversal invariant systems. 
 In the following, we denote the set of TRIM as $\HighSymmetryPoints$.
At these time reversal invariant points, the two partners of a Kramers pair live in the same fiber. As they are orthogonal and possess
 the same energy, the spectrum is necessarily always degenerate at these TRIM points (see Fig.~\ref{fig:TypicalTimeReversalSpectrum}). 
  We will see in the following that the constraints imposed by the presence of these Kramers partners around the valence Bloch bundle are at the origin of the Kane--Mele topological order.

\subsubsection{Inversion symmetry}
\label{sec:Z2:Inversion}

In the previous section, time-reversal symmetry was shown to relate fibers at $k$ and $-k$ around the Brillouin torus. 
Hence, a particular spatial symmetry plays an important role when time reversal is at work: parity symmetry. 
The space inversion (or parity)
%\footnote{There are no differences, and these words are used interchangeably.}
 $\Parity$ reverses the space coordinates, i.e. its action on coordinates is $\Parity \vect{x} = - \vect{x}$. 
% \footnote{Note that as spin is a pseudovector, it is not modified by parity. In two dimensions, parity transforms $(x,y)$ into $(-x, -y)$ as can do a rotation. However, unlike a rotation, it leaves spin invariant. In three dimension, the action on space coordinates of parity and rotation are completely different, as parity has determinant $-1$ (see \cite{Murakami2011})}. 
In an inversion symmetric crystal, the lattice is left globally invariant by the inversion operation $\Parity$. However, different 
sublattices are exchanged. 
%and $\Parity$ may act nontrivially on the Hilbert space of internal degrees of freedom, and it may be useful to use a basis whose elements have a well defined parity eigenvalue\footnote{The sublattice degree of freedom can also be considered as internal, but it is more natural not to.}. For example, spin is a pseudovector, and thus even under parity. A $s$~orbital is also even under parity\footnote{The parity of an orbital is given by the parity $(-1)^{\ell}$ of the associated spherical harmonic, $\ell$ being its angular momentum quantum number.} but a $p$~orbital is odd.
Moreover $\Parity$ may act nontrivially on the Hilbert space of the internal degrees of freedom of the electronic states, depending on the atomic orbitals chosen to build the Bloch bands. 
Hence the explicit form of the parity operator will depend on the considered atomic basis. In the following, we will consider two particular cases of historical importance (with the bases \eqref{eq:TensorProductBasis}):
\begin{itemize}
    \item the Kane--Mele model: a tight-binding model built out of identical atomic orbitals on a bipartite lattice. In this case, the parity operator is diagonal 
    in the space of atomic orbitals: its only action is to exchange the sublattices, so 
     $\Parity = \BandPauliMatrix{x} \Tensor \IdentityMatrix$ 
    (up to a global sign\footnote{In the case of graphene, the parity eigenvalues of the $p_z$ orbitals is $-1$.} that does not affect the topological properties); 
   
    \item the Bernevig--Hughes--Zhang (BHZ) model: a tight-binding model built out of  atomic orbitals with opposite parity eigenvalues, but on a Bravais  lattice. This corresponds to $\Parity = \BandPauliMatrix{z} \Tensor \IdentityMatrix$ (e.g., the Bernevig--Hughes--Zhang model).
\end{itemize}

\noindent In the first case, $\BandPauliMatrix{x}$ exchanges the sublattices; in the second case, $\BandPauliMatrix{z}$ implements the parity eigenvalues of s and p orbitals. In both cases, the spin remains obviously unaffected.

\subsubsection{Simple four-band models and symmetries}
\label{sec:Z2:FourBandModel}

 We are now in a position to define the simplest insulator with spin-dependent time-reversal bands. The
Kramers degeneracy imposes that they correspond to a pair of Kramers related bands below a gap, and 
pair of Kramers related bands above the gap (see Fig.~\ref{fig:TypicalTimeReversalSpectrum}). 
 Hence the simplest model describing an insulator of electrons with spin and time-reversal symmetry is a four-level model.
The Bloch Hamiltonian of a four-level system (e.g., a two-level system  with spin) is a $4 \times 4$ Hermitian matrix. As a basis for the vector space of $4 \times 4$ Hermitian matrix, we can choose  
to use the identity matrix $\IdentityMatrix$,  five Hermitian gamma matrices $(\GammaMatrix{a})_{1\leq a \leq 5}$ which 
obey the Clifford algebra $\anticommutator{\GammaMatrix{a}}{\GammaMatrix{b}} = 2 \delta_{a,b}$, 
and their commutators $\GammaMatrix{a}{b} = (2 \ii)^{-1} \commutator{\GammaMatrix{a}}{\GammaMatrix{b}}$, ten of which are independent  \cite{FuKane2007}. 
 In such a basis, the Bloch Hamiltonian is parameterized by real function $d_i(k),d_{ij}(k)$ as: 

\begin{equation}
H(k) = 
d_{0}(k) \, \IdentityMatrix
\> + \>
\sum_{i=1}^{5} d_{i}(k) \, \GammaMatrix{i}
\> + \>
\sum_{i>j} d_{i  j}(k) \, \GammaMatrix{i}{j}
\label{eq:GenericFourByFourHamiltonian}
\end{equation}

The gamma matrices are constructed as tensor products of Pauli matrices that represent the two-level systems associated with 
a first degree of freedom and the spin of electrons. 
For the Kane--Mele model, the bipartite lattice has two sublattices A and B, corresponding to this first degree of freedom. In the BHZ model, this two-level system consists of the two different orbitals s and p associated with each lattice site. The tensor product of a sublattice (or orbital) basis (A, B) and 
a spin basis $(\uparrow, \downarrow)$ provides the basis
\begin{equation}
 \mathrm{(A, B)} \Tensor (\uparrow, \downarrow) = \mathrm{(A \, \uparrow, A \, \downarrow, B \, \uparrow, B \, \downarrow)}
 \qquad
 \text{or}
 \qquad
 \mathrm{(s, p)} \Tensor (\uparrow, \downarrow) =  \mathrm{(s \, \uparrow, s \, \downarrow, p \, \uparrow, p \, \downarrow)}
 \label{eq:TensorProductBasis}
 \end{equation}
 \noindent for the four-level system. The sublattice (orbital) operators are expanded on the basis ($\BandPauliMatrix{i}$), and the spin ones are expanded on ($\SpinPauliMatrix{i}$), where $\BandPauliMatrix{i}$ and $\SpinPauliMatrix{i}$ are sublattice (orbital) and spin Pauli matrices, and the zeroth Pauli matrix is taken to be the identity matrix. With those choices, the time-reversal operator reads:
\begin{equation}
\TR = \ii \, ( \IdentityMatrix \Tensor \SpinPauliMatrix{y} ) \, \ComplexConjugate. 
\end{equation}

Several conventions are possible for the gamma matrices, and it is judicious to choose them so that the symmetries of the Hamiltonian reflect into simple condition on the functions $d_i(k),d_{ij}(k)$ \cite{FuKane2007}. 
The expression for the \Ztwo invariant will turn out to be simpler for parity-invariant systems: this motivate our choice 
to impose this symmetry. 
Following Fu and Kane \cite{FuKane2007}, we choose the first gamma matrix $\GammaMatrix{1}$ to correspond to 
 the parity operator: $\GammaMatrix{1} = \Parity$. 
 Therefore, $\GammaMatrix{1}$ is obviously even under $\Parity$ and also under $\TR$. 
This choice ensures that the other $\Gamma_{i}$ matrices are odd under parity, {\it i.e.} 
$\Parity \GammaMatrix{i} \Parity^{-1} = \eta_{i} \, \GammaMatrix{i}$ with $\eta_{1} = + 1$ and $\eta_{j} = - 1$ for $j \geq 2$. 
Similarly, we obtain for the $\Gamma_{ij}$ matrices 
 $\Parity \GammaMatrix{i}{j} \Parity^{-1} = \eta_{i} \eta_{j} \, \GammaMatrix{i}{j}$. 
 Let us now enforce the $\Gamma_{i\geq 2}$ matrices to be odd under time-reversal symmetry: 
$\TR \GammaMatrix{i} \TR^{-1} = \eta_{i} \, \GammaMatrix{i}$. Due to the presence of $i$ in their definition, the 
$\GammaMatrix{i}{j}$ now follow a different rule under time-reversal symmetry than under parity: 
$\TR \GammaMatrix{i}{j} \TR^{-1} = - \eta_{i} \eta_{j} \, \GammaMatrix{i}{j}$
 Hence, with this convention, both $\Parity$ and $\TR$ symmetries imply consistent conditions on the function $d_i(k)$: 
 $d_1(k)$ is an even function around the TRIM points\footnote{On the Brillouin torus, the odd or even behaviour of a function happens around any TRIM. Let us consider a function $f$ and suppose that we have $f(k) = f(-k)$ for all $k$. Let $\HighSymmetryPoint \in \HighSymmetryPoints$ be a time-reversal invariant point. We have then $\HighSymmetryPoint = - \HighSymmetryPoint$, so $f(\HighSymmetryPoint + k) = f(-\HighSymmetryPoint - k) = f(\HighSymmetryPoint - k)$ for any $k$. Hence, if $f$ is even, it is also ``even'' around any TRIM. It obviously also works for an odd function.}
:  $d_{1}(k) = d_{1}(-k)$, while the functions 
 $d_{i}$ ($i>1$) are odd, {\it i.e.} $d_{i}(k) = - d_{i}(-k)$. On the other hand, the parity conditions imposed on the 
 functions $d_{ij}(k)$ by $\Parity$ and $\TR$ symmetries are opposite to each other and cannot be simultaneously satisfied: the 
 $d_{ij}(k)$ must vanish. These constraints can equivalently be deduced from the behavior of the matrices under the 
   $\Parity \TR$ symmetry: with our choice, the $\Gamma_i$ are even 
 $(\Parity \TR) \; \GammaMatrix{i} \; (\Parity \TR)^{-1} = \GammaMatrix{i}$ while their
 commutators  are odd under $\Parity \TR$: 
 $(\Parity \TR) \; \GammaMatrix{i}{j} \; (\Parity \TR)^{-1} = -\GammaMatrix{i}{j}$.
 
 Hence with the above conventions, we have reduced our study of $\Parity \TR$ invariant four band insulators from the 
 general Hamiltonian \eqref{eq:GenericFourByFourHamiltonian}  to the  simpler  Hamiltonian: 
\begin{equation}
H(k) = 
d_{0}(k) \, \IdentityMatrix
\> + \>
\sum_{i=1}^{5} d_{i}(k) \, \GammaMatrix{i} . 
\label{eq:InversionSymmetricFourByFourHamiltonian}
\end{equation}

\noindent Note that because of the $\Parity \TR$ symmetry, the spectrum of such an Hamiltonian is everywhere degenerate  
(Fig.~\ref{fig:TypicalTimeReversalSpectrum}, dashed lines). In the general case, it reads: 
\begin{equation}
E_{\pm}(k) = 
d_{0}(k) \pm \sqrt{\sum_{i=1}^{5} d_{i}^2(k)}
\label{eq:InversionSymmetricSpectrum}
\end{equation}

\noindent In the following, we neglect the $d_{0}$ coefficient, which plays no role in the topological properties of the system. 

 We will now turn to the detailed study of topological properties of two such four-band Hamiltonians. We will use the notion of obstruction to illustrate the occurrence of topological order in the valence bands of these models.  
Before proceeding, let us stress that in the presence of time-reversal symmetry, 
the bundle of filled bands $\FilledBandsBundle$ is \emph{always} trivial as a vector bundle. Hence, there is always a global basis of eigenstates $\ket{u_{i}}_{1 \leq i \leq 2}$ of the valence bundle perfectly defined on the whole Brillouin torus. However, the valence bundle $\FilledBandsBundle$ is not always trivial when endowed with the additional structure imposed by time reversal symmetry. 
Hence topological order will manifest itself as an impossibility to continuously define Kramers pairs on the whole Brillouin torus when the insulator is nontrivial, that is to say, the global basis cannot satisfy $\TR \ket{u_{1}(k)} = \ket{u_{2}(-k)}$. Hence, special care has to be devoted to this Kramers constraints when determining the valence bands' eigenstates. 
The aim of the following section is to demonstrate the occurrence of such an obstruction, before describing more general expressions of the topological index.

\subsection{Atomic orbitals of identical parity:  Kane--Mele-like model}
\label{sec:Z2:GrapheneLikeModel}

\subsubsection{Minimal model}
\begin{figure}
\centering
\includegraphics[width=6cm]{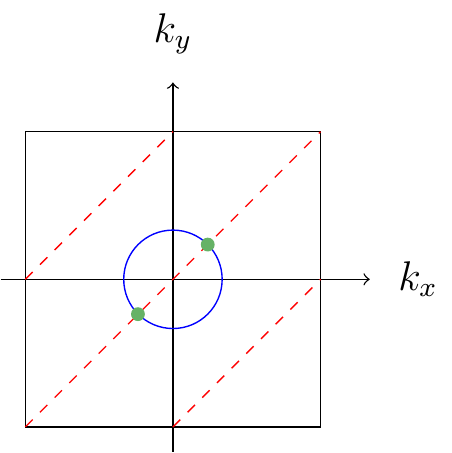}      
\caption{Examples of lines of zero of $d_1$ (continuous blue) and $d_2$ (dashed red) in the topological case, where $d_{1}$ is only negative (or only positive) at the origin. 
 For simplicity, these lines have been represented as straight lines, without loss of generality. 
At the intersections (green points) where $d_1=d_2=0$, singularities of the globally defined Kramers pairs of eigenvectors appear. These singularities cannot be removed by continuously deforming the parameters, unless the gap closes or the time-reversal invariance is broken.}
\label{fig:OnePointNegative}
\end{figure}

Let us consider a time-reversal invariant band insulator on an inversion symmetric bipartite lattice with spin, such as
 the Kane--Mele model introduced in \cite{KaneMele2005} and \cite{FuKane2007}. We work in the \enquote{sublattice tensor spin} basis $(A \,\uparrow, A \, \downarrow, B \, \uparrow, B \, \downarrow)$, where the parity operator only exchanges $A$ and $B$ sites: 
\begin{equation}
\Parity = \BandPauliMatrix{x} \Tensor \IdentityMatrix
\end{equation}
%%
%\noindent whereas the time-reversal operator still reads
%%
%\begin{equation}
% \TR = \ii \, ( \IdentityMatrix \Tensor \SpinPauliMatrix{y} ) \, \ComplexConjugate
%\end{equation}
%
\noindent The model is written in the form of eq.~\eqref{eq:InversionSymmetricFourByFourHamiltonian}, with the gamma matrices  chosen to be: 
%$\Gamma_1=\Parity$ and 
%
\begin{equation}\label{eq:GammaKaneMele}
\GammaMatrix{1} =\Parity = \BandPauliMatrix{x} \Tensor \IdentityMatrix
\qquad
\GammaMatrix{2} = \BandPauliMatrix{y} \Tensor \IdentityMatrix
\qquad
\GammaMatrix{3} = \BandPauliMatrix{z} \Tensor \SpinPauliMatrix{x}
\qquad
\GammaMatrix{4} = \BandPauliMatrix{z} \Tensor \SpinPauliMatrix{y}
\qquad
\GammaMatrix{5} = \BandPauliMatrix{z} \Tensor \SpinPauliMatrix{z}
\end{equation}
Following the discussion in the precious section \ref{sec:Z2:FourBandModel}, the parity and time-reversal constraints imply 
$d_1(k)$ to be an even function in the Brillouin torus, while the $d_{i\geq 2}(k)$ are odd functions. Hence, 
 all $d_{i}$ except $d_{1}$ vanish at the time reversal invariants points. Moreover, the  functions 
 $d_{i\geq 2}(k)$ should vanish around time-reversal invariant lines connecting those TRIM (see Fig.~\ref{fig:OnePointNegative}). 
 
  For the system to remain insulating, and due to the vanishing of the 
  $d_{i\geq 2}(\HighSymmetryPoint),$
 we must have $d_{1}(\HighSymmetryPoint) \neq 0$ for all TRIM $\HighSymmetryPoint \in \HighSymmetryPoints$. 
The quantities $d_{1}(\HighSymmetryPoint)$ correspond to the opposite of the parity eigenvalues $\xi(\lambda)$ 
of the bands at the TRIM 
\footnote{The energy of filled bands is always negative (with $d_{0} = 0$). Thus, at a TRIM $\HighSymmetryPoint$, the parity eigenvalue of the filled states is the opposite of the sign of the coefficient~$d_{1}(\HighSymmetryPoint)$. 
This is because
\newcommand{\ParityEV}{\ensuremath{\xi}}
$E(\HighSymmetryPoint) \, \ket{u} = d_{1}(\HighSymmetryPoint) \, \GammaMatrix{1} \, \ket{u} = d_{1}(\HighSymmetryPoint) \, \ParityEV(\HighSymmetryPoint) \, \ket{u}$
so $E(\HighSymmetryPoint) = d_{1}(\HighSymmetryPoint) \, \ParityEV(\HighSymmetryPoint)$. As the state is filled, we have $E(\HighSymmetryPoint) < 0$, so we get $\sign[\ParityEV(\HighSymmetryPoint)] = - \sign[d_{1}(\HighSymmetryPoint)]$.}.
In the following, we will try to convince the reader that the bulk invariant unveiled by Fu and Kane (see section~\ref{sec:Z2:SewingMatrixInvariant}):
\begin{equation}
\prod_{\HighSymmetryPoint \in \HighSymmetryPoints} \sign \, d_{1}(\HighSymmetryPoint)
\end{equation}
\noindent is indeed a topological index related to the obstruction to globally define Kramers pairs (especially of eigenvectors of the Bloch Hamiltonian).
Hence, for the model to display a topological insulating phase, the function $d_{1}(k)$ cannot take values of same sign at all the TRIM. Hence, this function should  vanish somewhere on the Brillouin torus. As $d_{1}$ is even, it will typically vanish on a time-reversal invariant loop around one or several TRIM. Hence, to  keep the gap open, at least two non-zero coefficients $d_{i\geq 2}$ are needed (see Fig.~\ref{fig:OnePointNegative}). 
As these functions $d_{i\geq 2}$ are odd, they vanish on time-reversal invariant curves connecting the TRIM which cross necessarily the loop where $d_{1}$ vanishes. 
With only one nonzero function $d_{a}$ in addition to $d_{1}$, there are always  crossing points where the gap vanishes, at the intersections of the $d_{1}=0$ loops and the $d_{i\geq 2}=0$ curves. Thus, to impose an insulating state, we need to consider a minimal model with at least two functions $d_{i\geq 2}$. 
% We can generically rule out the cases where the intersections of the zero lines are the same and the gap closes\footnote{We may have to turn on another coefficient to connect apparently different situations without closing the gap.}
To sum up, the simplest description of an insulating state with a possibly nontrivial topology is parameterized through 
eq.~\eqref{eq:InversionSymmetricFourByFourHamiltonian} by 
 three nonzero functions:  $d_{1}$ and two additional functions $d_{i\geq 2}$. 
 The simplest such example consists of choosing nonzero $d_{1}$, $d_{2}$, and $d_{5}$ values, as this choice preserves the spin quantum numbers. It is precisely the case considered by Fu and Kane \cite{FuKane2007} 
 to describe an inversion symmetric version of the Kane and Mele's model of graphene.

\subsubsection{Obstruction}

The filled normalized eigenstates with energy $-\norm{d} = -\sqrt{d_{1}^{2} + d_{2}^{2} + d_{5}^{2}}$ for the model defined in the previous section are, up to a phase\footnote{Notice that this expression only stands when $d_{4}=d_{3}=0$.}:
\begin{equation}
\ket{u_{1}} = 
\frac{1}{\mathcal{N}_{1}} \;
\begin{pmatrix}
    0 \\
    - d_{5} - \norm{d} \\
    0 \\
    d_{1} + \ii d_{2}
\end{pmatrix}
\qquad
\text{and}
\qquad
\ket{u_{2}} = 
\frac{1}{\mathcal{N}_{2}} \;
\begin{pmatrix}
    d_{5} - \norm{d} \\
    0 \\
    d_{1} + \ii d_{2} \\
    0
\end{pmatrix}, 
\label{eq:Z2:GrapheneLike:Eigenstates}
\end{equation}

\noindent where $\mathcal{N}_{j}(\vect{d})$ are positive coefficients ensuring that the vectors are normalized. Those vectors are obviously orthogonal, and thus form an orthonormal basis of the space of filled bands. We can easily verify that they also 
form Kramers pairs, {\it i.e.} that $\TR \ket{u_{1}(k)} = \ket{u_{2}(-k)}$.
To do so, let us  note that if $\ket{u_{j}[\vect{d}]}$ is an eigenvector at~$k$, the corresponding eigenvector at~$-k$ is obtained by changing the sign of all components of $\vect{d}$ except $d_{1}$.
Note that a naive point-wise diagonalisation of the Hamiltonian does not automatically provide Kramers pairs in the inversion symmetric case, in particular when $d_{3}$, $d_{4}$ are nonzero, {\it i.e.} when the spin projection $s_{z}$ is not conserved. The easiest and systematic procedure is then to remove the additional degeneracy by introducing an infinitesimal 
parity-breaking, time-reversal invariant perturbation, such as a constant $d_{12}$.

From \eqref{eq:Z2:GrapheneLike:Eigenstates}, we infer that the limit of these eigenstates is ill-defined when $d_{1},d_{2}\to 0$. 
To analyze this limit, we consider the polar decomposition $d_{1}+ \ii d_{2} = t \, \ee^{\ii \theta}$, to obtain: 
\begin{equation}
\ket{u_{1}} = 
\frac{1}{\mathcal{N}_{1}} \;
\begin{pmatrix}
    0 \\
    - d_{5} - \Abs{d_{5}} \sqrt{1 + (t/d_{5})^2} \\
    0 \\
    t \ee^{\ii \theta}
\end{pmatrix}
\qquad
\text{and}
\qquad
\ket{u_{2}} = 
\frac{1}{\mathcal{N}_{2}} \;
\begin{pmatrix}
    d_{5} - \Abs{d_{5}} \sqrt{1 + (t/d_{5})^2} \\
    0 \\
    t \ee^{\ii \theta} \\
    0
\end{pmatrix}. 
\end{equation}
\noindent In the limit $t \to 0$ (while keeping the vectors normalized), we obtain 
 (see \ref{app:sec:demoKMlikeEigenvectors} for details):
\begin{subequations}
\begin{equation}
\ket{u_{1}} \to 
\begin{pmatrix}
    0 \\
    - 1 \\
    0 \\
    0
\end{pmatrix}
\qquad
\text{and}
\qquad
\ket{u_{2}} \to 
\begin{pmatrix}
    0 \\
    0 \\
    \ee^{\ii \theta} \\
    0
\end{pmatrix}
\qquad
\qquad
\text{ for $d_{5} > 0$, }
\label{eq:Z2:GrapheneLike:LimitOfEigenvectorsAtPositived5}
\end{equation}
\noindent and 
\begin{equation}
\ket{u_{1}} \to 
\begin{pmatrix}
    0 \\
    0 \\
    0 \\
    \ee^{\ii \theta}
\end{pmatrix}
\qquad
\text{and}
\qquad
\ket{u_{2}} \to 
\begin{pmatrix}
    -1 \\
    0 \\
    0 \\
    0
\end{pmatrix}
\qquad
\qquad
\text{ for $d_{5} < 0$}.
\label{eq:Z2:GrapheneLike:LimitOfEigenvectorsAtNegatived5}
\end{equation}
\label{eq:Z2:GrapheneLike:LimitOfEigenvectors}
\end{subequations}
The phase $\theta$ is ill-defined when $t \to 0$, so one of these eigenstates is ill-defined 
at the points where $d_{1}=d_{2}=0$ are ill-defined. It is possible to show that such a singularity cannot be removed by a \UnitaryGroup{2} change of basis preserving the Kramers pairs structure (see also \cite{GrafPorta2012} for a related point of view). 
%Notice that a Kramers pair of states at $t\to 0$ is composed of e.g. $\ket{u_{1}}[d_5>0]$ and $\ket{u_{2}}[d_5<0]$ (or $\ket{u_{1}}[d_5<0]$ and $\ket{u_{2}}[d_5>0]$). These states have no relative phase, i.e. there is a common phase choice for Kramers partners. The \Ztwo obstruction is an obstruction to globally define such Kramers pairs.
%
% (i.e. under the constraint $\swm = J$ on the sewing matrix $\swm$ defined in section~\ref{par:SewingMatrix} p.~\pageref{par:SewingMatrix}).
%

\begin{figure}[ht]
\centering
\includegraphics[width=15cm]{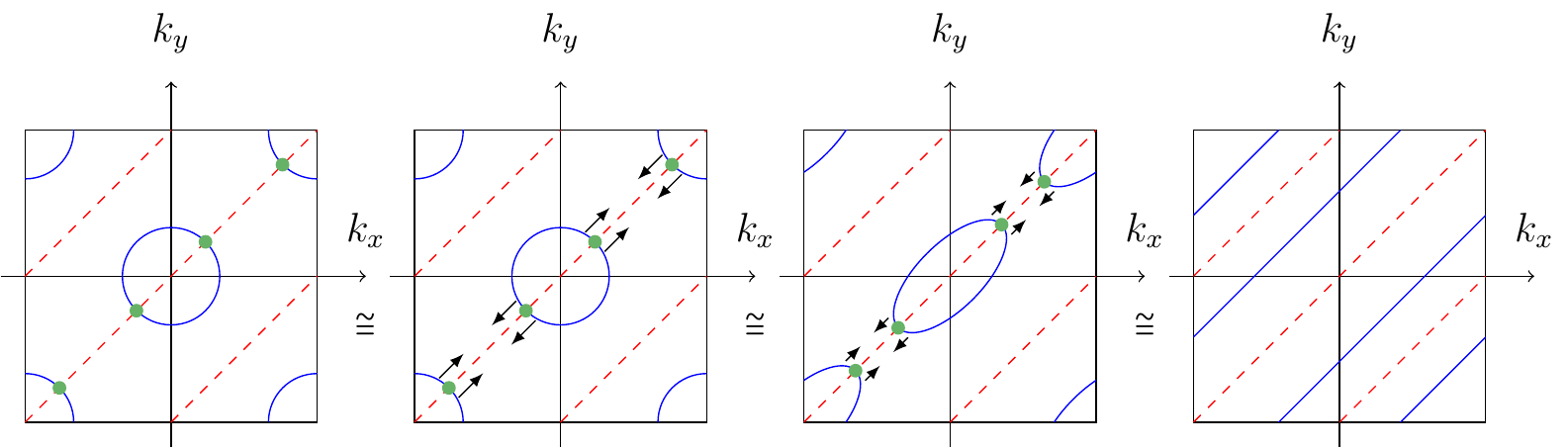}      
\caption{Zero lines of $d_1$ (continuous blue) and $d_2$ (dashed red) in the topological case. In this case, there are two TRIM with positive $d_{1}$ and two TRIM with negative ones. Although there are singularities (left-hand side), a smooth deformation (middle) of the Hamiltonian can remove them (right-hand side). This can be understood as the annihilation of vortices of $d_{1} + \ii d_{2}$, which automatically have opposite vorticities because of the geometry.}
\label{fig:TwoPointsNegative}
\end{figure}

 Let us now discuss the existence of points $k$ in the Brillouin torus where  both $d_{1}(k)$  and $d_{2}(k)$ vanish. 
When $d_1(k)$ takes values of constant sign on the whole Brillouin torus, such points cannot arise and there is no obstruction to define Kramers pairs of eigenstates from \eqref{eq:Z2:GrapheneLike:Eigenstates}. This corresponds to a topologically trivial insulator. If however $d_1(k)$ changes sign, there appear time reversal symmetric lines  along which $d_1(k)$ vanishes. If 
these lines do not encircle one of the TRIM, a smooth deformation of the Hamiltonian allows us to shrink and remove these lines without closing the gap. However, if one such line encircles a TRIM, the condition of $d_1(\HighSymmetryPoint)\neq 0$ at the TRIM points $\HighSymmetryPoint \in \HighSymmetryPoints$ 
implies that this line cannot be removed without closing the gap. 
In this last case, which corresponds to the situation where $d_1(k)$ takes values of 
opposite sign at the different TRIM, as the $d_{1}=0$ loops cannot cross the TRIM, there are always common zeros of $d_{1}$ and $d_{2}$ -- hence singularities for the eigenfunctions. This corresponds to an obstruction to globally define the Kramers pairs of eigenstate on the Brillouin torus, and thus to a topological non-trivial insulator (see Fig.~\ref{fig:TwoPointsNegative}). 

To properly relate the existence of this obstruction, {\it i.e.} the occurrence of the singularities (\ref{eq:Z2:GrapheneLike:LimitOfEigenvectors}), to topological properties of the phase, 
we study the effect of a continuous deformation of the Hamiltonian preserving time-reversal, parity invariance, and the existence of a gap. 
The lines where the odd functions $d_{i\geq 2}=0$ can be deformed provided they remain invariant under $k \to -k$, and that $d_{i}=0$ at each TRIM. One can globally change the sign of the $d_{i}$'s without altering the topology of the system. Hence, we can suppose that at least two TRIM have a positive $d_{1}$ without loss of generality. The reader can convince himself, e.g., by drawing little diagrams showing the zero lines of the coefficients, that the only cases where the singularities $d_{1} = d_{2}=0$ cannot be removed  
corresponds to the situation where the line $d_{1}(k)=0$ encircles only one of the TRIM 
(see Fig.~\ref{fig:OnePointNegative}), {\it i.e.} when 
$\prod_{\HighSymmetryPoint \in \HighSymmetryPoints} \sign \, d_{1}(\HighSymmetryPoint)<0$. 
In the cases where $d_{1}$ is negative at two different TRIM (Fig.~\ref{fig:TwoPointsNegative}), we can  remove 
the singularities by a smooth deformation (Fig.~\ref{fig:TwoPointsNegative}) or there is no singularity at all.
\begin{figure}[ht]
\centering
\includegraphics[width=10cm]{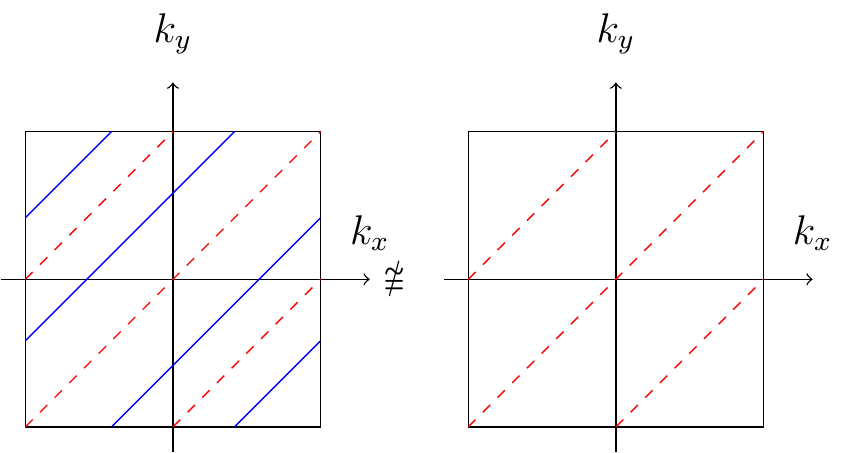}      
\caption{Zero lines of $d_1$ (continuous blue) and $d_2$ (dashed red) in two \Ztwo-trivial cases.  If we maintain the inversion symmetry, there is no way to deform a system where $d_{1}$ changes sign to a system where it has a constant sign.}
\label{fig:DeformationsWithParity}
\end{figure}
This direct obstruction  allows us to identify the quantity 
$\prod_{\HighSymmetryPoint \in \HighSymmetryPoints} \sign \, d_{1}(\HighSymmetryPoint)$ as a topological 
index for the insulating phases of this model. This is precisely the expression of the Kane--Mele invariant for parity symmetric Hamiltonian, as derived by Fu and 
Kane \cite{FuKane2007}. 
Moreover, the singularities of the globally defined Kramers pairs that we have identified, which occur at the Dirac points of the underlying graphene model, 
correspond to the vortices of the Pfaffian of the matrix introduced by Kane and Mele in their seminal article \cite{KaneMele2005}  (see section~\ref{sec:Z2:PfaffianInvariant}).

 We can notice that in the above argument, 
parity breaking is required to adiabatically deform a model where $d_{1}(k)$ changes sign twice in the Brillouin torus  to a model where $d_{1}$ has a constant sign (Fig.~\ref{fig:TwoPointsNegative} and \ref{fig:DeformationsWithParity}).  
Indeed, as all coefficients $d_{i\geq 2}$ are zero at the TRIM, it is not possible for $d_{1}$ to change sign at those points 
 without an additional parity breaking  function $d_{ij}$ maintaining the gap open. 
 Indeed, these models correspond to identical values of the \Ztwo Kane--Mele--Fu invariant. But additional finer
  topological classes appear when only parity symmetric deformation of the Hamiltonian are allowed. 
 This property is in agreement with the recent results of Alexandradinata {\it et al.} who have identified  an integer-valued 
 $\IntegerRing$ topological invariant classifies \emph{inversion}-symmetric topological insulators in two and three dimensions
 \cite{AlexandradinataDaiBernevig2012}. 

 The above discussion shows that the twist of a topologically nontrivial valence band structure originates from the mixing of orbitals with opposite parities in the Brillouin zone, or more generally eigenstates mutually orthogonal in different points of the Brillouin zone. 
 %This is to be contrasted with the usual situations of insulators bands  that can be smoothly connected to the atomic limit 
 %of identical parity bands. 
 
\subsubsection{Role of the dimension of the base space}

\begin{figure}[ht]
\centering
\includegraphics[width=12cm]{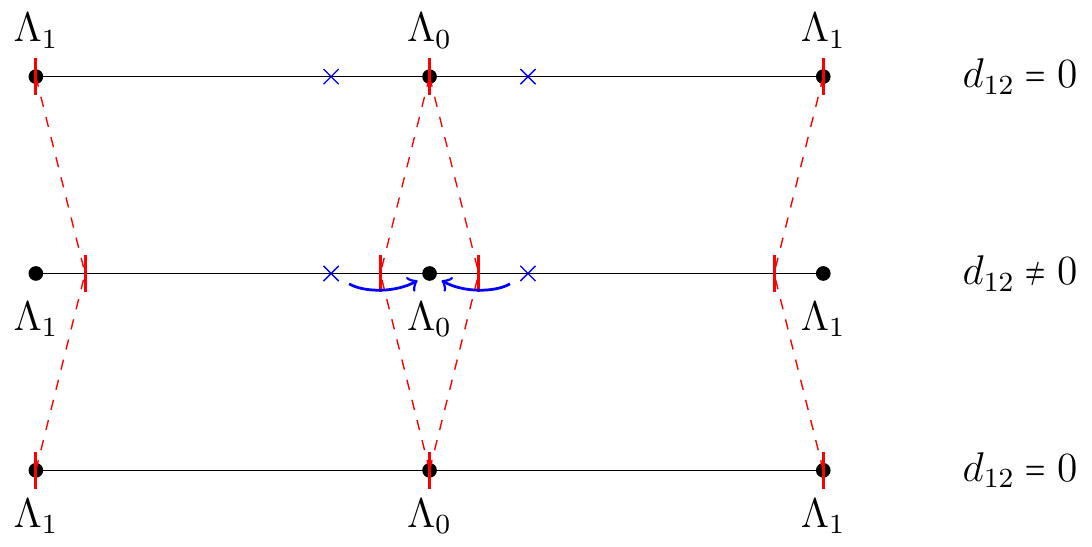}      
\caption{In one dimension, the sign of $d_{1}$ at the TRIM does not give rise to a topological invariant when inversion symmetry breaking perturbations are allowed. The blue cross represents the points where $d_{1}=0$ (accross which $d_{1}$ changes sign) and the black dots the points where $d_{2}=0$. Above, $d_{1 \, 2} = 0$, and the red vertical lines show the points where $d_{5}=0$. Below, we have switched on $d_{1 \, 2} \neq 0$, and the red lines show the points where $(d_{5} \pm d_{1 \, 2}) = 0$. The constraint to keep an open gap implies that these three kinds of points must never be at the same place: when $d_{1 \, 2} = 0$, it is not possible to deform the system so that $d_{1}$ has a constant sign without closing the gap. On the opposite, when  $d_{1 \, 2} \neq 0$, there is no problem to do it.}
\label{fig:OneDimensionalDeformation}
\end{figure}
\begin{figure}[ht]
\centering
\includegraphics[width=12cm]{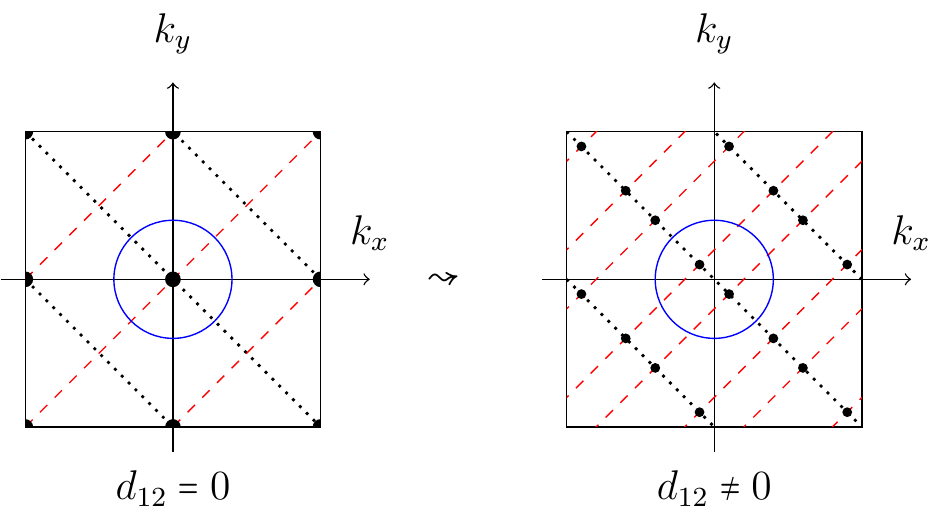}      
\caption{Two-dimensional case. The zero lines of $d_{1}$ (continuous blue), of $d_{5} \pm d_{1 \, 2}$ (dashed red) and of $d_{2}$ (dotted black) are drawn on the Brillouin zone. We start from the inversion-symmetric case where $d_{1 2}=0$ (left) and switch on $d_{1 2}$ (right). The points that the $d_{1}=0$ line cannot cross are marked with black circles. In this case, we see that it is not possible to deform $d_{1}$ into a constant value and to switch off $d_{1 2}$ without closing the gap.}
\label{fig:TwoDimensionalDeformation}
\end{figure}

The classification of time-reversal invariant insulators 
tells us that a \Ztwo invariant exists in two and three dimensions, but not in one dimension \cite{Kitaev2009,RyuSchnyderFurusakiLudwid2010,FreedMoore2013}. 
This result can be illustrated in the above discussion on this simple model with only $d_{1}(k)$, $d_{2}(k)$ and $d_{5}(k)$ functions. 
To proceed, 
 we need to add a small parity-breaking term such as an odd function $d_{1 \,2}$, as argued above. 
 In this case, the filled bands spectrum becomes:

\begin{equation}
    \sqrt{d_{1}^{2} + d_{2}^{2} + \left( d_{5} \pm d_{1 2} \right)^{2}}
\end{equation}

\noindent We thus have to consider zero lines of $d_{5} \pm d_{12}$ instead of $d_{2}$ when ensuring the gap does not close. 
We then realize that in one dimension, we can adiabatically deform $d_{1}(k)$ to a constant function 
(Fig.~\ref{fig:OneDimensionalDeformation}), while this is not possible 
in two dimensions (Fig.~\ref{fig:TwoDimensionalDeformation}). 
Indeed, in two dimensions, $d_{1}=0$ forms a loop around a TRIM (Fig.~\ref{fig:TwoDimensionalDeformation}), so that there are always points where $d_{1}=d_{5}=0$. The lines where $d_{5} \pm d_{1 2} = 0$ cannot cross these points without closing the gap.

\subsubsection{Time-reversal breaking.}

\begin{figure}[ht]
\centering
\centering
\includegraphics[width=6cm]{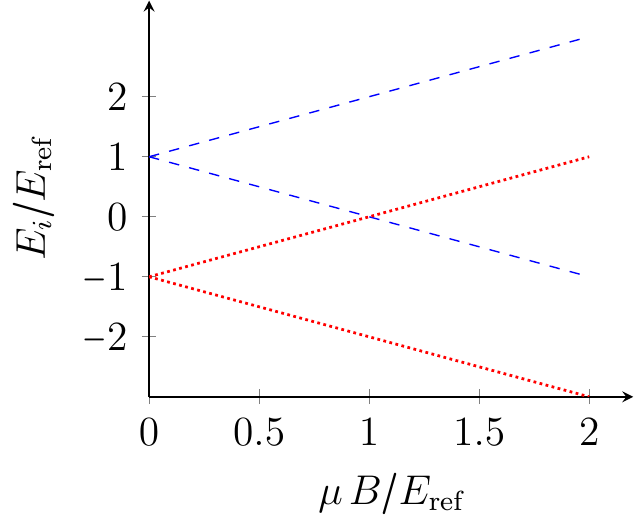}      
\caption{Zeeman spliting of the energy levels $E_{i}$ under a transverse magnetic field $B$. The levels originating from the filled bands at $B=0$ are drawn in red dotted lines, while the empty ones are drawn in blue dashed lines. At $B/(E_{\text{ref}}/\mu) = 1$, the crossing of bands enables a change in the topology of the system.}
\label{fig:Zeeman}
\end{figure}

Let us now consider the effect of a time-reversal breaking Zeeman term 
$H_{Z}= g B~ \IdentityMatrix \Tensor \SpinPauliMatrix{z}$, corresponding to a constant function 
$d_{3 4} = g \, B$ in addition to the previous functions $d_{1}$, $d_{2}$ and $d_{5}$. 
The global degeneracy of the eigenstates is now lifted, but the eigenstates \eqref{eq:Z2:GrapheneLike:Eigenstates} 
of the Hamiltonian are not modified. The full spectrum of the Bloch Hamiltonian is now:
\begin{equation}
- \norm{d} - d_{3 4}
\quad; \quad
- \norm{d} + d_{3 4}
\quad; \quad
\norm{d} - d_{3 4}
\quad; \quad
\norm{d} + d_{3 4}
\end{equation}

\noindent where $\norm{d} = \sqrt{d_1^2+d_2^2+d_5^2}$. It is represented in Fig.~\ref{fig:Zeeman}. 
As a consequence, it is not possible in this case to deform with this Zeeman perturbation 
a topological insulator into a trivial insulator without closing the gap.

\subsubsection{Edge states}

At the boundary  between a Kane--Mele topological insulator and a trivial insulator,
 a helical gapless edge states occur: the spin and the direction of these edge states are tight together. 
To understand the origin of these edge states we will proceed similarly to  the Chern insulator's discussion in Haldane's model 
in section \ref{sec:ChernSurfaceStates}. 
Let us consider the Kane--Mele model (\ref{eq:InversionSymmetricFourByFourHamiltonian},
\ref{eq:GammaKaneMele}) with only $d_1,d_2,d_5$ non-zero parameter functions. 
 In a trivial insulator, the parity eigenvalues $-d_{1}$ have a uniform sign, {\it e.g.} positive, at all TRIM, whereas in a nontrivial insulator it changes sign between the 
 TRIM : {\it e.g.} positive all TRIM except one $\HighSymmetryPoint_{0}$ where it is negative. 
 To  continuously describe an interface from a trivial to a \Ztwo topological phase without breaking 
 time-reversal symmetry requires a change of sign of $d_1$ at this particular TRIM $\HighSymmetryPoint_{0}$: this corresponds to a gap closing and 
 the appearance of a surface state. 
 This suggests that the low energy physics of the interface is captured by an analysis around the TRIM in Kane--Mele insulators.

We denote by $\HighSymmetryPoint_{i}$ ($i=0,\cdots,3$) the four TRIM in $d=2$. As an example, let us consider an interface at $y=0$ between a trivial insulator for $y>0$ where $d_{1}$ is positive at all TRIM, and a \Ztwo topological insulator 
for $y<0$ 
where only $d_{1}(\HighSymmetryPoint_{0})$ is negative.
 The dispersion relation at the $\HighSymmetryPoint_{0}$ point is the only one involving a sign change of $d_1$: we naturally focus on the dispersion around this point, while a smooth evolution of the dispersion is expected elsewhere on the Brillouin torus. 
  Let us now define 
 $m (y)=d_{1}[\HighSymmetryPoint_{0}](y)$:  we have always $m (y>0)>0$ and $m (y<0)<0$. 
 The linearized Hamiltonian around the TRIM $\HighSymmetryPoint_{0}$ reads, up to a rotation of the local coordinates on the Brillouin zone $(q_x,q_y)$: 
\begin{equation}
H_{\text{l}}(q) = q_x \GammaMatrix{5} - q_{y} \GammaMatrix{2} + m (y) \, \GammaMatrix{1} , 
\end{equation}
where we used the oddness of the  functions $d_{i\geq 2}$ around $\HighSymmetryPoint_{0}$. 
We have chosen local coordinates so that $d_{5}(q)=q_{x}$ and $d_{2}(q)=-q_{y}$, in order to simplify the calculations. To describe edge states, it is useful to block-diagonalize the Hamiltonian in the ``sublattice tensored with spin'' basis $\mathrm{(A \uparrow, B \uparrow, A \downarrow, B \downarrow)}$ in which it reads in real space representation (through the substitution  $q \to - \ii \nabla$): 
\begin{equation}
H_{\text{l}} = 
%\begin{pmatrix}
%q_{x} & m(y) + \ii q_{y} & 0 & 0 \\
%m(y) - \ii q_{y} & -q_{x} & 0 & 0 \\
%0 & 0 & - q_{x} & m(y) + \ii q_{y} \\
%0 & 0 & m (y)- \ii q_{y} & q_{x}
%\end{pmatrix} =
 \begin{pmatrix}
H_{\uparrow} & 0 \\
0 & H_{\downarrow}
\end{pmatrix}, 
\label{eq:EdgeStateMatrix}
\end{equation}
%
%When substituing $q$ with $- \ii \nabla$, we get
with
\begin{equation}
H_{\uparrow}= \begin{pmatrix}
- \ii \partial_{x} & m(y) + \partial_{y} \\
m(y) - \partial_{y} & \ii \partial_{x}
\end{pmatrix}
\qquad
\text{and}
\qquad
H_{\downarrow} = \begin{pmatrix}
+ \ii \partial_{x} & m(y) + \partial_{y} \\
m(y) - \partial_{y} & \ii \partial_{x}
\end{pmatrix} .
\label{eq:EdgeStateMatrixSubstitution}
\end{equation}
As discussed in the section \ref{sec:ChernSurfaceStates} for our choice of $m(y)$, the Schrödinger equation:
\begin{equation}
\begin{pmatrix}
\Op{H}_{\uparrow} & 0 \\
0 & \Op{H}_{\downarrow}
\end{pmatrix}
\,
\psi(x,y) = E \psi(x,y) 
\end{equation}
\noindent possesses solutions:
\begin{equation}
\psi_{q_{x},\uparrow}(x,y) \propto \ee^{- \ii q_{x} \, x} \, \exp \left[ - \int_{0}^{y} m(y') \, \dd{}y' \right] \; 
\begin{pmatrix}
  0 \\
  1 \\
  0 \\
  0
\end{pmatrix}
\end{equation}

\begin{equation}
\psi_{q_{x},\downarrow}(x,y) \propto \ee^{+ \ii q_{x} \, x} \, \exp \left[ - \int_{0}^{y} m(y') \, \dd{}y' \right] \; 
\begin{pmatrix}
  0 \\
  0 \\
  0 \\
  1
\end{pmatrix}, 
\end{equation}
one of which is a spin-up right-moving state, while the other one is a spin-down left-moving state. These states obviously constitute a Kramers pair of edge states. 
\begin{figure}[ht]
\centering  
\includegraphics[width=8cm]{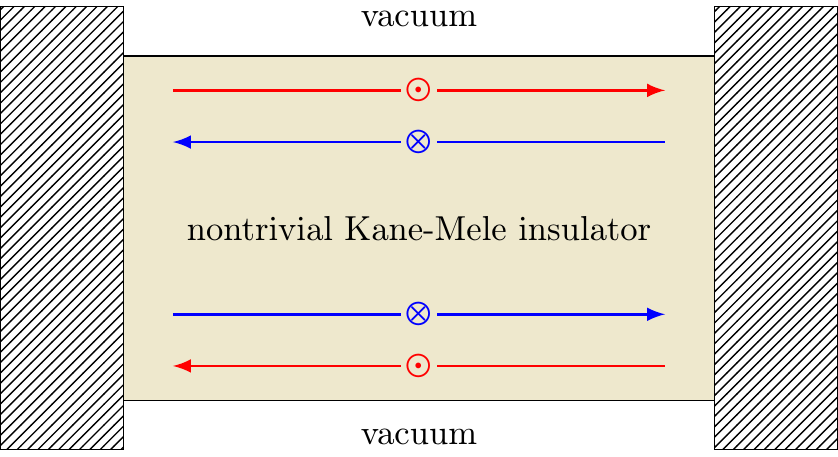}      
\caption{
Surface states of a Kane--Mele insulator. On each interface with a trivial insulator (e.g., vacuum), spin-up states ($\odot$, in red) and spin-down states ($\otimes$, blue) propagate in opposite direction.
}
\label{fig:Z2Edge}
\end{figure}
A schematic representation of such a pair of edge states is represented in Fig.~\ref{fig:Z2Edge}. 
This demonstrates the existence of helical edge states at the interface between a trivial and a topological insulating phase in the Kane--Mele model. 
A mathematical discussion on this bulk-edge correspondence that goes far beyond the present introduction can be found in 
\cite{GrafPorta2012}, while 
a pedagogical discussion of the existence of these edge states in \Ztwo insulators is presented in the book by Fradkin \cite{Fradkin}, in relation with previous work by \cite{JackiwRebbi1976}.

%%%%%%%%%%%%%%%%%%%%%%%%%%%%%%%%%%%%%%%%%
\subsection{Atomic orbitals of opposite parity:  Bernevig--Hughes--Zhang-like model}
\label{sec:Z2:BHZLikeModel}

Let us  consider a second time-reversal invariant band insulator, designed on 
an inversion symmetric lattice, but with two atomic orbitals of different parities per site (e.g. s and p orbitals). 
This is the case in the BHZ model introduced in \cite{BHZ2006}, as studied in \cite{FuKane2007}. We now work in the \enquote{orbitals tensor spin basis} $(s \, \uparrow, s \, \downarrow, p \, \uparrow, p \, \downarrow)$. The parity operator is  diagonal here, with two different eigenvalues, e.g., s and p orbitals, and reads: 

\begin{equation}
\Parity = \BandPauliMatrix{z} \Tensor \IdentityMatrix
\end{equation}

\noindent while the expression of the time-reversal operator obviously does not change. The gamma matrices are chosen as:
\begin{equation}
\GammaMatrix{1} = \BandPauliMatrix{z} \Tensor \IdentityMatrix 
\ ; \ 
\GammaMatrix{2} = \BandPauliMatrix{y} \Tensor \IdentityMatrix 
\ ; \ 
\GammaMatrix{3} = \BandPauliMatrix{x} \Tensor \SpinPauliMatrix{x} 
\ ; \ 
\GammaMatrix{4} = \BandPauliMatrix{x} \Tensor \SpinPauliMatrix{y} 
\ ; \ 
\GammaMatrix{5} = \BandPauliMatrix{x} \Tensor \SpinPauliMatrix{z}.
\end{equation}
\noindent The constraints on the coefficients of the Hamiltonian are the same as in section \ref{sec:Z2:GrapheneLikeModel}. However, the singularities of the Kramers eigenstates are different.
We consider the same set of nonzero  $d_{1}$, $d_{2}$ and $d_{5}$ functions that in section  \ref{sec:Z2:GrapheneLikeModel}.
Diagonalisation of the Hamiltonian, with special care to ensure Kramers degeneracy, 
yields the following eigenvectors for the filled bands with energy $-\norm{d} = -\sqrt{d_{1}^{2} + d_{2}^{2} + d_{5}^{2}}$:

\begin{equation}
\ket{u_{1}} = 
\frac{1}{\mathcal{N}_{1}} \;
\begin{pmatrix}
    0 \\
    \ii  \left(  \norm{d}- d_{1}  \right) \\
    0 \\
    d_{2} + \ii d_{5}
\end{pmatrix}
\qquad
\text{and}
\qquad
\ket{u_{2}} = 
\frac{1}{\mathcal{N}_{2}} \;
\begin{pmatrix}
    \ii  \left(  \norm{d} - d_{1}  \right) \displaystyle\frac{d_{2} + \ii d_{5}}{d_{2} - \ii d_{5}} \\
    0 \\
    d_{2} + \ii d_{5} \\
    0
\end{pmatrix} .
\label{eq:Z2:BHZLike:Eigenstates}
\end{equation}
For these states, 
the singularities appear when $d_{2}=d_{5}=0$. Through the polar decomposition $d_{2}+ \ii d_{5} = t \, \ee^{\ii \theta}$, we obtain the limit $t \to 0$:
\begin{equation}
\ket{u_{1}} = 
\frac{1}{\mathcal{N}_{1}} \;
\begin{pmatrix}
    0 \\
    0 \\
    0 \\
    1
\end{pmatrix}
\qquad
\text{and}
\qquad
\ket{u_{2}} = 
\frac{1}{\mathcal{N}_{2}} \;
\begin{pmatrix}
    0 \\
    0 \\
    1 \\
    0
\end{pmatrix}
\qquad
\qquad
\text{ for $d_{1} > 0$} , 
\end{equation}
and
\begin{equation}
\ket{u_{1}} = 
\frac{1}{\mathcal{N}_{1}} \;
\begin{pmatrix}
    0 \\
    \ii \ee^{- \ii \theta} \\
    0 \\
    0
\end{pmatrix}
\qquad
\text{and}
\qquad
\ket{u_{2}} = 
\frac{1}{\mathcal{N}_{2}} \;
\begin{pmatrix}
    \ii \ee^{- \ii \theta} \\
    0 \\
    0 \\
    0
\end{pmatrix}
\qquad
\qquad
\text{ for $d_{1} < 0$}.
\end{equation}
Note that when $d_{1} > 0$, the eigenvectors possess only p orbitals components, whereas when $d_{1} < 0$, they are only supported by s orbitals components. This is coherent with the idea 
of mixing the atomic bands with different parties in a \Ztwo insulator.

In the trivial case where $d_{1}$ is always positive, there are no singularities. In the trivial case where $d_{1}$ is always negative, we can remove the singularities that appear at all TRIM by Kramers-pairs-preserving \UnitaryGroup{2} transformations that get rid of the ill-defined phases of the states. However, in the topological case where $d_{1}$ changes sign, there are singularities at the TRIM that cannot be removed. Constructing the continuous transformations that allow us to identify the topological properties of this model requires a different analysis than in the previous model. This model-dependent obstruction identification definitely begs for 
a more formal approach to the \Ztwo topology, independent of the specificities of the models. We now turn to the discussion of such an approach.

%%%%%%%%%%%%%%%%%%%
\subsection{\Ztwo invariants}

\subsubsection{Introduction}

There exist several equivalent expressions of the \Ztwo invariant, especially in two dimensions, which are all useful for different purposes. Historically, the first expression discovered by Kane and Mele \cite{KaneMele2005}  is the ``Pfaffian invariant'' (see section~\ref{sec:Z2:PfaffianInvariant}), which highlights the topological nature of the invariant and its connection with a ``twist'' of the vector bundle where all bands are intertwined. 
The ``sewing matrix invariant'' (section~\ref{sec:Z2:SewingMatrixInvariant}) designed by Fu and Kane \cite{FuKane2006} paves the way to a more intrinsic understanding of the \Ztwo topology, as it does not require an immersion of the filled bands bundle in the larger trivial bundle of all bands. 
Moreover, it underlines the rigidity of the \Ztwo topology, which is only determined by the behaviour of Kramers pairs at the time-reversal invariant points.
This approach requires a continuous set of eigenstates on the Brillouin torus, which is unfortunately impractical numerically \cite{SoluyanovVanderbilt2011}; however, it leads to a deeper theoretical understanding of the \Ztwo topology in terms of Wilson loops as $SU(2)$ holonomies \cite{LeeRyu2008}, as well as to expressions of practical importance.
In particular, Fu, Kane, and Mele \cite{FuKaneMele2007} used this expression  to generalize the \Ztwo order to
three dimensions.
When inversion symmetry is present, Fu and Kane \cite{FuKane2007} have shown that this \Ztwo invariant acquires a very simple form as a function solely of the parity eigenvalues of the filled bands at the TRIM. 
Finally, the sewing matrix invariant is a useful starting point in the connection to the 
 K-theoretical point of view \cite{FreedMoore2013} of classification of time-reversal invariant fiber bundles.
The ``obstruction invariant'' (section~\ref{sec:Z2:ObstructionInvariant}), formulated by Roy \cite{Roy2009} and 
by Fu and Kane \cite{FuKane2006}, highlights the similarity with the Chern invariant. 
A similar point of view with homotopy arguments was used by Moore and Balents \cite{MooreBalents2007} to understand the \Ztwo topology in three dimensions.
Furthermore, the expression of the \Ztwo invariant as an obstruction is of practical importance in the numerical determination of the topological order \cite{SoluyanovVanderbilt2011}.
As shown in the previous section \ref{sec:Z2:GrapheneLikeModel},
this obstruction point of view arises when considering simple examples, where singular eigenvectors appear naturally.
Notice that other expressions of the Kane-Mele \Ztwo invariant have been discussed, e.g. based on a Chern-Simons topological effective field theory \cite{QiHughesZhang:2008,WangQiZhang:2010}.

\subsubsection{General considerations}

\paragraph{Time-reversal effective Brillouin zone}
\label{sec:EffectiveBrillouinZone}
\begin{figure}[ht]
\centering
\includegraphics[width=15cm]{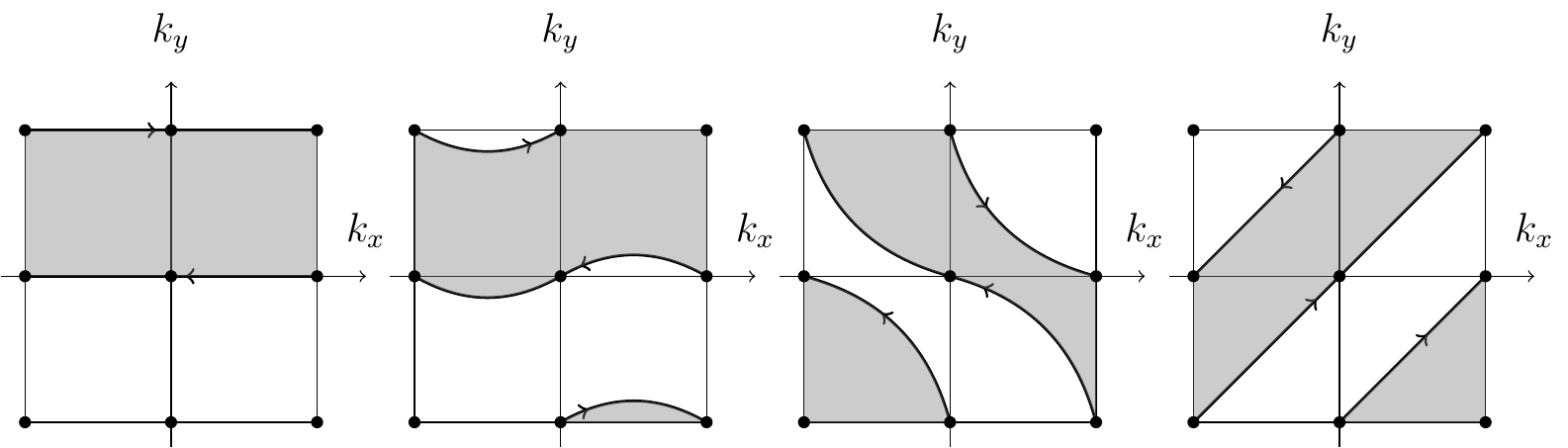}      
\caption{Examples of Effective Brillouin Zones (EBZ). The EBZ are filled in gray, and their boundaries are oriented thick lines (which represent closed curves on the torus). The TRIM are drawn as black circles. }
\label{fig:EBZExamples}
\end{figure}

As time reversal maps the fibers at $k$ and $-k$, there is a redundancy in the description of the system on the whole Brillouin torus when time-reversal constraints are enforced. 
To describe the properties of the system, it is possible to circumvent this redundancy by introducing a time-reversal effective Brillouin zone (EBZ) \cite{MooreBalents2007}. This EBZ consists of half of the Brillouin torus, keeping only one member of each Kramers pair $(k, -k),$ except at the boundary. 
Each half (with the boundary) defines an EBZ (Fig.~\ref{fig:EBZExamples}), which has the topology of a cylinder. 
The boundary of an EBZ consists of two homotopic time-reversal invariant closed curves connecting two TRIM. 
Notice that all TRIM are necessarily inside an EBZ.
As the choice of the boundary curves is free (e.g., they are not necessarily straight), there are many different EBZs. However,
 the precise choice of this effective Brillouin zone will not affect the discussion below. 

\paragraph{Matrix elements of the time-reversal operator}
\label{sec:TRMatrixElements}

From now on, let 
$(e_i(k))_{i=1}^{2m}$
 be a global basis of of the filled band fiber \FilledBandsFiber{k} at $k$, i.e. a collection of never-vanishing global sections of 
 the valence bundle that, at each point, form a basis of the valence fiber. Let us stress that in the general case, the $( e_{i}(k) )$ are not required to be eigenstates of the Hamiltonian: 
%, especially when the valence bands bundle is not trivial: 
they constitute a basis of the Hilbert  subspace spanned by the eigenstates. 
The most intuitive way to define matrix elements of the time-reversal operator is to define: 
\begin{equation}
\KMTRMatrix_{i j}(k) = \braket{e_{i}(k) | \TR e_{j}(k)}
\label{eq:KMTRMatrixDefinition}
\end{equation}
\noindent where $( e_{i}(k) )_{i}$ is a global basis of the filled bundle. This corresponds to the original formulation 
 of Kane and Mele in \cite{KaneMele2005}. This matrix is not unitary, but it is antisymmetric (because $\TR^{2} = -1$). As the number of filled bands is even, the Pfaffian (see~\ref{app:Pfaffian}) 
 of $\KMTRMatrix$ is always defined.
To define this matrix, it is necessary to calculate scalar products of vectors that live in different vector subspaces (the fibers of the valence band bundle at $k$ and $-k$). 
%This requires a scalar product defined on the whole Hilbert space, including filled and empty bands. 
This uses the trivialization as
$\Torus{2} \Cross \ComplexField^{2n}$ of the total bundle of Bloch states and the scalar product in $\ComplexField^{2n}$.
Hence, in this approach, the filled bands vector bundle is necessarily viewed as a subbundle of the topologically trivial vector bundle of all bands (which is always possible, see section~\ref{sec:TrivialityTotalBundle}).
%
% -k = trb k ?
A more intrinsic quantity, which does not require this immersion in a trivial vector bundle, consists in the sewing matrix, 
 defined by Fu and Kane in \cite{FuKane2006} as:
\begin{equation}
\swm_{i j}(k) = \braket{e_{i}(-k) | \TR e_{j}(k)} . 
\label{eq:SewingMatrixDefinition}
\end{equation}
%
%where $(e_{i}(k))_{i}$ is an orthonormal basis of the filled band fiber \FilledBandsFiber{k} at $k$. 
In this expression, the 
scalar product is intrinsic to the fiber at $-k$. 
The sewing matrix relates the vectors at $-k$ to the ones at $k$ by:
\begin{equation}
\TR e_{i}(k) = \sum_{j} \swm_{j i}(k) e_{j}(-k)
\qquad
\text{or}
\qquad
e_{i}(-k) = \sum_{j} \Conjugate{\swm}_{i j}(k) \TR e_{j}(k)
\label{eq:SewingMatrixTRExpand}
\end{equation}
This sewing matrix is unitary (see \ref{app:sec:PropertiesSewingMatrix}):
\begin{equation}
\Adjoint{\swm}(k) \swm(k) = \IdentityMatrix
\label{eq:SewingMatrixUnitarity}
\end{equation}
and has the property (see \ref{app:sec:PropertiesSewingMatrix}):
\begin{equation}
\swm(-k) = - \swm^{T}(k)
\label{eq:SewingMatrixTR}
\end{equation}
At the TRIM, and {\it a priori} only at those points, the sewing matrix is antisymmetric. For a system with only two filled bands, 
it takes the simple form:
\begin{equation}
\swm(\HighSymmetryPoint) = \begin{pmatrix}
0 & t(\HighSymmetryPoint) \\
- t(\HighSymmetryPoint) & 0
\end{pmatrix}
\label{eq:swf_at_HSP}
\end{equation}
with $\Abs{t(\HighSymmetryPoint)} = 1$ for a TRIM $\HighSymmetryPoint \in \HighSymmetryPoints$. Note that, as the sewing matrix is antisymmetric at the TRIM, its Pfaffian is well-defined at those points 
and $t(\lambda) = \Pf \swm(\HighSymmetryPoint)$.

\subsubsection{Pfaffian invariant}
\label{sec:Z2:PfaffianInvariant}
\begin{figure}[ht]
\centering
\includegraphics[width=10cm]{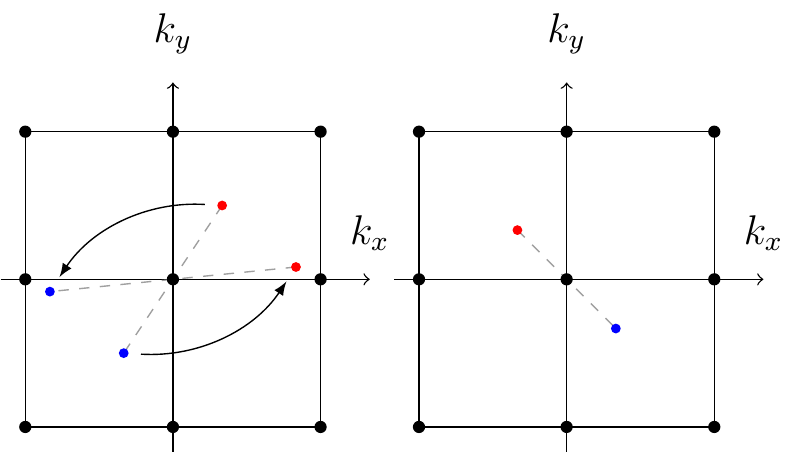}      
\caption{Vortices of the Pfaffian in a trivial case (left) and a topological case (right). Vortices are drawn as red (positive ones) and blue (negative ones) circles. In the trivial case, it is possible to move (arrows) the vortices without crossing the TRIM (black circles) to annihilate them. In the topological case, it is not possible.  }
\label{fig:VortexAnnihilationZ2}
\end{figure}

In their seminar paper \cite{KaneMele2005}, Kane and Mele argued that the Pfaffian $\Pf (\KMTRMatrix)$ of the matrix $\KMTRMatrix$ reveals a \Ztwo topological property of a time-reversal invariant insulator. 
This quantity tracks the orthogonality between Kramers related eigenspaces of the filled band. 
If at some points $k_0$ of the Brillouin torus, the Pfaffian vanishes and has complex vortices, the valence bands fiber at $k_0$ and 
its  time-reversed partner are orthogonal to each other. As we have seen, this can be achieved for example 
by a local band inversion in a BHZ-like scenario. If the vortices cannot be removed by a smooth deformation of the Hamiltonian, it is the sign of a nontrivial topological behaviour. 

Following Kane and Mele, let us assume that the vortices of $\Pf (\KMTRMatrix)$ are simple zeros, {\it i.e.} 
they are phase vortices with a vorticity~$\pm 1$\footnote{This vorticity is the winding number of the vortex.}. Because of eq.~\eqref{eq:SewingMatrixKMTRMRelation} and of the unitarity of the sewing matrix $\SewingMatrix$:
\begin{itemize}
    \item the vortices of $\Pf (\KMTRMatrix)$ come in time-reversal pairs, with one zero at $k$ and the other at $-k$, and the two elements of a pair have opposite vorticities;
    \item at the time-reversal invariant points $\HighSymmetryPoint \in \HighSymmetryPoints$, the Pfaffian has unit modulus, i.e. $\Abs{\Pf m(\HighSymmetryPoint)} = 1$.
\end{itemize}

\noindent As a consequence, if there are two pairs of vortices, through a continuous deformation of the Hamiltonian 
it is always possible to remove them through a simple merging. 
However, if there is only a single pair of vortices, it is not possible to remove it, as the only points where the vortices could annihilate, the TRIM, have always $\Abs{\Pf m(\HighSymmetryPoint)} = 1$ (Fig.~\ref{fig:VortexAnnihilationZ2}). 
This merging procedure imposes that only 
the parity of the number of pairs of vortices of $\Pf (\KMTRMatrix)$ in the Brillouin torus is an invariant, which corresponds to 
 the parity of the number of vortices in an effective Brillouin zone (with no vortices on its boundary): 
\begin{equation}
\ZtwoInvariantModTwo = \frac{1}{2 \pi \ii} \; \oint_{\partial\EBZ} \dd \log \Pf m   \quad \modTwo
\label{eq:Z2InvariantPfaffian}
\end{equation}
By construction, this is a  \Ztwo topological invariant.

\begin{figure}[ht]
\centering
\includegraphics[width=10cm]{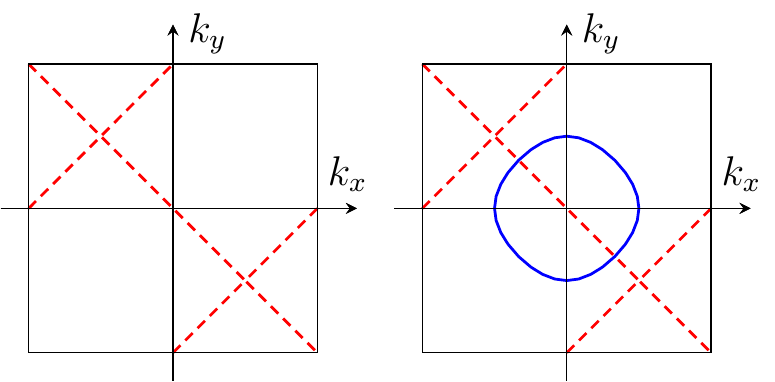}      
\caption{Zeros of the function $d_{1}$ (continuous blue line) and $d_{2}$ (dashed red lines) in the examples of fig.~26, for $\mu=-3$ (trivial, left) and $\mu=-1$ (topological, right). In the topological case, the common zeros of $d_{1}$ and $d_{2}$ correspond to the vortices of the Pfaffian (Fig.~26, left). }
\label{fig:Zerosd1d2}
\end{figure}

As an illustration, let us consider the model discussed in section \ref{sec:Z2:GrapheneLikeModel}. In this case, the Pfaffian is: 

\begin{equation}
\Pf (\KMTRMatrix) = \frac{d_{1} \, (d_{1} + \ii d_{2})}{\sqrt{(d_{1}^{2} + d_{2}^{2}) \, (d_{1}^{2} + d_{2}^{2} + d_{5}^{2})}}
\label{eq:GrapheneLikeModelPfaffian}
\end{equation}

\noindent and has vortices at the points where $d_{1}=d_{2}=0$. 
In Fig.~\ref{fig:Zerosd1d2}, we show the contours where $d_{1}=0$ and $d_{2}=0$ in the Brillouin zone in a 
specific example, both in a trivial and a topological cases. In Fig.~26, the phase (and the vortices, if any) of $\Pf (\KMTRMatrix)$ is/are shown in the same examples.

\begin{figure}
% @Mathematica: 2013-07-14-vortex-pfaffien-2.nb
% images compressées avec optipng -o7
\centering
\includegraphics[width=15cm]{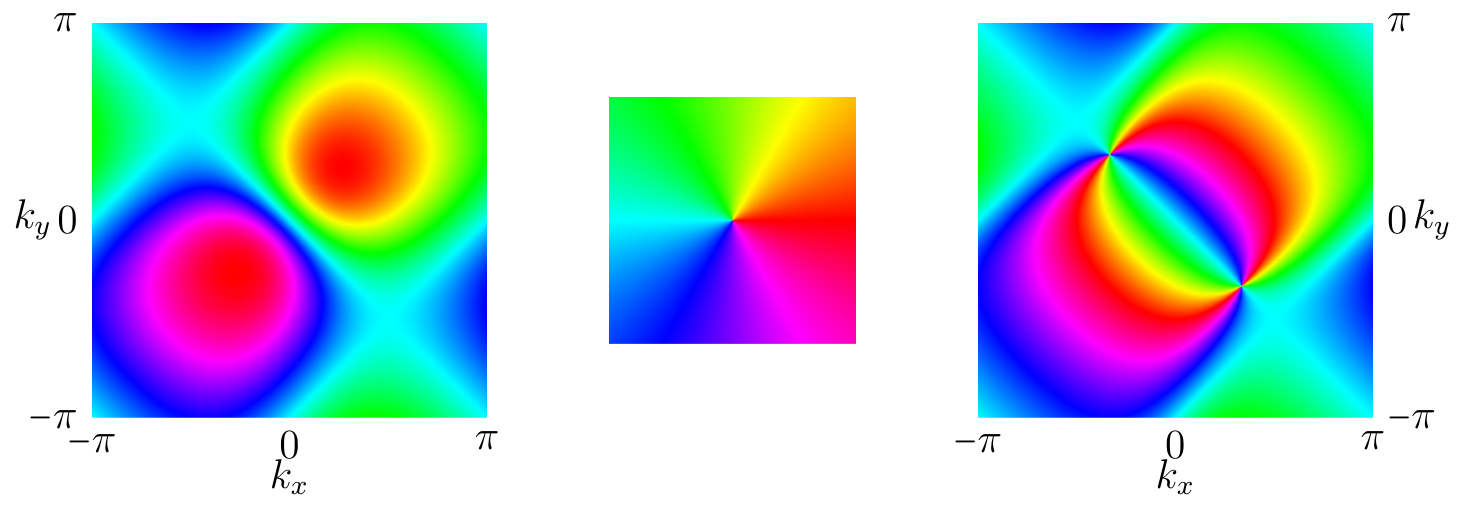}      

  \caption{Phase of the pfaffian $\Pf (\KMTRMatrix)$ on the Brillouin zone, in a trivial (left, $\mu=-3$) and topological (right, $\mu=-1$), for the model 
discussed in section~\ref{sec:Z2:GrapheneLikeModel} (p.~\pageref{sec:Z2:GrapheneLikeModel}) with 
$d_{1}(k_x,k_y) = \mu + \cos k_x + \cos k_y$, $d_{2}(k_x,k_y) = \sin k_x + \sin k_y$, 
$d_{5}(k_x,k_y) = \sin k_x - \sin k_y - \sin (k_x-k_y)$ (coefficients adapted from \cite{FuKane2007}). In the middle is displayed the phase color function we used. }
  \label{fig:PhasePfaffian}
\end{figure}

\subsubsection{Sewing matrix invariant}
\label{sec:Z2:SewingMatrixInvariant}

Another expression of the \Ztwo invariant, introduced by Fu and Kane \cite{FuKane2006},  uses the sewing matrix $w$ 
(see eq.~\eqref{eq:SewingMatrixDefinition})  as opposed to the $\KMTRMatrix$ matrix . This expression also requires a continuous basis of the filled bands bundle , but 
it allows in particular for an easy extension to three dimensions. 
 If it is the case, the \Ztwo invariant can be expressed as:
\begin{equation}
(-1)^{\nu}
=
\prod_{\HighSymmetryPoint \in \HighSymmetryPoints}
\frac{\Pf \SewingMatrix(\HighSymmetryPoint)}{\sqrt{\det \SewingMatrix(\HighSymmetryPoint)}}
\label{eq:Z2InvariantSewingMatrix}
\end{equation}
Notice that this expression in only meaningful provided that $\SewingMatrix$ is calculated from a continuous basis. If so, 
 the square root of $\det \SewingMatrix(\HighSymmetryPoint)$ 
 can be defined globally as it has no winding (see~\ref{app:NoWindingOfSewingMatrixDeterminant}). 
 The topological invariant then depends only on the behaviour of  $\SewingMatrix(\HighSymmetryPoint)$ at the TRIM. 
Following Fu and Kane, we can show that this expression of the \Ztwo invariant is equivalent to the Pfaffian invariant: this is done 
 in \ref{sec:DemoPfaffianToSewing}.

When there is an additional inversion symmetry (such as on section~\ref{sec:Z2:Inversion}), 
the expression~\eqref{eq:Z2InvariantSewingMatrix} can be rewritten in a particularly simple form
  \cite{FuKane2007} as the product of the parity eigenvalues of half the filled bands at all TRIM (Kramers partners sharing parity eigenvalues); in the case of a four-level system studied in sections~\ref{sec:Z2:GrapheneLikeModel}, \ref{sec:Z2:BHZLikeModel}, 
it reads:
\begin{equation}
(-1)^{\ZtwoInvariantModTwo} = \prod_{\HighSymmetryPoint \in \HighSymmetryPoints} \sign \, d_{1}(\HighSymmetryPoint)
\end{equation}

\subsubsection{The obstruction point of view}
\label{sec:Z2:ObstructionInvariant}

In sections~\ref{sec:Z2:GrapheneLikeModel} and \ref{sec:Z2:BHZLikeModel}, we have seen  examples of obstructions to ``continuously define Kramers pairs'' when the system is in a topologically nontrivial state. This property has been studied by Fu and Kane \cite{FuKane2006} who described it in the general case as an obstruction to the Stokes theorem under constraint, in a similar way as the Chern class is an obstruction to the Stokes theorem \emph{without} constraint. 

Indeed, as the Chern class is always trivial in a time-reversal invariant system (see section~\ref{sec:TrivialityTotalBundle}), 
there is no obstruction to continuously define the basis vector of the filled bands bundle. 
However, if we enforce a gauge which respects time-reversal invariance, which amounts to choosing the relative phase between Kramers partners to vanish, it is not always possible to define such a continuous basis anymore. 
In other words, this would correspond to an impossibility 
to choose a \emph{continuous} basis $(e_{i})_{i=1}^{2m}$ of the filled bands bundle such that $\TR e_{2i-1}(k)=e_{2i}(-k)$. 
Below, we shall limit our discussion to the case $m=1$ for simplicity. 
 Another form of the \Ztwo invariant that expresses  this point of view is \cite{FuKane2006}:
\begin{equation}
\ZtwoInvariantModTwo = \frac{1}{2 \pi} \; \left[
\oint_{\partial \EBZ} A - \int_{\EBZ} F
\right]
\modTwo
\end{equation}
where $A$ is the total Berry connection (the sum of the Berry connections of all bands) constructed from Kramers pairs, and $F= \dd A$ is the Berry curvature. In the trivial case, it is possible to define a continuous basis over the EBZ, the Stokes theorem can be applied, and $\ZtwoInvariantModTwo$ vanishes. However, in the nontrivial case, when $\ZtwoInvariantModTwo=1$, there is a topological obstruction to do so. 

The equivalence between this expression and the previous ones was demonstrated in \cite{FuKane2006}, and is reviewed in \cite[§10.5]{Bernevig}. Notice that whereas the  $\IntegerRing$ Chern obstruction is a \UnitaryGroup{1} obstruction, hence described by a winding number, the \Ztwo obstruction is a \SpecialUnitaryGroup{2} obstruction, as pointed out by Lee and Ryu \cite{LeeRyu2008}. Moore and Balents \cite{MooreBalents2007} discuss the obstruction point of view using homotopy theory arguments. They show that the \Ztwo invariant is a ``Chern parity'' and generalize it to three dimensions.

%%%%%
\subsection{An intrinsic point of view on the \Ztwo invariant}

%\subsubsection{Introduction}

The aim of this section is to relate the sewing matrix expression \eqref{eq:Z2InvariantSewingMatrix} of the \Ztwo invariant 
to an index recently introduced by Freed and Moore  \cite{FreedMoore2013}. In doing so, we will gain another interpretation of this 
\Ztwo invariant, \emph{intrinsic} to a fiber bundle with a time-reversal symmetric structure. 
This point of view uses a type of an orientation on this fiber bundle
over the TRIM's defined through the introduction of the so-called
determinant line bundle. The determinant line bundle possesses over TRIM
an oriented real structure. The latter originates from a quaternionic
structure in the valence band bundle imposed by the time-reversal symmetry \cite{AvronSadunSegertSimon1989}.
The real orientations of the determinant bundle at different TRIM
are constrained by the topology of the whole time-reversal symmetric
vector bundle in a way described by the Kane--Mele invariant.

\subsubsection{Quaternionic structure and determinant bundle}

At the time-reversal invariant points, the filled bands fiber $\FilledBandsFiber{\lambda}$ is equipped with a quaternionic structure \cite{FreedMoore2013}. 
Indeed, the time-reversal operator $\TR$ acts on this fiber, endowing it with an anti-linear anti-involution that plays the role of 
a quaternionic element 
$\jj = \TR$ (see~\ref{app:QuaternionicVectorSpaces}).  With the imaginary unit $\ii$, the third quaternionic element is $\kk =  \ii \TR$. 
The constitutive relations of quaternions are then satisfied: $\jj \ii = \ii \TR  = \kk = -  \TR \ii = -  \jj \ii$, etc.

From the $n$-dimensional vector bundle $\FilledBandsBundle$ (with $n=2$), we construct the associated determinant bundle as the $n^{\text{th}}$ exterior power $\ExteriorPower{n} \FilledBandsBundle$. By construction, it is a (complex) line bundle on the torus $\Torus{2}$.
The global basis of $\FilledBandsBundle$ provides a global (single element) basis of $\ExteriorPower{n} \FilledBandsBundle$, 
{\it e.g.}, for $n=2$:
\begin{equation}
s(k) = e_{1}(k) \wedge e_{2}(k)
\end{equation}
which is indeed always nonzero.
Time-reversal acts on the determinant bundle $\ExteriorPower{2} \FilledBandsBundle$ through the operator 
$\ExteriorPower{2} \TR$ defined as:
\begin{align}
\ExteriorPower{2} \TR 
\;:\; \ExteriorPower{2} \FilledBandsFiber{k}
& \to \ExteriorPower{2} \FilledBandsFiber{-k} 
\nonumber \\
  a \wedge b & \mapsto \TR a \wedge \TR b . 
\label{eq:TRActionOnDetBundle}
\end{align}

%
%\subsubsection{Real part}
%\label{sec:RealPart}

At the TRIM, the operator $\ExteriorPower{2} \TR$, which is anti-linear involutive map  from the local fiber onto itself, acts similarly to a ``complex conjugation'', and induces a  real structure on the determinant bundle fiber as well as a \emph{natural orientation} \cite{FreedMoore2013}. 
 Indeed, at these TRIM, the filled bands' fiber $\FilledBandsBundle_{\HighSymmetryPoint}$ has a quaternionic basis, denoted 
 $(e_{1})$ (therefore  a complex basis $(e_{1},\TR e_{1})$, as $\TR e_{1}$ is orthogonal to $e_{1}$), which induces the basis $e_{1} \wedge \TR e_{1}$ of the fiber $\ExteriorPower{2} \FilledBandsBundle_{\HighSymmetryPoint}$ at $\HighSymmetryPoint$ of the determinant bundle $\ExteriorPower{2} \FilledBandsBundle$. This element satisfies: 
\begin{equation}
\ExteriorPower{2}\TR \left[ e_{1} \wedge \TR e_{1} \right] = e_{1} \wedge \TR e_{1}
\end{equation}
\noindent and is therefore called a real element of the fiber $\ExteriorPower{2} \FilledBandsBundle_{\HighSymmetryPoint}$. On the contrary,
\begin{equation}
\ExteriorPower{2}\TR \left[ \ii \, e_{1} \wedge \TR e_{1} \right] = - \ii e_{1} \wedge \TR e_{1}
\end{equation}
is a purely imaginary element. We can define the real part and the imaginary part in the same manner that for complex numbers, but 
with $\ExteriorPower{2} \TR$ playing the role of the complex conjugation.

Moreover, we can define a natural orientation for the real elements on this fiber:  
let us choose by convention $e_{1} \wedge \TR e_{1}$ as a \emph{positive} element, 
along with all vectors in $\ExteriorPower{2} \FilledBandsBundle_{\HighSymmetryPoint}$ that are proportional with a positive coefficient. The elements proportional to this vector with a negative coefficient are called negative elements.
These definitions of orientation do not depend on the choice of the quaternionic basis. Indeed, if we consider another element: 
\begin{equation}
f_{1} = \lambda e_{1} + \mu \TR e_{1}
\qquad
(\lambda \text{ et } \mu \in \ComplexField)
\end{equation}
\noindent the corresponding element on the determinant fiber reads: 
\begin{align*}
f_{1} \wedge \TR f_{1}
&=
\left(
\lambda e_{1} + \mu \TR e_{1}
\right)
\wedge
\TR
\left(
\lambda e_{1} + \mu \TR e_{1}
\right) \\
&=
\left(
\lambda e_{1} + \mu \TR e_{1}
\right)
\wedge
\left(
\lambda^{\star} \TR e_{1} + \mu^{\star} \TR \TR e_{1}
\right) \\
&=
\Abs{\lambda}^{2} e_{1} \wedge \TR e_{1} 
- \Abs{\mu}^{2} \TR e_{1} \wedge e_{1} \\
&=
\left[ \Abs{\lambda}^{2} + \Abs{\mu}^{2} \right]
e_{1} \wedge \TR e_{1} . 
\end{align*}
\noindent The proportionality coefficient between $e_{1} \wedge \TR e_{1}$ and $f_{1} \wedge \TR f_{1}$ being real and positive, 
 the choice real element and positive element (resp. imaginary, negative) deduced 
 from both choices of a basis will be identical: this provides a well-defined natural orientation, and associated real subspace. 
Note that the  natural orientation of the real part of the determinant fiber at the TRIM  
directly originates from the underlying quaternionic structure of the filled bands fiber. 

\subsubsection{Kane--Mele invariant}

Let us consider a nowhere vanishing section $\sigma$ of the determinant bundle $\ExteriorPower{2} \FilledBandsBundle$ previously defined. This section is said to be time-reversal covariant if:
\begin{equation}
\ExteriorPower{2} \TR \, \sigma(k) = \sigma(\trb k)
\end{equation}
In the following, we will only consider time-reversal covariant sections $\sigma$. 
For a TRIM $\HighSymmetryPoint$,  $\sigma(\HighSymmetryPoint)$ is a real element as:
\begin{equation}
\ExteriorPower{2}\TR \sigma(\HighSymmetryPoint) = \sigma(\trb \, \HighSymmetryPoint) = \sigma(\HighSymmetryPoint)
\end{equation}
The sign of $\sigma(\HighSymmetryPoint)$ is therefore well defined. Following 
Freed and Moore, let us define the following index \cite{FreedMoore2013}: 
\begin{equation}
\ZtwoInvariantIPKM = \prod_{\HighSymmetryPoint \in \HighSymmetryPoints}
\sign[\sigma(\HighSymmetryPoint)], 
\label{eq:Z2inDeterminant}
\end{equation}
which will turn out to be the Kane--Mele invariant. Indeed, it takes only values $\ZtwoInvariantIPKM = \pm 1$ and is a 
\Ztwo quantity. Moreover,  
 $\ZtwoInvariantIPKM$ can be shown to be independent of the chosen section $\sigma$. 
 %, {\it i.e.}
 %  it is $\UnitaryGroup{2}$ gauge invariant, whereas 
%analogous products on restricted number (one, two, or three) TRIM are \emph{not} invariant. 
 The demonstration of the invariance of  $\ZtwoInvariantIPKM$ on the chosen section $\sigma$ is presented in \ref{app:determinant}.

Moreover, for a carefully chosen section $\sigma$, the Freed--Moore \Ztwo invariant \eqref{eq:Z2inDeterminant} can be expressed as \cite{FreedMoore2013}
\begin{equation}
\ZtwoInvariantIPKM = (-1)^{\ZtwoInvariantModTwo} = \prod_{\HighSymmetryPoint \in \HighSymmetryPoints}
\frac{\sqrt{\det \swf(\HighSymmetryPoint)}}{\Pf \swf(\HighSymmetryPoint)}, 
\label{eq:KMencoreunefois}
\end{equation}
where   the sewing function is defined as the determinant of the sewing matrix (up to a complex conjugation):
\begin{equation}
\swf(k) = \det \Conjugate{\swm}(k)
\end{equation}
 The demonstration of this equality is done in \ref{sec:FreedMooreKaneMele}. 
The above formulation \eqref{eq:KMencoreunefois} 
is nothing but the Fu and Kane \cite{FuKane2006} expression \eqref{eq:Z2InvariantSewingMatrix} for the Kane--Mele invariant. 
This demonstrates that the Freed--Moore invariant \eqref{eq:Z2inDeterminant} is indeed another expression of the Kane--Mele \Ztwo invariant.

%%%%%

\section{Conclusion}

 In this introduction to the notion of topological order in insulators, we have discussed the examples of Chern and \Ztwo Kane--Mele insulators. 
 In a first part, we have studied a simple two-band example of Chern insulator, illustrated by the Haldane model. 
 The occurrence of topological order was shown to manifests itself in an obstruction to define the eigenstates of the fill band in a continuous manner on the Brillouin zone. The index allowing one to determine the underlying topological nature of an insulator was shown to be the standard first Chern number, whose various expressions were discussed. 
  In the following part, we discussed the  \Ztwo topological 
  ordering in insulators described by time-reversal symmetric spin $\frac12$ Hamiltonian. 
 To develop a better understanding of \Ztwo topological insulators, we have considered two classes of two-dimensional models 
 that account for the \Ztwo topological insulators discovered so far: the Kane--Mele and Bernevig--Hughes--Zhang models. 
   In those examples, we have shown explicitly that, similarly to the case of Chern insulators, 
 an obstruction to define Kramers pairs of states on the Brillouin torus arises in the nontrivial phase of a \Ztwo insulator. 
This obstruction consists of point singularities of the vectors forming Kramers pairs, which cannot be removed without closing the gap. We argued that this obstruction has indeed a topological meaning. These examples provide an intuitive justification of the expression of the \Ztwo invariant in an inversion symmetric system, and help us to understand the mechanism underlying the \Ztwo topology. We then reviewed the different equivalent expressions of the \Ztwo invariant.

 By choice and for pedagogical reasons, many aspects of this active field have been overlooked. We hope that this introduction will be a gateway towards their understanding. We can mention a few directions, including the extensions of the notion of topological insulator to three dimensions \cite{FuKaneMele2007,MooreBalents2007,Roy2009b}, the topological ordering of various types of superconductors 
 \cite{Bernevig,HasanKane2010,Qi:2011,RyuSchnyderFurusakiLudwid2010}, the interplay between topological order and crystalline symmetries \cite{Fu:2011} 
 and topological ordering in gapless and interacting phases \cite{Turner:2013}.

{\it Acknowledgments.} Our understanding of the subject of this manuscript has enormously benefitted from fruitful and numerous discussions with our colleague Krzysztof Gawędzki, and we thank him for these as well as his careful reading of this manuscript. We also benefitted from interesting discussions with G.-M. Graf about the notion of obstruction in \Ztwo topological insulators. 
This work was supported by an ANR grant Blanc-2010 IsoTop. 

\appendix

\def\appendixname{Appendix~}
%\begin{appendices}

\section{Pfaffian of a matrix}
\label{app:Pfaffian}

The pfaffian $\Pf (A)$ of a $2 n \times 2n$ skew-symmetric matrix $A$ is the quantity defined by \cite{BourbakiALGCh9}:
\begin{equation}
  \Pf A = \frac{1}{2^{n} \, n!} \; \sum_{\sigma \in \PermutationGroup{2n}} \, \sign(\sigma) \, \prod_{i=1}^{n} \, a_{\sigma(2i-1) ,\, \sigma(2i)}
\end{equation}

\noindent where $\PermutationGroup{2n}$ is the permutation group of $2n$ elements. The Pfaffian is related to the determinant by $(\Pf A)^{2} = \det A$. 
%We define the pfaffian of an odd-dimensional square matrix to be zero. 
For $2\times 2$ skew-symmetric matrices,  
\begin{equation}
  \Pf \begin{pmatrix}
    0 & a \\
    -a & 0
  \end{pmatrix} = a
\end{equation}
Other useful properties of the Pfaffian include $\Pf(\Transpose{A}) = (-1)^n \,\Pf(A)$, $\Pf(\lambda A) = \lambda^n \, \Pf(A)$ for $\lambda \in \ComplexField$ and, for any $2n \times 2n$ square matrix $B$, the identity $\Pf(B \, A \, \Transpose{B}) = \det(B) \, \Pf(A)$.
%
%\subsection{Useful properties of anti-unitary operators}
%\label{app:AntiunitaryOperators}
%
%Let $\Theta$ be an anti-unitary operator. Then \cite{LeBellac},
%
%\begin{equation}
%    \braket{\Theta \phi | \Theta \chi} = \braket{\chi | \phi } = \Conjugate{\braket{\chi | \phi}}
%\end{equation}
%
%\noindent We have always $\Theta^{2} = \eta \IdentityMatrix$ with $\eta = \pm 1$ according to the total spin of the system. Hence, with $\chi = \Theta \psi$, we have
%
%\begin{equation}
%    \braket{\Theta \phi | \chi} = \eta \, \braket{\Theta \chi | \phi}
%\end{equation}

\section{Quaternionic vector spaces}
\label{app:QuaternionicVectorSpaces}

\subsection{Quaternions}

The quaternion ring $\QuaternionRing$ is a division ring (or skew field), built by enhancing the reals 
with three symbols $\ii, \jj, \kk$ that satisfy: 
\begin{equation}
\ii^2 = \jj^2 = \kk^2 = \ii \jj \kk = -1
\end{equation}
so that a generic quaternion reads
\begin{equation}
q = a + b \ii + c \jj + d \kk
\end{equation}
where $a,b,c,d$ are real. 
An associative, but noncommutative multiplication on $\QuaternionRing$ can be defined, 
as well as a conjugation where the sign of the symbols $\ii,\jj,\kk$ is reversed, from which a modulus and an inverse can be defined.

\subsection{Vector spaces and quaternions}

Vector spaces over quaternions are defined naturally. 
%, and, as well as a complex vector space, 
%can be seen as a twice-dimensional real vector space,
Three points of view on a quaternionic vector space $E$ are useful \cite{Lawson,Darling}:
\begin{itemize}
\item $E$ is a $n$-dimensional quaternionic vector space;
\item $E$ is a $2n$-dimensional complex vector space, endowed with 
%\footnote{In our case, it is the case with $J=\TR$, and one can convince oneself thatif $J$ acts like $\jj$, then $-\ii J$ plays the role of $\kk$. } 
a $\ComplexField$-antilinear operation $J \;:\; E \to E$ satisfying $J^{2} = - 1$ (called a \enquote{quaternionic structure map} (see \cite{AvronSadunSegertSimon1989}));
\item $E$ is a $4n$-dimensional real vector space endowed with three maps $I$, $J$, $K$ satisfying $I^2 = J^2 = K^2 = -1$ and $I J = - J I = K$. Those applications correspond to the multiplication by $\ii$, $\jj$, and $\kk$.
\end{itemize}

Conversely, if one considers a $2n$-dimensional complex vector space, a map such as $J$ endows it with a quaternionic structure \cite{FreedMoore2013}. 
In this case, we call a quaternionic basis an indexed family $(e_{i})_{i \in I}$ of vectors that form a basis of $E$ as a quaternionic vector space (i.e. using linear combinations with $\QuaternionRing$-valued coefficients). That is to say, we ask $(e_{i}, J\,e_{i})_{i \in I}$ to be a basis of $E$ as a complex vector space.

\section{The degree of a map and the winding number}
\label{app:IndexTheoryWindingNumber}

 We briefly recall the definitions and properties of the degree of a map \cite{GuilleminPollack,DubrovinFomenkoNovikovII,Flanders,DincaMawhin,BleeckerBooss}. 
Let us consider a $(n-1)$-dimensional, compact, oriented manifold $M$ and an application $f$ from $M$ to the punctured space $\RealField^{n} - p$ (the space $\RealField^{n}$ where some point $p$ has been removed), whose image $N = f(M)$ is a $n-1$-dimensional compact (thus closed), simply connected in an immersed
surface.
The natural projection $\pi$, defined by:

\begin{align}
\pi \;: \; & (\RealField^{n} - p ) \to S^{n-1} 
\nonumber \\
& y \to \frac{y - p}{\norm{y - p}}
\end{align}

\noindent defines a retractation from $\RealField^{n} - p$ to $S^{n-1}$, and those two spaces are thus homotopically equivalent, i.e. $\RealField^{n} - p \simeq S^{n-1}$.
The pre-image by $\pi$ of a point $a \in S^{n-1}$ is the ray $\Delta(a)$ coming from $p$ in $\RealField^{n} - p$, the point $p$ being excluded from $\Delta(a)$.
%
%\begin{figure}
%\centering
%\includegraphics[width=4cm]{Figures/CRAS-figure-35}      
%\caption{Representation of the ray $\Delta(a)$.}
%\end{figure}

The natural projection $\pi$ induces a map $\pi_{N}: N \to S^{n-1}$, which simply projects the closed surface $N$ on the sphere $S^{n-1}$.
Hence, $\pi \circ f: M \to S^{n-1}$ is a map between same-dimensional manifolds, and one can thus define its degree
%\footnote{It is called degree of a (continuous) map, Kronecker degree, Brouwer index, or topological degree, or simply degree - and also winding number. There are some differences between those quantities, which vary through authors and era, but they are all the same in the present case.} 
$\Degree(\pi \circ f)$ (see \cite{GuilleminPollack}) as the ``intersection number'', defined from the finite set $(\pi \circ f)^{-1}( \lbrace a \rbrace )$ by assigning an orientation number $\pm 1$ to each of its points, according to whether $\pi \circ f$ preserves ($+1$) or not ($-1$) the orientation at this point (that means the orientation number is the sign of the determinant of the tangent map on this point), and by summing those orientation numbers, the result does \emph{not} depend on the chosen point $a \in S^{n-1}$.

In the case considered in this review, this degree is precisely
%\footnote{There are several pathological cases seem to wreck our description (e.g. when a ray $\Delta$ crosses a self-intersection point of $N$, or is tangent to $N$, etc.). However, the degree of a map is an homotopy invariant, so that all those special cases can be avoided by slighlty modifying the surface $N$ to make them disappear. } 
the number of intersection of $N$ with a ray $\Delta(a) = \pi^{-1}( \lbrace y \rbrace )$,
 the orientation number being given by the sign of the scalar product of the radial unit vector $\hat{u}_{r}(\Delta)$ with the normal vector at the intersection, and does not depend on the choice of the ray.
This amounts to determining the pre-image $(\pi \circ f)^{-1}( \lbrace a \rbrace )$ in two times.
\begin{figure}[ht]
\centering
\includegraphics[width=6cm]{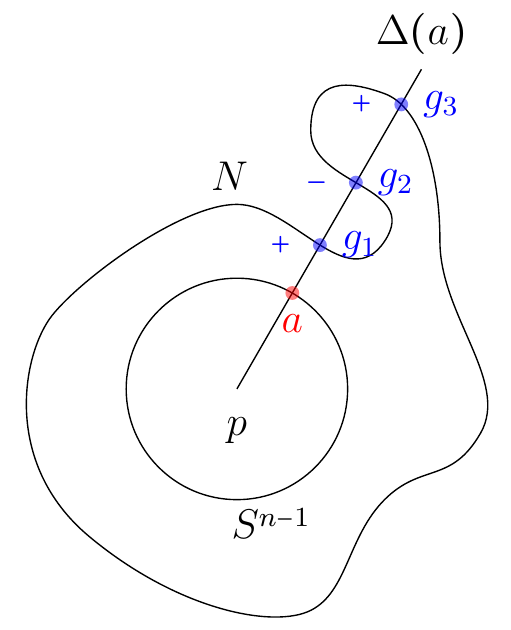}      
\caption{Projection of $N$ on the sphere by the map $\pi$. In the depicted case,  $\Delta(a) \cap N = \pi^{-1}(a) = \left\lbrace g_{1}, g_{2}, g_{3} \right\rbrace$, with orientation numbers $(+1, -1, +1)$, so that the degree of $N$ with respect to $p$ is $1$.}
\label{fig:projection_N_to_Sphere}
\end{figure}

\begin{figure}[h]
\centering
\includegraphics[width=10cm]{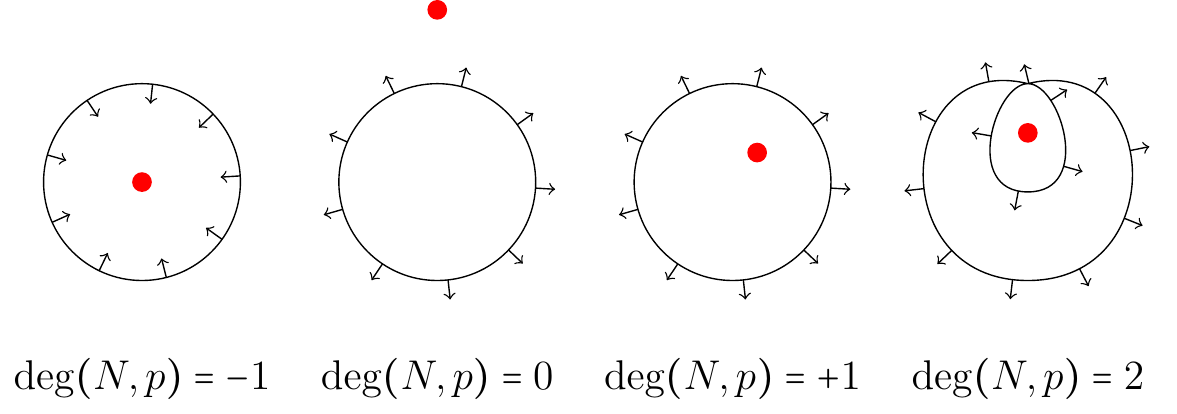}      
\caption{Examples in dimension $n=2$ of curves $N$ with different winding numbers with respect to the red point~$p$.}
\label{fig:WindingNumberExamples}
\end{figure}

Moreover, two homotopic maps have the same degree, and in the case $M \to S^{n-1}$ at hand, conversely, two maps with the same degree are homotopic  \cite{GuilleminPollack,DubrovinFomenkoNovikovII}. 
As a consequence, the degree of $\pi \circ f$ is the homotopy class of the surface $N$ in the homotopy group:

\begin{equation}
\pi_{n-1}(\RealField^{n} - p) \simeq \pi_{n-1}(S^{n-1}) \simeq \IntegerRing
\end{equation}

\noindent that is to say, it is the number of times the surface $N$ warps around the point $p$ (it is the \emph{engulfing number}, a generalization of the winding number obtained for $n=2$).
Often, it appears more useful to consider a given surface $N$, and move it with respect to the point $p$. 
In the following, we will thus slightly abuse the notations by calling ``degree of $f$ with respect to $p$'' or ``degree of $N$ with respect to $p$'' the degree of $\pi \circ f: M \to N$ (figs. \ref{fig:projection_N_to_Sphere} and \ref{fig:WindingNumberExamples}). The degree being a homotopy invariant, it can only change when the point $p$ passes through the surface $N$ (where it is not defined).

The engulfing number is also given \cite{Flanders} by the Kronecker integral:
\begin{equation}
\Degree(f) = \frac{1}{A_{n-1}} \, \int_{M} f^{\star} \tau
\end{equation}
where $\tau$ is the $(n-1)$-form:
\begin{equation}
\tau = \frac{1}{r^{n}} \, \star r \dd r
= \frac{1}{r^{n}} \, \sum_{i=1}^{n} (-1)^{i-1} \, x_{i} \, \dd x_{1} \cdots \except{\dd x_{j}} \cdots \dd x_{n}
\end{equation}

The $(n-1$)-form $\, \star r \dd r$ (where $\star$ is Hodge's star) restricted to the sphere $S^{n-1}$ is the volume form on the sphere. Integrated, it gives the area $A_{n-1}$ of the $n-1$-sphere.
The term decorated with a hat $\except{\phantom{a}}$ is omitted. 
By expliciting the pullback when $p=0$, we obtain: 
\begin{equation}
\Degree(f,0) = \frac{1}{A_{n-1}} \,
\int_{M} \left[
\sum_{j=1}^{n} (-1)^{j-1} \, \frac{f_{j}}{\Norm{f}^n} \,
\dd f_{1} \wedge \cdots \wedge \except{\dd f_{j}} \wedge \cdots \wedge \dd f_{n}
\right]
\label{eq:KroneckerIndexDefinition}
\end{equation}
When $n=3$ (where $A_{2}= 4 \pi$), the formula \eqref{eq:KroneckerIndexDefinition} can be written:
\begin{equation}
\Degree(f,0) = \frac{1}{4 \pi} \, 
\int_{M} \frac{1}{2} \epsilon^{i j k} \, 
\norm{f}^{-3} \,f_{i} \, \dd f_j \wedge \dd f_k . 
\end{equation}
If we intepret the term $f/\norm{f}^3$ as the magnetic field of the monopole  located 
at $p=0$ (or, in the general case, at $p$),  the integral on the right-hand side
can be seen as the magnetic flux through the surface $N$.

%
%\textbf{References :} 
%\begin{itemize}
%\item \cite{GuilleminPollack} defines clearly the degree with differential topology and intersection theory, and shortly explains the formulation with integrals
%\item \cite{DubrovinFomenkoNovikovII} has a more analytical (i.e. less geometric) approach than \cite{GuilleminPollack}, but further explains the relationship with integrals
%\item \cite{Flanders} carefully expressed the degree as the Kronecker integral, using usual properties (see e.g. \cite{DubrovinFomenkoNovikovII})
%\item \cite{DincaMawhin} has a very analytical approach, but shows the relation between several \enquote{degrees}
%\item \cite[ch.11 §1]{BleeckerBooss} gives an excellent review of the different methods to compute the winding number, but no proof
%\end{itemize}
%% Fulton ?
%% Hatcher ?

\section{Winding number in the complex plane}
\label{app:sec:WindingNumberComplexPlane}

Let us consider a closed curve $\Gamma: \Sphere{1} \to \ComplexField \setminus\{0\}$ in the complex plane without zero. The winding number of $\Gamma$ can be expressed \cite{Fulton} as the complex integral: 
\begin{equation}
    W[\Gamma] = \deg(\Gamma,0) = \frac{1}{2 \pi \ii} \; \int_{\Gamma} \frac{\dd z}{z} = \frac{1}{2 \pi \ii} \; \int_{\Gamma} \dd \log z, 
\end{equation}
and $W[\Gamma] \in \IntegerRing$ is an integer. 
Now, let us consider a closed curve $\mathcal{C}$ on a manifold $X$, and a continuous complex-valued function $f: X \to \ComplexField$. 
We can similarly define the winding number of the curve $\Gamma = f(\mathcal{C}) = f \circ \mathcal{C}$ if it does not pass through zero. 

A $n$-th root of $f$ exists iff $W[f(\mathcal{C})] \in n \IntegerRing$ for any closed curve $\mathcal{C}$. Furthermore, a logarithm of $f$ exists iff $W[f(\mathcal{C})] = 0$ for any closed curve $\mathcal{C}$, i.e. iff $f$ \enquote{has no winding} 
(see {\it e.g.} \cite{Ullrich}).
%
%\subsection{Complex logarithm, square root, and integrals}
%
%To sum up, the complex logarithm is \emph{a priori} only defined modulo $2 \pi \ii$, because if $z = \ee^{\ii \theta}$, we would like to have $\log z = \ii \theta$, but as $z = \ee^{\ii (\theta + 2 \pi)}$, we should also have $\log z = 2 \pi \ii$. From a mathematical point of view, the complex logarithm is ill-defined (it does not exist) on the complex plane. However, if a function $f$ has no winding, it is possible to globally define its logarithm $\log f$, that is, to have a continuous choice of this \enquote{modulo $2 \pi \ii$} on the whole domain of $f$. A consequence of this global definition is that the integral of $\dd \log f$ depends only on the start and end points, without any \enquote{modulo $2 \pi$}. On the contrary, the same integral of a function with a winding will depend on the integration path ; if we do not specify this path, the integral is only defined modulo $2 \pi \ii$.
%

\section{Singular eigenvectors in the Kane--Mele model}
\label{app:sec:demoKMlikeEigenvectors}

In this appendix, we derive the singular behaviours \eqref{eq:Z2:GrapheneLike:LimitOfEigenvectorsAtPositived5} 
 and \eqref{eq:Z2:GrapheneLike:LimitOfEigenvectorsAtNegatived5} of the eigenstates for the Kane--Mele model. 
We obviously have to explicitely include the normalisation when taking the limit $t/d_{5} \to 0$. 
By using the square root series expansion:
\begin{equation}
\sqrt{1 + (t/d_{5})^2} \simeq 1 + \frac{t^{2}}{2 \, d_{5}^{2}},
\end{equation}
we obtain: 
\begin{equation}
\ket{u_{1}} = 
\frac{1}{\sqrt{t^2 + \Abs{d_{5}}^2 \left( - \sign(d_{5}) - 1 - t^{2} / (2 \, d_{5}^{2}) \right)^2 }} \;
\begin{pmatrix}
    0 \\
    \Abs{d_{5}} \left( - \sign(d_{5}) - 1 - \displaystyle\frac{t^{2}}{2 \, d_{5}^{2}} \right) \\
    0 \\
    t \ee^{\ii \theta}
\end{pmatrix}.
\end{equation}
\noindent When $d_{5} > 0$, the limit $t\to 0$ reads: 
\begin{equation}
\ket{u_{1}(d_5>0)} = 
\frac{1}{\sqrt{t^2 + \Abs{d_{5}}^2 \left( - 2 - t^{2} / (2 \, d_{5}^{2}) \right)^2 }} \;
\begin{pmatrix}
    0 \\
    \Abs{d_{5}} \left( - 2 - \displaystyle\frac{t^{2}}{2 \, d_{5}^{2}} \right) \\
    0 \\
    t \ee^{\ii \theta}
\end{pmatrix}
\to
\frac{1}{\sqrt{4 \Abs{d_{5}}^2}} \;
\begin{pmatrix}
    0 \\
    - 2 \Abs{d_{5}}  \\
    0 \\
    0
\end{pmatrix},
\end{equation} 
whereas for $d_{5} < 0$ we have: 
\begin{align}
\ket{u_{1}(d_{5}<0)} 
%&= 
%\frac{1}{\sqrt{t^2 + \Abs{d_{5}}^2 \left(t^{2} / (2 \, d_{5}^{2}) \right)^2 }} \;
%\begin{pmatrix}
%    0 \\
%    \Abs{d_{5}} \left(\displaystyle\frac{t^{2}}{2 \, d_{5}^{2}} \right) \\
%    0 \\
%    t \ee^{\ii \theta}
%\end{pmatrix} \\
%&=
%\frac{1}{t \; \sqrt{1 +  t^{2} / (4 \, \Abs{d_{5}}^2) }} \;
%\begin{pmatrix}
%    0 \\
%    \Abs{d_{5}} \left(\displaystyle\frac{t^{2}}{2 \, d_{5}^{2}} \right) \\
%    0 \\
%    t \ee^{\ii \theta}
%\end{pmatrix} \\
&=
\frac{1}{\sqrt{1 +  t^{2} / (4 \, \Abs{d_{5}}^2) }} \;
\begin{pmatrix}
    0 \\
    \Abs{d_{5}} \left(\displaystyle\frac{t}{2 \, d_{5}^{2}} \right) \\
    0 \\
    \ee^{\ii \theta}
\end{pmatrix} 
\to
\begin{pmatrix}
    0 \\
    0 \\
    0 \\
    \ee^{\ii \theta}
\end{pmatrix}
\end{align}
We proceed similarly for $\ket{u_{2}}$.

\section{Properties of the sewing matrix}
\label{app:sec:PropertiesSewingMatrix}

Let us first prove the unitarity eq.~\eqref{eq:SewingMatrixUnitarity} of $\swm(k)$:
\begin{subequations}
\begin{align}
\left[ \Adjoint{\swm}(k) \cdot \swm(k) \right]_{ij}
&= 
\sum_{n} \left(\Adjoint{\swm} \right)_{in}(k) \left(\swm \right)_{nj} (k) 
=
\sum_{n} \Conjugate{\swm}_{ni}(k) \, \swm_{nj}(k) \\
&=
\sum_{n} 
\braket{\TR e_{i}(k) | e_{n}(-k)} \, \braket{e_{n}(-k) | \TR e_{j}(k)} \\
&=
\braket{\TR e_{i}(k) | \TR e_{j}(k)} 
=
\braket{e_{j}(k) | e_{i}(k)} 
= \KroneckerDelta_{ji} = \KroneckerDelta_{ij} . 
\end{align}
\end{subequations}
Two relations between the $w$ and $m$ matrices are useful in relating different expressions of the \Ztwo invariant. First let 
us note that at the time-reversal invariant points $\HighSymmetryPoint \in \HighSymmetryPoints$, the matrices $\SewingMatrix$ and $\KMTRMatrix$ coincide: 
\begin{equation}
  \SewingMatrix(\HighSymmetryPoint) = \KMTRMatrix(\HighSymmetryPoint)
  \qquad
  \text{($\HighSymmetryPoint \in \HighSymmetryPoints$)}
\end{equation}
Moreover, the sewing matrix relates $\KMTRMatrix(k)$ to $\KMTRMatrix(-k)$ by the relation \cite{FuKane2006}:
\begin{equation}
\KMTRMatrix(-k) = \SewingMatrix(k) \cdot \Conjugate{\KMTRMatrix}(k) \cdot \Transpose{\SewingMatrix}(k)
\label{eq:SewingMatrixKMTRMRelation}
\end{equation}
and therefore, using the identity $\Pf(B \, A \, \Transpose{B}) = \det{B} \, \Pf(A)$ for a $2n \times 2n$ antisymetric matrix $A$ and any $2n \times 2n$ matrix $B$, we get:
\begin{equation}
\det \SewingMatrix(k) = \frac{\Pf (\KMTRMatrix)(k)}{\Conjugate{\left( \Pf (\KMTRMatrix)(-k) \right)}}
\label{eq:SewingMatrixDeterminantKMTRMatrixPfaffian}
\end{equation}
Note that the determinant of the sewing matrix has no winding (see \ref{app:NoWindingOfSewingMatrixDeterminant}).

Let us now prove equation~\eqref{eq:SewingMatrixTR}: 
\begin{subequations}
\begin{align}
\Transpose{\swm}_{i j}(-k) 
&= \braket{e_{j}(k) | \TR e_{i}(-k)} 
= - \braket{e_{i}(k) | \TR e_{j}(k)} 
= - \swm_{i j}(k)
\end{align}
\end{subequations}
\noindent using the anti-unitarity of the time-reversal operator $\TR$ and the fact that $\TR^{2} = -\IdentityMatrix$ 

\section{Winding of the  the sewing matrix determinant}
\label{app:NoWindingOfSewingMatrixDeterminant}

Let us now turn to the determination of the winding of the determinant of the sewing matrix. 
From the unitarity \eqref{eq:SewingMatrixUnitarity} and the behavior under time-reversal \eqref{eq:SewingMatrixTR} of the sewing matrix, we deduce that its deteminant has modulus 
$\Abs{\det \swm(k)} = 1$. As the sewing matrix $\swm$ is an even-sized square matrix, we have:
\begin{equation}
\det \swm(\trb k) \equiv \det \swm(-k) = \det \swm(k)
\end{equation}
\noindent These properties imply that $\det \swm$ has no winding: its winding number, which is the holonomy of the differential form:
\begin{equation}
\omega[\swf] = \frac{1}{2 \pi \ii} \, \frac{\dd \swf}{\swf}
\end{equation}
\noindent along any loop $\mathcal{C}$ is zero:
\begin{equation}
W[f, \mathcal{C}] \equiv \frac{1}{2 \pi \ii} \oint_{\mathcal{C}} \dd \log \swf \stackrel{\shortdownarrow}{=} 0 .
\end{equation}
Indeed, if $\mathcal{C}$ is a time-reversal invariant loop, we write $\swf(k) = |\swf(k)| \ee^{\ii \phi(k)}$ so that $k \mapsto \phi(k)$ is even whereas $k \mapsto \nabla_{k} \phi(k)$ is odd, so the integral of $\dd \swf/\swf = \ii \; \dd \phi$ on the time-reversal invariant loop $\mathcal{C}$ is zero. Moreover, any loop $\mathcal{C}$ can be replaced by a time-reversal invariant loop $\mathcal{C}'$ that lies in the same homotopy class, so this property is always true.

Hence, the logarithm and the square root of $\swf$ can be globally defined. This property 
is crucial for the \Ztwo invariant to be well defined.

\section{From the Pfaffian invariant to the sewing matrix invariant}
\label{sec:DemoPfaffianToSewing}

In the following, we start from the expression \eqref{eq:Z2InvariantPfaffian} of the \Ztwo invariant and show (following \cite{FuKane2006}) that it is equivalent to the expression \eqref{eq:Z2InvariantSewingMatrix}. 
First, we recall that $\swm$ and $\KMTRMatrix$ coincide at the TRIM. 
We then split the integral in \eqref{eq:Z2InvariantPfaffian} into several integrals on paths connecting TRIM. 
Through the use of Stokes' theorem these integrals will simplify to values only at the TRIM. 
This procedure has to be done carefully in order to keep quantities defined modulo $2$ (as opposed to meaningless modulo $1$ integer quantities).

Notice that a complex logarithm is only defined modulo $2 \pi \ii$, and that a square root is only defined up to a sign. However, the logarithm and the square root of a continuous function $f$ with no winding can be globally defined 
\begin{figure}
\centering
\includegraphics[width=15cm]{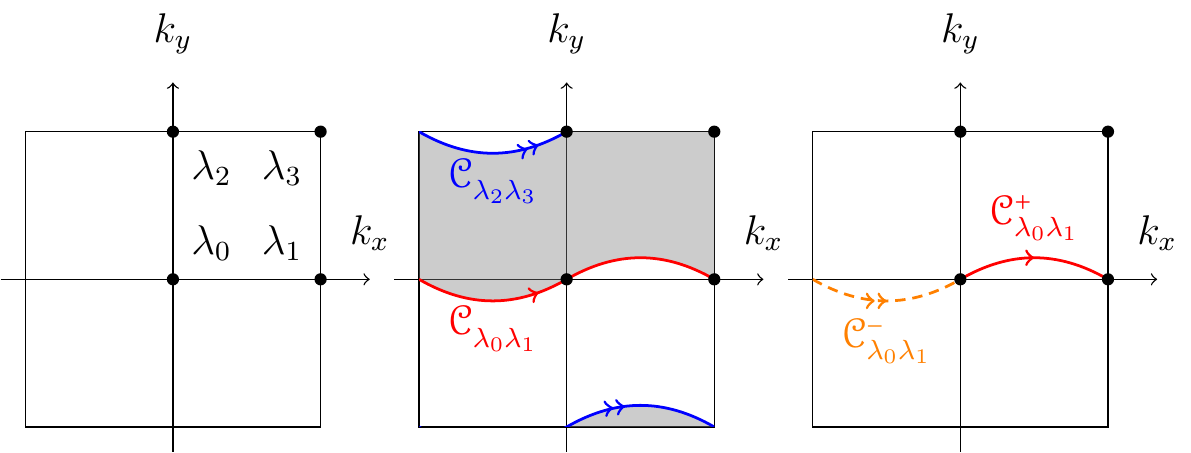}      
\caption{We denote by $\HighSymmetryPoint_{i}$ with $i=0,1,2,3$ the four TRIM in two dimension (left). Note that we use a specific indexing to have a concrete picture, but any indexing leads to the same result. The boundary of an EBZ consists, up to the orientation, of two time-reversal invariant curves connecting the TRIM (middle). We can decompose any time-reversal invariant curve $\TRIClosedCurve{}$ connecting two TRIM in two curves $\TRIClosedCurve[+]{}$ and $\TRIClosedCurve[-]{}$, images of one another by time-reversal, and each connecting the two TRIM (right).}
\label{fig:BZandEBZforMW}
\end{figure}

Let us start from the expression of the pfaffian invariant \eqref{eq:Z2InvariantPfaffian}: 
\begin{equation}
\ZtwoInvariantModTwo = \frac{1}{2 \pi \ii} \; \oint_{\partial\EBZ} \dd \log \Pf m  \quad \modTwo, 
\end{equation}
and rewrite it as 
\begin{equation}
  \ZtwoInvariantModTwo = \frac{1}{2 \pi \ii} \; \PTRC(\partial \EBZ)  \quad \modTwo
  \label{eq:Z2InvariantAsIdEBZ}
\end{equation}
where we defined the following quantity, for any closed loop $\TRIClosedCurve{}$ along which $Pf(m)$ does not vanish: 
\begin{equation}
\PTRC(\TRIClosedCurve{}) = \int_{\TRIClosedCurve{}} \dd \log \Pf m . 
\end{equation}
\noindent The boundary of the cylindrical Effective Brillouin Zone (EBZ) is a disjoint union of two time-reversal invariant 
closed loops with opposite orientations (see Fig.~\ref{fig:BZandEBZforMW}). 
As $\ZtwoInvariantModTwo$ is only defined modulo $2$, its global sign is of no importance, and the orientation of the EBZ is therefore irrelevant. Hence, with a correct relative orientation of the curves, we have: 
\begin{equation}
(-1)^{\ZtwoInvariantModTwo} = \ee^{\PTRC(\partial \EBZ) / 2}
=
\exp \left[ \frac{1}{2} \;  \left(
\PTRC(\TRIClosedCurve{\HighSymmetryPoint_{0} \, \HighSymmetryPoint_{1}})
-
\PTRC(\TRIClosedCurve{\HighSymmetryPoint_{2} \, \HighSymmetryPoint_{3}})
\right)
\right]
\end{equation}
As $\TRIClosedCurve{\HighSymmetryPoint_{i} \, \HighSymmetryPoint_{j}}$ is globally time-reversal invariant, we can
split it into two parts $\TRIClosedCurve[+]{\HighSymmetryPoint_{0} \, \HighSymmetryPoint_{1}}$ and $\TRIClosedCurve[-]{\HighSymmetryPoint_{0} \, \HighSymmetryPoint_{1}}$, images of each other by $k \to -k$ (but with the same orientation). 
Therefore,
\begin{equation}
\PTRC(\TRIClosedCurve{\HighSymmetryPoint_{i} \, \HighSymmetryPoint_{j}})
=
\PTRC(\TRIClosedCurve[+]{\HighSymmetryPoint_{i} \, \HighSymmetryPoint_{j}})
+
\PTRC(\TRIClosedCurve[-]{\HighSymmetryPoint_{i} \, \HighSymmetryPoint_{j}})
=
2 \PTRC(\TRIClosedCurve[+]{\HighSymmetryPoint_{i} \, \HighSymmetryPoint_{j}})
+
\left(
\PTRC(\TRIClosedCurve[-]{\HighSymmetryPoint_{i} \, \HighSymmetryPoint_{j}})
-
\PTRC(\TRIClosedCurve[+]{\HighSymmetryPoint_{i} \, \HighSymmetryPoint_{j}})
\right) . 
\label{eq:DemoMW:IntegralDecomposition}
\end{equation}
\noindent We integrate the first term, and using that at a TRIM $\HighSymmetryPoint$, $m(\HighSymmetryPoint) = \SewingMatrix(\HighSymmetryPoint)$, we get: 
\begin{equation}
\exp \left[
%1/2
%2 
\PTRC(\TRIClosedCurve[+]{\HighSymmetryPoint_{i} \, \HighSymmetryPoint_{j}})
\right]
=
\exp \left[
% 1/2
%2 \; 
\int_{\TRIClosedCurve[+]{\HighSymmetryPoint_{i} \, \HighSymmetryPoint_{j}}} \dd \log \Pf m(k) 
\right]
=
% 1/2
%2 \log 
\frac{\Pf m(\HighSymmetryPoint_{j})}{\Pf m(\HighSymmetryPoint_{i})} 
=
% 1/2
%2 \log 
\frac{\Pf \SewingMatrix(\HighSymmetryPoint_{j})}{\Pf \SewingMatrix(\HighSymmetryPoint_{i})}
\label{eq:DemoMW:Pf}
\end{equation}

\noindent because $\partial \TRIClosedCurve[+]{\HighSymmetryPoint_{i} \, \HighSymmetryPoint_{j}} = \HighSymmetryPoint_{j} - \HighSymmetryPoint_{j}$.
Using the fact that the two curves \TRIClosedCurve[+]{\HighSymmetryPoint_{i} \, \HighSymmetryPoint_{j}} and \TRIClosedCurve[-]{\HighSymmetryPoint_{i} \, \HighSymmetryPoint_{j}} are images of each other by time-reversal (up to the orientation), we 
can write: 
\begin{subequations}
\label{eq:DemoMW:OppositeCurvesIntegrals}
\begin{align}
\PTRC(\TRIClosedCurve[+]{\HighSymmetryPoint_{i} \, \HighSymmetryPoint_{j}})
-
\PTRC(\TRIClosedCurve[-]{\HighSymmetryPoint_{i} \, \HighSymmetryPoint_{j}})
&=
\int_{\TRIClosedCurve[+]{\HighSymmetryPoint_{i} \, \HighSymmetryPoint_{j}}} \nabla_{k} \log \Pf m(k) \cdot \dd k
-
\int_{\TRIClosedCurve[-]{\HighSymmetryPoint_{i} \, \HighSymmetryPoint_{j}}} \nabla_{k} \log \Pf m(k) \cdot \dd k \\
&=
\int_{\TRIClosedCurve[+]{\HighSymmetryPoint_{i} \, \HighSymmetryPoint_{j}}} \nabla_{k} \left( \log \Pf m(k) + \log \Pf m(-k) \right) \cdot \dd k \\
&=
\int_{\TRIClosedCurve[+]{\HighSymmetryPoint_{i} \, \HighSymmetryPoint_{j}}} \nabla_{k} \left( \log \Pf m(k) - \log \Pf \Conjugate{m}(-k) \right) \cdot \dd k
\end{align}
\end{subequations}

\noindent In the last step, we used, {\it via} a polar decomposition $\Pf m(k) = \rho(k) \; \ee^{\ii \theta(k)}$:
\begin{equation}
\log \Pf m(k) + \Pf m(-k) = 2 \log \rho(k) + \ii \left( \theta(k) + \theta(-k) \right)
\end{equation}
\noindent whereas
\begin{equation}
\log \Pf m(k) - \Pf \Conjugate{m}(-k) = \ii \left( \theta(k) + \theta(-k) \right) .
\end{equation}
\noindent The modulus of a complex number being real, it yields: 
\begin{equation}
\int_{\TRIClosedCurve[+]{\HighSymmetryPoint_{i} \, \HighSymmetryPoint_{j}}} \log \rho(k) = \log \rho(\HighSymmetryPoint_{j}) - \log \rho(\HighSymmetryPoint_{i}) = 0
\end{equation}
\noindent since the Pfaffian has modulus one at the TRIM. 
From \eqref{eq:SewingMatrixDeterminantKMTRMatrixPfaffian}, we deduce: 
\begin{equation}
\log \det \SewingMatrix(k) = \log \Pf m(k) - \log \Pf \Conjugate{m}(-k)
\label{eq:DemoMW:SewingMatrixKMMatrix}
\end{equation}
\noindent so that, with \eqref{eq:DemoMW:OppositeCurvesIntegrals} and \eqref{eq:DemoMW:SewingMatrixKMMatrix},
\begin{equation}
\exp \left[ \frac{1}{2} \left(
\PTRC(\TRIClosedCurve[+]{\HighSymmetryPoint_{i} \, \HighSymmetryPoint_{j}})
-
\PTRC(\TRIClosedCurve[-]{\HighSymmetryPoint_{i} \, \HighSymmetryPoint_{j}})
\right)
\right]
=
\exp \left[ \frac{1}{2}
\int_{\TRIClosedCurve[+]{\HighSymmetryPoint_{i} \, \HighSymmetryPoint_{j}}} \dd \log \det \SewingMatrix(k)
\right]
=
%2 \log 
\frac{\sqrt{\det \SewingMatrix(\HighSymmetryPoint_{j})}}{\sqrt{\det \SewingMatrix(\HighSymmetryPoint_{i})}}
\label{eq:DemoMW:SqrtDet}
\end{equation}

\noindent At this stage, it is necessary to continuously define the square root of $\det \swm$ so that the following formula is 
properly defined. This is only possible because $\det \swm$ has no winding provided the base vectors are continuously defined.
Collecting \eqref{eq:DemoMW:SqrtDet} and \eqref{eq:DemoMW:Pf} into \eqref{eq:DemoMW:IntegralDecomposition}, we finally obtain: 
\begin{equation}
\exp \left[ \frac{1}{2} \, \PTRC(\TRIClosedCurve{\HighSymmetryPoint_{i} \, \HighSymmetryPoint_{j}}) \right]
=
%2 \; \log \left(
\frac{\sqrt{\det \SewingMatrix(\HighSymmetryPoint_{i})}}{\sqrt{\det \SewingMatrix(\HighSymmetryPoint_{j})}}
\frac{\Pf \SewingMatrix(\HighSymmetryPoint_{j})}{\Pf \SewingMatrix(\HighSymmetryPoint_{i})}
%\right)
=
%2 \; \log \left(
\frac{\Pf \SewingMatrix(\HighSymmetryPoint_{i})}{\sqrt{\det \SewingMatrix(\HighSymmetryPoint_{i})}}
\frac{\Pf \SewingMatrix(\HighSymmetryPoint_{j})}{\sqrt{\det \SewingMatrix(\HighSymmetryPoint_{j})}}
%\right)
\end{equation}
%
%Note as $\Pf^2 = \det$, each ratio is $\pm 1$, so we can invert them to get a simpler expression. 
Therefore, the \Ztwo invariant \eqref{eq:Z2InvariantAsIdEBZ} can be written as: 
%
% \begin{equation}
% \ZtwoInvariantModTwo = \frac{1}{2 \pi \ii} \PTRC(\partial \EBZ)
% =
% \frac{1}{\pi \ii} \; \log \left(
% \prod_{\HighSymmetryPoint \in \HighSymmetryPoints}
% \frac{\Pf \SewingMatrix(\HighSymmetryPoint)}{\sqrt{\det \SewingMatrix(\HighSymmetryPoint)}}
% \right)
% \end{equation}
%
% \noindent that is
%
\begin{equation}
(-1)^{\ZtwoInvariantModTwo} = \ee^{\PTRC(\partial \EBZ) / 2}
=
\prod_{\HighSymmetryPoint \in \HighSymmetryPoints}
\frac{\Pf \SewingMatrix(\HighSymmetryPoint)}{\sqrt{\det \SewingMatrix(\HighSymmetryPoint)}}
\end{equation}
which is the desired result:

\section{Independence from the determinant bundle section of the \Ztwo Freed--Moore index}
\label{app:determinant}

 In this appendix, following \cite{FreedMoore2013}, we show the independence of the $\ZtwoInvariantIPKM$ index defined in eq.~\eqref{eq:Z2inDeterminant} 
 on the chosen section $\sigma$.  
Let us consider two time-reversal covariant nowhere vanishing global sections $\sigma$ et $\sigma'$. As the determinant bundle is a line bundle, $\sigma$ and $\sigma'$ are proportional: here is a nowhere vanishing smooth complex function $j$ so that:
\begin{equation}
\sigma'(k) = j(k) \, \sigma(k) . 
\end{equation}
Hence,
\begin{equation}
\ZtwoInvariantIPKM' = \prod_{\HighSymmetryPoint \in \HighSymmetryPoints}
\sign[\sigma'(\HighSymmetryPoint)]
= \prod_{\HighSymmetryPoint \in \HighSymmetryPoints}
\sign[j(\HighSymmetryPoint) \, \sigma(\HighSymmetryPoint)]
= \prod_{\HighSymmetryPoint \in \HighSymmetryPoints}
\sign[j(\HighSymmetryPoint)] \ZtwoInvariantIPKM . 
\end{equation}
\noindent Our purpose is thus to show that:
\begin{equation}
L = \prod_{\HighSymmetryPoint \in \HighSymmetryPoints}
\sign[j(\HighSymmetryPoint)] = 1
\end{equation}
To proceed, we will rewrite $L$ as the product of two identical holonomies %\footnote{Or \enquote{monodromies}.}
  (on homotopic loops). These holonomies correspond to the parity of a winding number, and thus take  values $\pm 1$. 
  However, as there are two identical ones, we obtain  $J=(\pm1)^{2}=1$. We also see that in the most general case, one \emph{can} change the sign of the product on two TRIM, which is therefore not invariant.
 Let us first deduce that:
\begin{equation}
j(-k) = \Conjugate{j(k)}
\label{eq:jConjugateSymmetry}
\end{equation}
from the covariance of  $\sigma$ and $\sigma'$: 
\begin{equation}
j(-k) \, \sigma(-k) = \sigma'(-k) = \ExteriorPower{2} \TR \, \sigma'(k)
= \ExteriorPower{2} \TR \left[ j(k) \sigma(k) \right]
\Conjugate{j(k)} \, \ExteriorPower{2} \TR \, \sigma(k)
= \Conjugate{j(k)} \, \sigma(-k) .
\end{equation}
We also note that $j(\HighSymmetryPoint) \in \RealField$, so we have $ \sign j(\HighSymmetryPoint)^{-1} = \sign j(\HighSymmetryPoint)$ and, with the notations of \ref{sec:DemoPfaffianToSewing}, we have:
\begin{equation}
L = \sign \left[
j(\HighSymmetryPoint_{0}) \, 
j(\HighSymmetryPoint_{1}) \, 
j(\HighSymmetryPoint_{2}) \, 
j(\HighSymmetryPoint_{3}) \, 
\right]
=  \sign \left[
\frac{j(\HighSymmetryPoint_{1})}{j(\HighSymmetryPoint_{0})}
\,
\frac{j(\HighSymmetryPoint_{3})}{j(\HighSymmetryPoint_{2})}
\right]
\end{equation}

Let ${\cal C}^+_{\lambda_i,\lambda_j}$ be a path connecting
$\lambda_i$ to $\lambda_j$ (see Fig.~\ref{fig:BZandEBZforMW}), and ${\cal C}^-_{\lambda_i,
\lambda_j}$ be its reoriented time reversed version. Together,
${\cal C}^-_{\lambda_i,\lambda_j}$, followed by ${\cal C}^+_{\lambda_i,
\lambda_j}$, forms a closed loop around the Brillouin torus. We
shall write $j(k)=\rho(k) \ee^{ \ii \phi(k)}$ with $\rho(k)>0$ and the real
phase $\phi(k)$ that may be multivalued on the Brillouin torus. We have:
\begin{equation}
\sign\left[\frac{j(\lambda_j)}{j\lambda_i}\right]
=\ee^{\ii \int_{{\cal C}^+_{\lambda_i,\lambda_j}}d\phi(k) } .
\end{equation}
On the other hand, by the change of variables $k\to-k$,
\begin{equation}
\int\limits_{{\cal C}^+_{\lambda_i,\lambda_j}} \dd\phi(k)
=-\int\limits_{{\cal C}^-_{\lambda_i,\lambda_j}} \dd\phi(-k) , 
\end{equation}
where the overall minus sign is due to reorientation of
${\cal C}^-_{\lambda_i,\lambda_j}$ as running from $\lambda_i$ to $\lambda_j$.
But the symmetry \eqref{eq:jConjugateSymmetry} implies that $d\phi(-k)=-d\phi(k)$. Hence:
\begin{equation}
\int\limits_{{\cal C}^+_{\lambda_i,\lambda_j}}\dd\phi(k)
=\int\limits_{{\cal C}^-_{\lambda_i,\lambda_j}}\dd\phi(k)=
\frac{1}{2}\int\limits_{{\cal C}_{\lambda_i,\lambda_j}}  \dd\phi(k)=
\pi W[j,{\cal C}_{\lambda_i,\lambda_j}] , 
\end{equation}
where $W[j,{\cal C}_{\lambda_i,\lambda_j}]$ is the winding number
of $j(k)$ along the loop ${\cal C}_{\lambda_i,\lambda_j}$.
We infer that:
\begin{equation}
\sign\left[\frac{j(\lambda_j)}{j\lambda_i}\right]
=\ee^{ \ii \pi W[j,{\cal C}_{\lambda_i,\lambda_j}]} , 
\end{equation}
and hence that
\begin{equation}
\sign\left[\frac{j(\lambda_1)}{j\lambda_0}\frac{j(\lambda_3)}{j(\lambda_2)}\right]
=\ee^{ \ii \pi(W[j,{\cal C}_{\lambda_0,\lambda_1}]+W[j,{\cal C}_{\lambda_2,\lambda_3}])} . 
\end{equation}
The two (integer) winding numbers on the right-hand side are equal
as the loops ${\cal C}_{\lambda_0,\lambda_1}$ and ${\cal C}_{\lambda_2,\lambda_3}$ on the Brilloin zone are homotopic. Hence the sign in question is $+$,
so that the index \eqref{eq:KMencoreunefois}
 does not depend on the choice of section $\sigma$.

\section{From Freed--Moore to Kane--Mele invariants}
\label{sec:FreedMooreKaneMele}

  We now identify explicitly a global time-reversal covariant section which allows us to identify the Freed--Moore invariant~\eqref{eq:Z2inDeterminant}
  with 
 Fu and Kane's expression  of the \Ztwo invariant.
Let us consider the global basis of $\ExteriorPower{2} \FilledBandsBundle$ defined by:
\begin{equation}
s(k) = e_{1}(k) \wedge e_{2}(k) . 
\end{equation}
As the determinant bundle is a line bundle, $\ExteriorPower{2} \TR s(k)$ is proportional to $s(-k)$. Let $\swf(k)$ be the proportionality coefficient, so that:
\begin{equation}
\ExteriorPower{2} \TR s(k) = \swf(k) \, s(-k)
\end{equation}
We call $\swf$ the sewing function, for reasons to appear below. 
By using the expression \eqref{eq:SewingMatrixTRExpand} 
in terms of the sewing matrix \eqref{eq:SewingMatrixDefinition},
 we obtain: 
\begin{align}
\ExteriorPower{2} \TR s(k)
&= \TR e_{1}(k) \wedge \TR e_{2}(k) 
\nonumber \\
&= \left( \swm_{1 1}(k) e_{1}(-k) + \swm_{2 1}(k) e_{2}(-k) \right)
\wedge
\left( \swm_{1 2}(k) e_{1}(-k) + \swm_{2 2}(k) e_{2}(-k) \right) 
\nonumber \\
&= 
\swm_{1 1}(k) \, \swm_{2 2}(k) \; e_{1}(-k) \wedge e_{2}(-k)
+
\swm_{2 1}(k) \, \swm_{1 2}(k) \; e_{2}(-k) \wedge e_{1}(-k) 
\nonumber \\
&= \left(
\swm_{1 1}(k) \, \swm_{2 2}(k) - \swm_{1 2}(k) \, \swm_{2 1}(k) 
\right) \; e_{1}(-k) \wedge e_{2}(-k) 
\nonumber \\
&= \det(\swm) \; e_{1}(-k) \wedge e_{2}(-k) . 
\end{align}
 Hence  the sewing function appears as the determinant of the sewing matrix (up to a complex conjugation):
\begin{equation}
\swf(k) = \det \swm(k)
\label{eq:swf_swm}
\end{equation}
This sewing function has no winding (see \ref{app:determinant},\ref{app:NoWindingOfSewingMatrixDeterminant}). We thus consider safely 
its square root $\swf^{1/2}$. 
As $\swf(-k)=\swf(k)$, we have $\swf^{1/2}(-k) = \pm \swf^{1/2}(k)$. Evaluating this expression at the TRIM, the global signs is fixed to be $+1$. 
Therefore,
\begin{equation}
\swf^{1/2}(-k) = \swf^{1/2}(k) . 
\end{equation}
\noindent We now define a new section:
\begin{equation}
\sigma(k) = \swf^{1/2}(k) s(k). 
\end{equation}
As $|\swf^{1/2}(k)| = 1$, $(\swf^{1/2}(k))^{\star} = (\swf^{1/2}(k))^{-1}$, it satisfies: 
\begin{subequations}
\begin{align}
\ExteriorPower{2} \TR \sigma(k)
%&= \ExteriorPower{2} \TR \left[
%\swf^{1/2}(k) s(k)
%\right] \\
&= (\swf^{1/2}(k))^{\star} \; \ExteriorPower{2} \TR s(k) 
 \\
&= (\swf^{1/2}(k))^{-1} \; f(k) s(-k) \\
&= \swf^{1/2}(k) \, s(-k) \\
&= \swf^{1/2}(-k) \, s(-k) \\
&= \sigma(-k)
\end{align}
\end{subequations}
Let us finally notice that: 
\begin{subequations}
\begin{align}
e_1(\HighSymmetryPoint) \wedge \TR e_1(\HighSymmetryPoint)
&= e_1(\HighSymmetryPoint) \wedge \left(
\swm_{1 1}(\HighSymmetryPoint) e_{1}(\trb \HighSymmetryPoint) + 
\swm_{2 1}(\HighSymmetryPoint) e_{2}(\trb \HighSymmetryPoint)
\right) \\
&= e_1(\HighSymmetryPoint) \wedge \left(
\swm_{1 1}(\HighSymmetryPoint) e_{1}(\HighSymmetryPoint) + 
\swm_{2 1}(\HighSymmetryPoint) e_{2}(\HighSymmetryPoint)
\right) \\
&= \swm_{2 1}(\HighSymmetryPoint) \, e_1(\HighSymmetryPoint) \wedge e_{2}(\trb \HighSymmetryPoint)
\\
&= - t(\HighSymmetryPoint) \, s(\HighSymmetryPoint)
\end{align}
\end{subequations}
where $t = \swm_{2 1} =  \Pf(\swm)$ is the Pfaffian of the sewing matrix, which is defined at the TRIM (see \eqref{eq:swf_at_HSP}). Hence,
\begin{align}
\sigma(\HighSymmetryPoint)
&\equiv \swf^{1/2}(\HighSymmetryPoint) s(\HighSymmetryPoint) 
= 
- \frac{\swf^{1/2}(\HighSymmetryPoint)}{t(\HighSymmetryPoint)}
\, e_1(\HighSymmetryPoint) \wedge \TR e_1(\HighSymmetryPoint)
\end{align}
Using equations \eqref{eq:swf_at_HSP} and \eqref{eq:swf_swm}, we obtain:
\begin{equation}
\left[t(\HighSymmetryPoint)\right]^{2} = \left[\Pf \swm(\HighSymmetryPoint)\right]^{2} = \det \swm(\HighSymmetryPoint) = f(\HighSymmetryPoint)
\end{equation}
so we have $\swf^{1/2}(\HighSymmetryPoint)/t(\HighSymmetryPoint)= \pm 1$.
Thereby, the ratio $\swf^{1/2}(\HighSymmetryPoint)/t(\HighSymmetryPoint)$ gives the sign of the global time-reversal covariant section~$\sigma$ at the TRIM~$\HighSymmetryPoint \in \HighSymmetryPoints$ (up to the overall minus sign).

At a TRIM  $\HighSymmetryPoint$,  $\sigma(\HighSymmetryPoint)$ is a real element whose sign is unambiguously determined, here by:
\begin{equation}
\sign[\sigma(\HighSymmetryPoint)] 
=-  \sign\left[ \frac{\swf^{1/2}(\HighSymmetryPoint)}{t(\HighSymmetryPoint)} \right]
=-  \frac{\swf^{1/2}(\HighSymmetryPoint)}{t(\HighSymmetryPoint)}
= - \frac{\sqrt{\det \swm(\HighSymmetryPoint)}}{\Pf \swm(\HighSymmetryPoint)}
\end{equation}
When $\sigma$ is constructed as explained in last paragraph, the \Ztwo invariant is indeed \cite{FreedMoore2013}:
\begin{equation}
\ZtwoInvariantIPKM = (-1)^{\ZtwoInvariantModTwo} = \prod_{\HighSymmetryPoint \in \HighSymmetryPoints}
\frac{\sqrt{\det \swm(\HighSymmetryPoint)}}{\Pf \swm(\HighSymmetryPoint)}
\end{equation}
This is obviously equivalent to the Fu and Kane \cite{FuKane2006} expression \eqref{eq:Z2InvariantSewingMatrix}.

% ceci est censé être le style des CRAS Physique d'après le pdf Elsevier
\bibliographystyle{CRAS_with_doi_eprint}
\bibliography{bibliographieCRAS}

\begin{thebibliography}{60}
\expandafter\ifx\csname natexlab\endcsname\relax\def\natexlab#1{#1}\fi
\providecommand{\url}[1]{\texttt{#1}}
\providecommand{\href}[2]{#2}
\providecommand{\path}[1]{#1}
\providecommand{\DOIprefix}{doi:}
\providecommand{\ArXivprefix}{arXiv:}
\providecommand{\URLprefix}{URL: }
\providecommand{\Pubmedprefix}{pmid:}
\providecommand{\doi}[1]{\href{http://dx.doi.org/#1}{\path{#1}}}
\providecommand{\Pubmed}[1]{\href{pmid:#1}{\path{#1}}}
\providecommand{\bibinfo}[2]{#2}
\ifx\xfnm\relax \def\xfnm[#1]{\unskip,\space#1}\fi
%Type = Book
\bibitem[{Nakahara(2003)}]{Nakahara}
\bibinfo{author}{M.~Nakahara}, \bibinfo{title}{Geometry, Topology and Physics,
  Second Edition}, \bibinfo{edition}{2} ed., \bibinfo{publisher}{Taylor \&
  Francis}, \bibinfo{year}{2003}.
%Type = Article
\bibitem[{Klitzing et~al.(1980)Klitzing, Dorda, and
  Pepper}]{KlitzingDordaPepper1980}
\bibinfo{author}{K.~v. Klitzing}, \bibinfo{author}{G.~Dorda},
  \bibinfo{author}{M.~Pepper},
\newblock \bibinfo{title}{New method for high-accuracy determination of the
  fine-structure constant based on quantized {Hall} resistance},
\newblock \bibinfo{journal}{Phys. Rev. Lett.} \bibinfo{volume}{45}
  (\bibinfo{year}{1980}) \bibinfo{pages}{494--497}.
  \DOIprefix\doi{10.1103/PhysRevLett.45.494}.
%Type = Book
\bibitem[{Dou{\c c}ot et~al.(2004)Dou{\c c}ot, Duplantier, Pasquier, and
  Rivasseau}]{Doucot:2004}
\bibinfo{editor}{B.~Dou{\c c}ot}, \bibinfo{editor}{B.~Duplantier},
  \bibinfo{editor}{V.~Pasquier}, \bibinfo{editor}{V.~Rivasseau} (Eds.),
  \bibinfo{title}{The Quantum Hall Effect, Poincar{\'e} Seminar},
  \bibinfo{publisher}{Birkh{\"a}user}, \bibinfo{year}{2004}.
%Type = Article
\bibitem[{Thouless et~al.(1982)Thouless, Kohmoto, Nightingale, and den
  Nijs}]{ThoulessKohmotoNightingaleNijs1982}
\bibinfo{author}{D.~J. Thouless}, \bibinfo{author}{M.~Kohmoto},
  \bibinfo{author}{M.~P. Nightingale}, \bibinfo{author}{M.~den Nijs},
\newblock \bibinfo{title}{Quantized {Hall} conductance in a two-dimensional
  periodic potential},
\newblock \bibinfo{journal}{Phys. Rev. Lett.} \bibinfo{volume}{49}
  (\bibinfo{year}{1982}) \bibinfo{pages}{405--408}.
  \DOIprefix\doi{10.1103/PhysRevLett.49.405}.
%Type = Article
\bibitem[{Haldane(1988)}]{Haldane88}
\bibinfo{author}{F.~D.~M. Haldane},
\newblock \bibinfo{title}{Model for a quantum {Hall} effect without landau
  levels: Condensed-matter realization of the "parity anomaly"},
\newblock \bibinfo{journal}{Phys. Rev. Lett.} \bibinfo{volume}{61}
  (\bibinfo{year}{1988}) \bibinfo{pages}{2015--2018}.
  \DOIprefix\doi{10.1103/PhysRevLett.61.2015}.
%Type = Article
\bibitem[{Chang et~al.(2013)Chang, Zhang, Feng, Shen, Zhang, Guo, Li, Ou, Wei,
  Wang, Ji, Feng, Ji, Chen, Jia, Dai, Fang, Zhang, He, Wang, Lu, Ma, and
  Xue}]{Chang:2013}
\bibinfo{author}{C.-Z. Chang}, \bibinfo{author}{J.~Zhang},
  \bibinfo{author}{X.~Feng}, \bibinfo{author}{J.~Shen},
  \bibinfo{author}{Z.~Zhang}, \bibinfo{author}{M.~Guo},
  \bibinfo{author}{K.~Li}, \bibinfo{author}{Y.~Ou}, \bibinfo{author}{P.~Wei},
  \bibinfo{author}{L.-L. Wang}, \bibinfo{author}{Z.-Q. Ji},
  \bibinfo{author}{Y.~Feng}, \bibinfo{author}{S.~Ji},
  \bibinfo{author}{X.~Chen}, \bibinfo{author}{J.~Jia},
  \bibinfo{author}{X.~Dai}, \bibinfo{author}{Z.~Fang}, \bibinfo{author}{S.-C.
  Zhang}, \bibinfo{author}{K.~He}, \bibinfo{author}{Y.~Wang},
  \bibinfo{author}{L.~Lu}, \bibinfo{author}{X.-C. Ma}, \bibinfo{author}{Q.-K.
  Xue},
\newblock \bibinfo{title}{Experimental observation of the quantum anomalous
  {Hall} effect in a magnetic topological insulator},
\newblock \bibinfo{journal}{Science} \bibinfo{volume}{340}
  (\bibinfo{year}{2013}) \bibinfo{pages}{167--170}.
  \DOIprefix\doi{10.1126/science.1234414}.
%Type = Article
\bibitem[{Kane and Mele(2005)}]{KaneMele2005}
\bibinfo{author}{C.~L. Kane}, \bibinfo{author}{E.~J. Mele},
\newblock \bibinfo{title}{${Z}_{2}$ topological order and the quantum spin
  {Hall} effect},
\newblock \bibinfo{journal}{Phys. Rev. Lett.} \bibinfo{volume}{95}
  (\bibinfo{year}{2005}) \bibinfo{pages}{146802}.
  \DOIprefix\doi{10.1103/PhysRevLett.95.146802}.
  \href{http://arxiv.org/abs/cond-mat/0506581}{\tt arXiv:cond-mat/0506581}.
%Type = Article
\bibitem[{{Kane} and {Mele}(2005)}]{KaneMele2005B}
\bibinfo{author}{C.~L. {Kane}}, \bibinfo{author}{E.~J. {Mele}},
\newblock \bibinfo{title}{Quantum spin hall effect in graphene},
\newblock \bibinfo{journal}{Phys. Rev. Lett.} \bibinfo{volume}{95}
  (\bibinfo{year}{2005}) \bibinfo{pages}{226801}.
  \DOIprefix\doi{10.1103/PhysRevLett.95.226801}.
  \href{http://arxiv.org/abs/cond-mat/0411737}{\tt arXiv:cond-mat/0411737}.
%Type = Article
\bibitem[{{Bernevig} et~al.(2006){Bernevig}, {Hughes}, and {Zhang}}]{BHZ2006}
\bibinfo{author}{B.~A. {Bernevig}}, \bibinfo{author}{T.~L. {Hughes}},
  \bibinfo{author}{S.-C. {Zhang}},
\newblock \bibinfo{title}{{Quantum Spin Hall Effect and Topological Phase
  Transition in HgTe Quantum Wells}},
\newblock \bibinfo{journal}{Science} \bibinfo{volume}{314}
  (\bibinfo{year}{2006}) \bibinfo{pages}{1757--1761}.
  \DOIprefix\doi{10.1126/science.1133734}.
  \href{http://arxiv.org/abs/cond-mat/0611399}{\tt arXiv:cond-mat/0611399}.
%Type = Article
\bibitem[{K{\"o}nig et~al.(2007)K{\"o}nig, Wiedmann, Br{\"u}ne, Roth, Buhmann,
  Molenkamp, Qi, and Zhang}]{Konig2007}
\bibinfo{author}{M.~K{\"o}nig}, \bibinfo{author}{S.~Wiedmann},
  \bibinfo{author}{C.~Br{\"u}ne}, \bibinfo{author}{A.~Roth},
  \bibinfo{author}{H.~Buhmann}, \bibinfo{author}{L.~W. Molenkamp},
  \bibinfo{author}{X.-L. Qi}, \bibinfo{author}{S.-C. Zhang},
\newblock \bibinfo{title}{Quantum spin {Hall} insulator state in {HgTe} quantum
  wells},
\newblock \bibinfo{journal}{Science} \bibinfo{volume}{318}
  (\bibinfo{year}{2007}) \bibinfo{pages}{766--770}.
  \DOIprefix\doi{10.1126/science.1148047}.
  \href{http://arxiv.org/abs/0710.0582}{\tt arXiv:0710.0582}.
%Type = Article
\bibitem[{Roth et~al.(2009)Roth, Br{\"u}ne, Buhmann, Molenkamp, Maciejko, Qi,
  and Zhang}]{Roth:2009}
\bibinfo{author}{A.~Roth}, \bibinfo{author}{C.~Br{\"u}ne},
  \bibinfo{author}{H.~Buhmann}, \bibinfo{author}{L.~W. Molenkamp},
  \bibinfo{author}{J.~Maciejko}, \bibinfo{author}{X.-L. Qi},
  \bibinfo{author}{S.-C. Zhang},
\newblock \bibinfo{title}{Nonlocal transport in the quantum spin {H}all state},
\newblock \bibinfo{journal}{Science} \bibinfo{volume}{325}
  (\bibinfo{year}{2009}) \bibinfo{pages}{294}.
  \DOIprefix\doi{10.1126/science.1174736}.
  \href{http://arxiv.org/abs/0905.0365}{\tt arXiv:0905.0365}.
%Type = Article
\bibitem[{{Fu} et~al.(2007){Fu}, {Kane}, and {Mele}}]{FuKaneMele2007}
\bibinfo{author}{L.~{Fu}}, \bibinfo{author}{C.~L. {Kane}},
  \bibinfo{author}{E.~J. {Mele}},
\newblock \bibinfo{title}{Topological insulators in three dimensions},
\newblock \bibinfo{journal}{Phys. Rev. Lett.} \bibinfo{volume}{98}
  (\bibinfo{year}{2007}) \bibinfo{pages}{106803}.
  \DOIprefix\doi{10.1103/PhysRevLett.98.106803}.
  \href{http://arxiv.org/abs/cond-mat/0607699}{\tt arXiv:cond-mat/0607699}.
%Type = Article
\bibitem[{{Moore} and {Balents}(2007)}]{MooreBalents2007}
\bibinfo{author}{J.~E. {Moore}}, \bibinfo{author}{L.~{Balents}},
\newblock \bibinfo{title}{Topological invariants of time-reversal-invariant
  band structures},
\newblock \bibinfo{journal}{Phys. Rev. B} \bibinfo{volume}{75}
  (\bibinfo{year}{2007}) \bibinfo{pages}{121306}.
  \DOIprefix\doi{10.1103/PhysRevB.75.121306}.
  \href{http://arxiv.org/abs/cond-mat/0607314}{\tt arXiv:cond-mat/0607314}.
%Type = Article
\bibitem[{{Roy}(2009)}]{Roy2009b}
\bibinfo{author}{R.~{Roy}},
\newblock \bibinfo{title}{Topological phases and the quantum spin hall effect
  in three dimensions},
\newblock \bibinfo{journal}{Phys. Rev. B} \bibinfo{volume}{79}
  (\bibinfo{year}{2009}) \bibinfo{pages}{195322}.
  \DOIprefix\doi{10.1103/PhysRevB.79.195322}.
  \href{http://arxiv.org/abs/cond-mat/0607531}{\tt arXiv:cond-mat/0607531}.
%Type = Article
\bibitem[{{Hasan} and {Kane}(2010)}]{HasanKane2010}
\bibinfo{author}{M.~Z. {Hasan}}, \bibinfo{author}{C.~L. {Kane}},
\newblock \bibinfo{title}{Colloquium: Topological insulators},
\newblock \bibinfo{journal}{Rev. Mod. Phys.} \bibinfo{volume}{82}
  (\bibinfo{year}{2010}) \bibinfo{pages}{3045--3067}.
  \DOIprefix\doi{10.1103/RevModPhys.82.3045}.
  \href{http://arxiv.org/abs/1002.3895}{\tt arXiv:1002.3895}.
%Type = Article
\bibitem[{Qi and Zhang(2011)}]{Qi:2011}
\bibinfo{author}{X.-L. Qi}, \bibinfo{author}{S.-C. Zhang},
\newblock \bibinfo{title}{Topological insulators and superconductors},
\newblock \bibinfo{journal}{Rev. Mod. Phys.} \bibinfo{volume}{83}
  (\bibinfo{year}{2011}) \bibinfo{pages}{1057}.
  \DOIprefix\doi{10.1103/RevModPhys.83.1057}.
  \href{http://arxiv.org/abs/1008.2026}{\tt arXiv:1008.2026}.
%Type = Book
\bibitem[{Bernevig and Hughes(2013)}]{Bernevig}
\bibinfo{author}{B.~A. Bernevig}, \bibinfo{author}{T.~L. Hughes},
  \bibinfo{title}{Topological Insulators and Topological Superconductors},
  \bibinfo{publisher}{Princeton University Press}, \bibinfo{year}{2013}.
%Type = Book
\bibitem[{Volovik(2003)}]{Volovik:2003}
\bibinfo{author}{G.~E. Volovik}, \bibinfo{title}{The Universe in a Helium
  Droplet}, \bibinfo{publisher}{Oxford University Press}, \bibinfo{year}{2003}.
%Type = Book
\bibitem[{Wen(2004)}]{Wen2004}
\bibinfo{author}{X.-G. Wen}, \bibinfo{title}{Quantum field theory of many-body
  systems}, \bibinfo{publisher}{Oxford University Press}, \bibinfo{year}{2004}.
%Type = Book
\bibitem[{Baez and Muniain(1994)}]{Baez}
\bibinfo{author}{J.~C. Baez}, \bibinfo{author}{J.~P. Muniain},
  \bibinfo{title}{Gauge Fields, Knots, and Gravity}, \bibinfo{publisher}{World
  Scientific}, \bibinfo{year}{1994}.
%Type = Book
\bibitem[{Hatcher(2003)}]{HatcherVBKT}
\bibinfo{author}{A.~Hatcher}, \bibinfo{title}{Vector bundles and K-theory},
  \bibinfo{year}{2003}. \URLprefix
  \url{http://www.math.cornell.edu/~hatcher/VBKT/VBpage.html}.
%Type = Article
\bibitem[{Simon(1983)}]{Simon1983}
\bibinfo{author}{B.~Simon},
\newblock \bibinfo{title}{Holonomy, the quantum adiabatic theorem, and
  {Berry}'s phase},
\newblock \bibinfo{journal}{Phys. Rev. Lett.} \bibinfo{volume}{51}
  (\bibinfo{year}{1983}) \bibinfo{pages}{2167--2170}.
  \DOIprefix\doi{10.1103/PhysRevLett.51.2167}.
%Type = Article
\bibitem[{Panati(2007)}]{Panati2007}
\bibinfo{author}{G.~Panati},
\newblock \bibinfo{title}{Triviality of bloch and bloch-dirac bundles},
\newblock \bibinfo{journal}{Ann. H. Poincar{\'e}} \bibinfo{volume}{8}
  (\bibinfo{year}{2007}) \bibinfo{pages}{995--1011}.
  \DOIprefix\doi{10.1007/s00023-007-0326-8}.
  \href{http://arxiv.org/abs/math-ph/0601034}{\tt arXiv:math-ph/0601034}.
%Type = Article
\bibitem[{Bena and Montambaux(2009)}]{BenaMontambaux2009}
\bibinfo{author}{C.~Bena}, \bibinfo{author}{G.~Montambaux},
\newblock \bibinfo{title}{Remarks on the tight-binding model of graphene},
\newblock \bibinfo{journal}{New Journal of Physics} \bibinfo{volume}{11}
  (\bibinfo{year}{2009}) \bibinfo{pages}{095003}.
  \DOIprefix\doi{10.1088/1367-2630/11/9/095003}.
  \href{http://arxiv.org/abs/0712.0765}{\tt arXiv:0712.0765}.
%Type = Article
\bibitem[{{Avron} et~al.(1983){Avron}, {Seiler}, and
  {Simon}}]{AvronSeilerSimon1983}
\bibinfo{author}{J.~E. {Avron}}, \bibinfo{author}{R.~{Seiler}},
  \bibinfo{author}{B.~{Simon}},
\newblock \bibinfo{title}{{Homotopy and Quantization in Condensed Matter
  Physics}},
\newblock \bibinfo{journal}{Phys. Rev. Lett.} \bibinfo{volume}{51}
  (\bibinfo{year}{1983}) \bibinfo{pages}{51--53}.
  \DOIprefix\doi{10.1103/PhysRevLett.51.51}.
%Type = Article
\bibitem[{Dirac(1931)}]{Dirac1931}
\bibinfo{author}{P.~A.~M. Dirac},
\newblock \bibinfo{title}{Quantised singularities in the electromagnetic
  field},
\newblock \bibinfo{journal}{Proceedings of the Royal Society of London. Series
  A, Containing Papers of a Mathematical and Physical Character}
  \bibinfo{volume}{133} (\bibinfo{year}{1931}) \bibinfo{pages}{pp. 60--72}.
%Type = Book
\bibitem[{{Chru{\'s}ci{\'n}ski} and
  {Jamio{\l}kowski}(2004)}]{ChrusscinnskiJamiolkowski}
\bibinfo{author}{D.~{Chru{\'s}ci{\'n}ski}},
  \bibinfo{author}{A.~{Jamio{\l}kowski}}, \bibinfo{title}{Geometric Phases in
  Classical and Quantum Mechanics}, Progress in Mathematical Physics,
  \bibinfo{publisher}{Birkh{\"a}user}, \bibinfo{year}{2004}.
%Type = Article
\bibitem[{Berry(1984)}]{Berry1984}
\bibinfo{author}{M.~V. Berry},
\newblock \bibinfo{title}{Quantal phase factors accompanying adiabatic
  changes},
\newblock \bibinfo{journal}{Proceedings of the Royal Society of London. A.
  Mathematical and Physical Sciences} \bibinfo{volume}{392}
  (\bibinfo{year}{1984}) \bibinfo{pages}{45--57}.
  \DOIprefix\doi{10.1098/rspa.1984.0023}.
%Type = Article
\bibitem[{{Sticlet} et~al.(2012){Sticlet}, {Pi{\'e}chon}, {Fuchs}, {Kalugin},
  and {Simon}}]{SticletPiechonFuchsKukuginSimon2012}
\bibinfo{author}{D.~{Sticlet}}, \bibinfo{author}{F.~{Pi{\'e}chon}},
  \bibinfo{author}{J.-N. {Fuchs}}, \bibinfo{author}{P.~{Kalugin}},
  \bibinfo{author}{P.~{Simon}},
\newblock \bibinfo{title}{Geometrical engineering of a two-band chern insulator
  in two dimensions with arbitrary topological index},
\newblock \bibinfo{journal}{Phys. Rev. B} \bibinfo{volume}{85}
  (\bibinfo{year}{2012}) \bibinfo{pages}{165456}.
  \DOIprefix\doi{10.1103/PhysRevB.85.165456}.
  \href{http://arxiv.org/abs/1201.6613}{\tt arXiv:1201.6613}.
%Type = Book
\bibitem[{Fradkin(2013)}]{Fradkin}
\bibinfo{author}{E.~Fradkin}, \bibinfo{title}{Field Theories of Condensed
  Matter Physics}, \bibinfo{edition}{2} ed., \bibinfo{publisher}{Cambridge
  University Press}, \bibinfo{year}{2013}.
%Type = Article
\bibitem[{Nielsen and Ninomiya(1981)}]{Nielsen:1981}
\bibinfo{author}{H.~B. Nielsen}, \bibinfo{author}{M.~Ninomiya},
\newblock \bibinfo{title}{No go theorem for regularizing chiral fermions},
\newblock \bibinfo{journal}{Phys. Lett.} \bibinfo{volume}{105}
  (\bibinfo{year}{1981}) \bibinfo{pages}{219}.
  \DOIprefix\doi{http://dx.doi.org/10.1016/0370-2693(81)91026-1}.
%Type = Book
\bibitem[{Le~Bellac(2012)}]{LeBellac}
\bibinfo{author}{M.~Le~Bellac}, \bibinfo{title}{Quantum Physics},
  \bibinfo{publisher}{Cambridge University Press}, \bibinfo{year}{2012}.
%Type = Book
\bibitem[{Sakurai(1993)}]{SakuraiModernQM}
\bibinfo{author}{J.~J. Sakurai}, \bibinfo{title}{Modern Quantum Mechanics},
  \bibinfo{edition}{1} ed., \bibinfo{publisher}{Addison Wesley},
  \bibinfo{year}{1993}.
%Type = Book
\bibitem[{Madelung(1996)}]{Madelung}
\bibinfo{author}{O.~Madelung}, \bibinfo{title}{Introduction to Solid-State
  Theory}, Springer Series in Solid-State Sciences,
  \bibinfo{publisher}{Springer}, \bibinfo{year}{1996}.
%Type = Article
\bibitem[{Fu and Kane(2007)}]{FuKane2007}
\bibinfo{author}{L.~Fu}, \bibinfo{author}{C.~L. Kane},
\newblock \bibinfo{title}{Topological insulators with inversion symmetry},
\newblock \bibinfo{journal}{Phys. Rev. B} \bibinfo{volume}{76}
  (\bibinfo{year}{2007}) \bibinfo{pages}{045302}.
  \DOIprefix\doi{10.1103/PhysRevB.76.045302}.
  \href{http://arxiv.org/abs/cond-mat/0611341}{\tt arXiv:cond-mat/0611341}.
%Type = Unpublished
\bibitem[{{Graf} and {Porta}(2012)}]{GrafPorta2012}
\bibinfo{author}{G.~M. {Graf}}, \bibinfo{author}{M.~{Porta}},
  \bibinfo{title}{{Bulk-edge correspondence for two-dimensional topological
  insulators}}, \bibinfo{year}{2012}.
  \href{http://arxiv.org/abs/1207.5989v2}{\tt arXiv:1207.5989v2}.
%Type = Unpublished
\bibitem[{{Alexandradinata} et~al.(2012){Alexandradinata}, {Dai}, and
  {Bernevig}}]{AlexandradinataDaiBernevig2012}
\bibinfo{author}{A.~{Alexandradinata}}, \bibinfo{author}{X.~{Dai}},
  \bibinfo{author}{B.~A. {Bernevig}}, \bibinfo{title}{Wilson-loop
  characterization of inversion-symmetric topological insulators},
  \bibinfo{year}{2012}. \href{http://arxiv.org/abs/1208.4234}{\tt
  arXiv:1208.4234}.
%Type = Inproceedings
\bibitem[{{Kitaev}(2009)}]{Kitaev2009}
\bibinfo{author}{A.~{Kitaev}},
\newblock \bibinfo{title}{Periodic table for topological insulators and
  superconductors},
\newblock in: \bibinfo{editor}{V.~{Lebedev}}, \bibinfo{editor}{M.~{Feigel'Man}}
  (Eds.), \bibinfo{booktitle}{American Institute of Physics Conference Series},
  volume \bibinfo{volume}{1134} of \textit{\bibinfo{series}{American Institute
  of Physics Conference Series}}, \bibinfo{year}{2009}, pp.
  \bibinfo{pages}{22--30}. \DOIprefix\doi{10.1063/1.3149495}.
  \href{http://arxiv.org/abs/0901.2686}{\tt arXiv:0901.2686}.
%Type = Article
\bibitem[{{Ryu} et~al.(2010){Ryu}, {Schnyder}, {Furusaki}, and
  {Ludwig}}]{RyuSchnyderFurusakiLudwid2010}
\bibinfo{author}{S.~{Ryu}}, \bibinfo{author}{A.~P. {Schnyder}},
  \bibinfo{author}{A.~{Furusaki}}, \bibinfo{author}{A.~W.~W. {Ludwig}},
\newblock \bibinfo{title}{Topological insulators and superconductors: tenfold
  way and dimensional hierarchy},
\newblock \bibinfo{journal}{New Journal of Physics} \bibinfo{volume}{12}
  (\bibinfo{year}{2010}) \bibinfo{pages}{065010}.
  \DOIprefix\doi{10.1088/1367-2630/12/6/065010}.
  \href{http://arxiv.org/abs/0912.2157}{\tt arXiv:0912.2157}.
%Type = Article
\bibitem[{Freed and Moore(2013)}]{FreedMoore2013}
\bibinfo{author}{D.~S. Freed}, \bibinfo{author}{G.~W. Moore},
\newblock \bibinfo{title}{Twisted equivariant matter},
\newblock \bibinfo{journal}{Ann. H. Poincar{\'e}}  (\bibinfo{year}{2013})
  \bibinfo{pages}{1--97}. \DOIprefix\doi{10.1007/s00023-013-0236-x}.
  \href{http://arxiv.org/abs/1208.5055}{\tt arXiv:1208.5055}.
%Type = Article
\bibitem[{Jackiw and Rebbi(1976)}]{JackiwRebbi1976}
\bibinfo{author}{R.~Jackiw}, \bibinfo{author}{C.~Rebbi},
\newblock \bibinfo{title}{Solitons with fermion number {1/2}},
\newblock \bibinfo{journal}{Phys. Rev. D} \bibinfo{volume}{13}
  (\bibinfo{year}{1976}) \bibinfo{pages}{3398--3409}.
  \DOIprefix\doi{10.1103/PhysRevD.13.3398}.
%Type = Article
\bibitem[{Fu and Kane(2006)}]{FuKane2006}
\bibinfo{author}{L.~Fu}, \bibinfo{author}{C.~L. Kane},
\newblock \bibinfo{title}{Time reversal polarization and a ${Z}_{2}$ adiabatic
  spin pump},
\newblock \bibinfo{journal}{Phys. Rev. B} \bibinfo{volume}{74}
  (\bibinfo{year}{2006}) \bibinfo{pages}{195312}.
  \DOIprefix\doi{10.1103/PhysRevB.74.195312}.
  \href{http://arxiv.org/abs/cond-mat/0606336}{\tt arXiv:cond-mat/0606336}.
%Type = Article
\bibitem[{{Soluyanov} and {Vanderbilt}(2011)}]{SoluyanovVanderbilt2011}
\bibinfo{author}{A.~A. {Soluyanov}}, \bibinfo{author}{D.~{Vanderbilt}},
\newblock \bibinfo{title}{Wannier representation of {Z$_{2}$} topological
  insulators},
\newblock \bibinfo{journal}{Phys. Rev. B} \bibinfo{volume}{83}
  (\bibinfo{year}{2011}) \bibinfo{pages}{035108}.
  \DOIprefix\doi{10.1103/PhysRevB.83.035108}.
  \href{http://arxiv.org/abs/1009.1415}{\tt arXiv:1009.1415}.
%Type = Article
\bibitem[{Lee and Ryu(2008)}]{LeeRyu2008}
\bibinfo{author}{S.-S. Lee}, \bibinfo{author}{S.~Ryu},
\newblock \bibinfo{title}{Many-body generalization of the z$_{2}$ topological
  invariant for the quantum spin {Hall} effect},
\newblock \bibinfo{journal}{Phys. Rev. Lett.} \bibinfo{volume}{100}
  (\bibinfo{year}{2008}) \bibinfo{pages}{186807}.
  \DOIprefix\doi{10.1103/PhysRevLett.100.186807}.
  \href{http://arxiv.org/abs/0708.1639}{\tt arXiv:0708.1639}.
%Type = Article
\bibitem[{{Roy}(2009)}]{Roy2009}
\bibinfo{author}{R.~{Roy}},
\newblock \bibinfo{title}{{Z$_{2}$ classification of quantum spin Hall systems:
  An approach using time-reversal invariance}},
\newblock \bibinfo{journal}{Phys. Rev. B} \bibinfo{volume}{79}
  (\bibinfo{year}{2009}) \bibinfo{pages}{195321}.
  \DOIprefix\doi{10.1103/PhysRevB.79.195321}.
  \href{http://arxiv.org/abs/cond-mat/0604211}{\tt arXiv:cond-mat/0604211}.
%Type = Article
\bibitem[{Qi et~al.(2008)Qi, Hughes, and Zhang}]{QiHughesZhang:2008}
\bibinfo{author}{X.-L. Qi}, \bibinfo{author}{T.~L. Hughes},
  \bibinfo{author}{S.-C. Zhang},
\newblock \bibinfo{title}{Topological field theory of time-reversal invariant
  insulators},
\newblock \bibinfo{journal}{Phys. Rev. B} \bibinfo{volume}{78}
  (\bibinfo{year}{2008}) \bibinfo{pages}{195424}.
  \DOIprefix\doi{10.1103/PhysRevB.78.195424}.
  \href{http://arxiv.org/abs/0802.3537}{\tt arXiv:0802.3537}.
%Type = Article
\bibitem[{Wang et~al.(2010)Wang, Qi, and Zhang}]{WangQiZhang:2010}
\bibinfo{author}{Z.~Wang}, \bibinfo{author}{X.-L. Qi}, \bibinfo{author}{S.-C.
  Zhang},
\newblock \bibinfo{title}{Equivalent topological invariants of topological
  insulators},
\newblock \bibinfo{journal}{New Journal of Physics} \bibinfo{volume}{12}
  (\bibinfo{year}{2010}) \bibinfo{pages}{065007}.
  \DOIprefix\doi{doi:10.1088/1367-2630/12/6/065007}.
  \href{http://arxiv.org/abs/0910.5954}{\tt arXiv:0910.5954}.
%Type = Article
\bibitem[{{Avron} et~al.(1989){Avron}, {Sadun}, {Segert}, and
  {Simon}}]{AvronSadunSegertSimon1989}
\bibinfo{author}{J.~E. {Avron}}, \bibinfo{author}{L.~{Sadun}},
  \bibinfo{author}{J.~{Segert}}, \bibinfo{author}{B.~{Simon}},
\newblock \bibinfo{title}{Chern numbers, quaternions, and {Berry}'s phases in
  {Fermi} systems},
\newblock \bibinfo{journal}{Comm. Math. Phys.} \bibinfo{volume}{124}
  (\bibinfo{year}{1989}) \bibinfo{pages}{595--627}.
  \DOIprefix\doi{10.1007/BF01218452}.
%Type = Article
\bibitem[{Fu(2011)}]{Fu:2011}
\bibinfo{author}{L.~Fu},
\newblock \bibinfo{title}{Topological crystalline insulators},
\newblock \bibinfo{journal}{Phys. Rev. Lett.} \bibinfo{volume}{106}
  (\bibinfo{year}{2011}) \bibinfo{pages}{106802}.
  \DOIprefix\doi{10.1103/PhysRevLett.106.106802}.
  \href{http://arxiv.org/abs/1010.1802}{\tt arXiv:1010.1802}.
%Type = Unpublished
\bibitem[{Turner and Vishwanath(2013)}]{Turner:2013}
\bibinfo{author}{A.~M. Turner}, \bibinfo{author}{A.~Vishwanath},
  \bibinfo{title}{Beyond band insulators: Topology of semi-metals and
  interacting phases}, \bibinfo{year}{2013}.
  \href{http://arxiv.org/abs/1301.0330}{\tt arXiv:1301.0330}.
%Type = Book
\bibitem[{Bourbaki(2006)}]{BourbakiALGCh9}
\bibinfo{author}{N.~Bourbaki}, \bibinfo{title}{Alg{\`e}bre : Chapitre 9},
  \bibinfo{edition}{1} ed., \bibinfo{publisher}{Springer},
  \bibinfo{year}{2006}.
%Type = Book
\bibitem[{Lawson(1985)}]{Lawson}
\bibinfo{author}{J.~H.~B. Lawson}, \bibinfo{title}{The Theory of Gauge Fields
  in Four Dimensions}, \bibinfo{publisher}{American Mathematical Society},
  \bibinfo{year}{1985}.
%Type = Book
\bibitem[{Darling(1994)}]{Darling}
\bibinfo{author}{R.~W.~R. Darling}, \bibinfo{title}{Differential Forms and
  Connections}, \bibinfo{publisher}{Cambridge University Press},
  \bibinfo{year}{1994}.
%Type = Book
\bibitem[{Guillemin and Pollack(1974)}]{GuilleminPollack}
\bibinfo{author}{V.~Guillemin}, \bibinfo{author}{A.~Pollack},
  \bibinfo{title}{Differential Topology}, \bibinfo{edition}{reprint (2010)}
  ed., \bibinfo{publisher}{American Mathematical Society},
  \bibinfo{year}{1974}.
%Type = Book
\bibitem[{Dubrovin et~al.(1985)Dubrovin, Fomenko, and
  Novikov}]{DubrovinFomenkoNovikovII}
\bibinfo{author}{B.~Dubrovin}, \bibinfo{author}{A.~Fomenko},
  \bibinfo{author}{S.~Novikov}, \bibinfo{title}{Modern Geometry - Methods and
  Applications: Part II: The Geometry and Topology of Manifolds},
  \bibinfo{edition}{1985} ed., \bibinfo{publisher}{Springer},
  \bibinfo{year}{1985}.
%Type = Book
\bibitem[{Flanders(1989)}]{Flanders}
\bibinfo{author}{H.~Flanders}, \bibinfo{title}{Differential forms with
  applications to the physical sciences}, Mathematics in science and
  engineering, v. 11, \bibinfo{edition}{2ed.} ed., \bibinfo{publisher}{New
  York, Academic Press}, \bibinfo{year}{1989}.
%Type = Book
\bibitem[{Dinca and Mawhin(2009)}]{DincaMawhin}
\bibinfo{author}{G.~Dinca}, \bibinfo{author}{J.~Mawhin},
  \bibinfo{title}{Brouwer Degree and Applications}, \bibinfo{year}{2009}.
  \URLprefix
  \url{http://www.ljll.math.upmc.fr/~smets/ULM/Brouwer_Degree_and_applications.pdf},
  \bibinfo{note}{to be published}.
%Type = Book
\bibitem[{Bleecker and Booss-Bavnbek(2004)}]{BleeckerBooss}
\bibinfo{author}{D.~Bleecker}, \bibinfo{author}{B.~Booss-Bavnbek},
  \bibinfo{title}{Index Theory with Applications to Mathematics and Physics},
  \bibinfo{year}{2004}. \URLprefix
  \url{http://milne.ruc.dk/~Booss/A-S-Index-Book/BlckBss_2012_04_25.pdf},
  \bibinfo{note}{preprint}.
%Type = Book
\bibitem[{Fulton(2008)}]{Fulton}
\bibinfo{author}{W.~Fulton}, \bibinfo{title}{Algebraic Topology: A First
  Course}, \bibinfo{publisher}{Springer}, \bibinfo{year}{2008}.
%Type = Book
\bibitem[{Ullrich(2008)}]{Ullrich}
\bibinfo{author}{D.~Ullrich}, \bibinfo{title}{Complex Made Simple}, Graduate
  Studies in Mathematics, \bibinfo{publisher}{American Mathematical Society},
  \bibinfo{year}{2008}.

\end{thebibliography}

\end{document}